\documentclass{my_class}
\usepackage{hyperref}
\usepackage{cleveref}
\usepackage{cuted}
\usepackage[normalem]{ulem}
\DeclareGraphicsExtensions{.jpg, .eps, .ps, .pdf}
\graphicspath{{../invited_cpc/figs/}{.}}

\newcommand{\ib}[1]{{\color{blue}#1}}

\title{\raggedright
 The Illusion of Simplicity: The Dramatic Failure of Koopmans Theorem for Antioxidants in Solvents --- The Ascorbic Acid Paradigm
}

\author{
\begin{minipage}{\textwidth}
  Ioan B\^aldea\textsuperscript{*}
\end{minipage}
}

\newcommand{\affiliation}{
\begin{itemize}
\item[{}]
\item[{*}] Dr. Ioan B\^aldea\\
Theoretical Chemistry, Heidelberg University, Im Neuenheimer Feld 229, D-69120 Heidelberg, Germany,
E-mail: ioan.baldea@pci.uni-heidelberg.de
\end{itemize}
}

\renewcommand{\abstract}{
  Antioxidants operate in biological environments where solvent effects dramatically alter their redox properties. Using ascorbic acid (vitamin C) as a paradigmatic example, we present a comprehensive quantum-chemical investigation of its global
  chemical reactivity indices---ionization potential, electron affinity, HOMO-LUMO gap, hardness, softness, electronegativity, electrophilicity, and electrodonating/accepting powers---computed at the compound chemistry CBS-QB3 and various DFT levels in vacuo and across six solvents. The results demonstrate that solvation stabilizes charged species so strongly that reactivity indices shift by several electronvolts, following a roughly Born-like dependence on dielectric constant. Most importantly, we show unequivocally that Koopmans' theorem, often used to estimate these indices from orbital energies, fails catastrophically in solution: it predicts solvent-independent values that are qualitatively and quantitatively wrong, missing the essential physics of dielectric screening and geometric relaxation. We therefore conclude that Koopmans' theorem must be abandoned for antioxidant studies in condensed phases; adiabatic calculations with solvent are mandatory for meaningful predictions.
}

  \newcommand{\keywords}{
      Antioxidants
      \textbullet Solvent Effects and Global Reactivity Indices
      \textbullet Ascorbic Acid
      \textbullet Koopmans Theorem
      \textbullet Compound Chemistry Models and Density Functional Theory
  }

\begin{document}

\twocolumn[\vspace{-1.5cm}\maketitle\vspace{-1cm}]
\small{\begin{shaded}
		\noindent\abstract
	\end{shaded}
}

\begin{figure} [!b]
\begin{minipage}[t]{\columnwidth}{\rule{\columnwidth}{1pt}\footnotesize{\textsf{\affiliation}}}\end{minipage}
\end{figure}

\section{Introduction}
\label{sec:intro}

Reactivity descriptors are crucial for understanding antioxidants, which protect biological systems
by neutralizing free radicals.
Global chemical reactivity indices---such as chemical hardness ($\eta$) and softness ($\sigma$), electrophilicity ($\omega$),
electronegativity ($\chi$) and related chemical potential ($\mu = - \chi$),
electroaccepting and electrodonating powers ($\omega^{\pm}$)---provide valuable  insights into the electron donation and acceptance capabilities of molecules.\cite{LaraGuzman:18,Galano:19,Galano:25a,Galano:25b}

All these quantities are expressed in terms of the ionization potential (IP) and electron affinity (EA),
for which Koopmans' theorem \cite{Koopmans:33} seems to be particularly appealing
because of its minimal computational cost.
It only implies the determination of the highest occupied and lowest unoccupied molecular orbital (HOMO, LUMO)---whose energies
with reversed sign are claimed to approximate IP and EA---
accounting for the motion of an electron in the average self-consistent field (SCF) of the other electrons.
Routinely, this is done within various flavors of the DFT---with extension through Janak's theorem \cite{Slater:72a,Janak:78}---,
which nowadays replaced the Hartree-Fock (HF) approximation originally suggested by Koopmans \cite{Koopmans:33}---
for a neutral molecule, which is typically a closed-shell system.
This obviates the dilemma facing conceptual DFT of choosing between unrestricted
or restricted open shell methods posed by considering cations or anions.
Because the latter are typically open shell systems, they often require
SCF quadratic convergence approaches which are more computationally demanding than standard SCF procedures.

Koopmans' theorem,\cite{Koopmans:33} while historically important for interpreting Hartree-Fock (HF) the highest occupied (HOMO) and lowest unoccupied
(LUMO) orbital energies as ionization potentials (IP) and electron affinities (EA), suffers from several fundamental approximations:

Koopmans theorem is based on a frozen orbital approximation. It
assumes no electron relaxation upon ionization/electron attachment. In reality,
e.g., the remaining electrons after electron removal rearrange, lowering the total energy of the ionized state.

Furthermore, it neglects geometry relaxation.
Molecular geometry changes after ionization/electron attachment are ignored.
Bond lengths and angles typically adjust, affecting total energies beyond the frozen-orbital picture.

Nevertheless, Koopmans theorem continues to be very often utilized for studies on antioxidants, where, in addition to the aforementioned shortcomings, it faces an extra difficulty: the vacuum reference state is physically inappropriate. In biological contexts, the electron escaping after ionization is released into the solvent, where its solvation enthalpy differs significantly from that in vacuo.

The theorem inherently assumes the electron is added to or removed from vacuum. In realistic antioxidant action, however, the ionization process in solution involves the equilibrium:
\[
\text{AO(solv)} \rightarrow \text{AO}^{+}\text{(solv)} + e^{-}\text{(solv)},
\]
where \( e^{-}\text{(solv)} \) denotes the solvated electron. Koopmans' theorem, in contrast, describes:
\[
\text{AO(gas)} \rightarrow \text{AO}^{+}\text{(gas)} + e^{-}\text{(vac)},
\]
thereby omitting solvent reorganization, dielectric screening, and the solvation free energies of both the ion and the electron.

This is a critical aspect to consider particularly because antioxidants primarily function in aqueous or polar environments,
which can significantly alter reactivity indices through stabilization of charged species and modification of orbital energies.\cite{Tomasi:05,Baldea:2013b}

The lack of systematic quantum chemical investigations on the solvents' impact of chemical reactivity in general,
and critical analysis of the applicability of the Koopmans theorem to molecules in solvents
in particular were the main reasons that triggered the present study.

To illustrate these critical issues, we select ascorbic acid (\ce{AscH}, \ce{C6H8O6}, vitamin C) as a paradigmatic case study. This choice is motivated by its dual role as a prototypical natural antioxidant of paramount biological importance and the surprising scarcity of systematic studies on its reactivity indices across different environments.

\section{Methodology}

\subsection{Thermodynamic Descriptors}

Calculations with GAUSSIAN 16 \cite{g16} were performed using the CBS-QB3 compound chemistry model
and DFT using Truhlar's M062x \cite{Truhlar:06,Truhlar:08}  and
the hybrid B3LYP exchange correlation functional \cite{Parr:88,Becke:88,Becke:93a,Frisch:94}
in combination with various
Pople basis sets (6-311++G(3df,3pd), 6-311+G(3df,3pd),
6-311G\\(3df,3pd), and 6-31+G(d,p)).\cite{Petersson:88,Petersson:91}

To preserve this comparability with the majority of existing work, we adopted IEFPCM for all geometry optimizations, harmonic frequency calculations, and evaluation of thermodynamic properties. IEFPCM treats the solvent as a continuum dielectric and computes electrostatic solute-solvent interactions self-consistently through an apparent surface charge distribution on a molecular cavity. While the more recent Truhlar's SMD model \cite{Truhlar:08a, Truhlar:08b,Truhlar:09} generally improves agreement with experimental solvation free energies by adding parameterized non-electrostatic terms (cavitation, dispersion, and short-range solvent structure), we retained IEFPCM to ensure direct alignment with the bulk of prior antioxidant studies.
Parenthetically, similar to the case of molecules containing aromatic rings studied earlier,\cite{Baldea:2013b,Baldea:2022f,Baldea:2024b}
we did not find notable trend differences between IEFPCM and SMD.\cite{Truhlar:08a, Truhlar:08b,Truhlar:09}

Unlike many literature studies, full geometry and frequency calculations were carried out
for each method (CBS-QB3, B3LYP, M06-2X),
basis sets (6-311++G(3df,3pd), 6-311+G(3df,3pd), 6-311G(3df,3pd), and 6-31+G(d,p)),
environment (vacuum plus the five solvents listed above), and redox state (neutral, cation, anion).
  For instance, we did not use the optimized geometry at M06-2X/6-31+G(d,p) and zero point corrections computed in vacuo for obtaining IP and EA in vacuo from single-point M06-2X/6-311++G(3df,3pd) calculations, nor for estimating their values in water from single-point M06-2X/6-31+G(d,p)/IEFPCM calculations.

Computations for cation (\ce{AscH+}) and anion (\ce{AscH-}), which are open shell systems, were carried out using unrestricted spin methods (UCBS-QB3, UB3LYP, UM062X). Spin contamination was never an issue: annihilation of the first spin contaminant yielded values $\sim 0.76$ before annihilation (versus the exact 0.75), while those after were invariably 0.7500.

Adiabatic IP and EA were calculated as enthalpy ($H$) differences at the corresponding optimum geometries
($\mathbf{R}_{0\varepsilon}, \mathbf{R}_{\pm \varepsilon}$, where $\varepsilon = \text{vacuum, solvent}$)

\begin{strip}
\begin{eqnarray}
  \label{eq-IP}
  \mbox{\small IP}_{\varepsilon} & \equiv & \mbox{\small IP}_{\varepsilon}^{\text{adiab}} = H\left(\left . \ce{{Asc}H^{\bullet +}}\right\vert {\mathbf{R_{\varepsilon}^+},\varepsilon}\right) + H\left(\left . \ce{e-}\right\vert {\varepsilon}\right) -
  H\left(\left . \ce{{Asc}H}\right\vert {\mathbf{R_{\varepsilon}^0},\varepsilon}\right) \\
  \label{eq-EA}
  \mbox{\small EA}_{\varepsilon} & \equiv & \mbox{\small EA}_{\varepsilon}^{\text{adiab}} =  H\left(\left . \ce{{Asc}H}\right\vert {\mathbf{R_{\varepsilon}^0},\varepsilon}\right) +
  H\left(\left . \ce{e-}\right\vert {\varepsilon}\right) -
  H\left(\left . \ce{{Asc}H^{\bullet -}}\right\vert {\mathbf{R_{\varepsilon}^-},\varepsilon}\right) \\
  \label{eq-Eg}
E_{g,\varepsilon} & \equiv & E_{g \varepsilon}^{\text{adiab}} = \mbox{\small IP}_{\varepsilon} - \mbox{\small EA}_{\varepsilon}
= H\left(\left . \ce{{Asc}H^{\bullet +}}\right\vert {\mathbf{R_{\varepsilon}^+},\varepsilon}\right) +
H\left(\left . \ce{{Asc}H^{\bullet -}}\right\vert {\mathbf{R_{\varepsilon}^-},\varepsilon}\right)
- 2  H\left(\left . \ce{{Asc}H}\right\vert {\mathbf{R_{\varepsilon}^0},\varepsilon}\right)
\end{eqnarray}
\end{strip}

\begin{strip}
\begin{eqnarray}
  \label{eq-IP_vert}
  \mbox{\small IP}_{\varepsilon}^{\text{vert}} & = & H\left(\left . \ce{{Asc}H^{\bullet +}}\right\vert {\mathbf{R_{\varepsilon}^0},\varepsilon}\right) + H\left(\left . \ce{e-}\right\vert {\varepsilon}\right) -
  H\left(\left . \ce{{Asc}H}\right\vert {\mathbf{R_{\varepsilon}^0},\varepsilon}\right) \\
    \label{eq-EA_vert}
  \mbox{\small EA}_{\varepsilon}^{\text{vert}} & = & H\left(\left . \ce{{Asc}H}\right\vert {\mathbf{R_{\varepsilon}^0},\varepsilon}\right) +
  H\left(\left . \ce{e-}\right\vert {\varepsilon}\right) -
  H\left(\left . \ce{{Asc}H^{\bullet -}}\right\vert {\mathbf{R_{\varepsilon}^0},\varepsilon}\right) \\
    \label{eq-Eg_vert}
E_{g,\varepsilon}^{\text{vert}} & = & \mbox{\small IP}_{\varepsilon}^{\text{vert}} - \mbox{\small EA}_{\varepsilon}^{\text{vert}}
= H\left(\left . \ce{{Asc}H^{\bullet 0}}\right\vert {\mathbf{R_{\varepsilon}^0},\varepsilon}\right) +
H\left(\left . \ce{{Asc}H^{\bullet -}}\right\vert {\mathbf{R_{\varepsilon}^0},\varepsilon}\right)
- 2  H\left(\left . \ce{{Asc}H}\right\vert {\mathbf{R_{\varepsilon}^0},\varepsilon}\right)
\end{eqnarray}
\end{strip}
where
\begin{equation}
  \label{eq-e-sol}
  H\left(\left . \ce{e-}\right\vert {\varepsilon}\right) =
  H_{\text{vacuum}}\left(\left . \ce{e-}\right\vert {\varepsilon}\right) + \Delta H_{\text{sol}}\left(\left . \ce{e-}\right\vert {\varepsilon}\right)
\end{equation}

For electron, in our calculations we set the vacuum enthalpy to
$ H_{\text{vacuum}}\left(\left . \ce{e-}\right\vert {\varepsilon}\right) = 3.1351$\,kJ/mol \cite{Fifen:13}
and used literature values of $ \Delta H_{\text{sol}}\left(\left . \ce{e-}\right\vert {\varepsilon}\right) $ for the solvents considered.\cite{Markovic:16}

For the discussion that follows, it is important to note that neither the adiabatic $E_{g,\varepsilon}$ nor the vertical $E_{g,\varepsilon}^{\text{vert}}$
depend on the electron's solvation enthalpy $\Delta H_{\text{sol}}\left(\left . \ce{e-}\right\vert {\varepsilon}\right)$
(cf.~eqs.~(\ref{eq-Eg}) and \ref{eq-Eg_vert})).
All quantities include corrections due to zero-point motion and thermal corrections at 298.15\,K.

To compute
$H\left(\left . \ce{{Asc}H^{\bullet \pm}}\right\vert {\mathbf{R_{\varepsilon}^0},\varepsilon}\right)$
zero-point corrections at the corresponding optimum geometries ($\mathbf{R_{\varepsilon}^{\pm},\varepsilon}$) were added
to the SCF electronic energy obtained from single-point calculations at neutral's optimum $\mathbf{R_{\varepsilon}^0}$; frequency calculations
can only be done at the pertaining energy minimum.

For completeness, we also computed the (``adiabatic'') Gibbs free energies of solvation
\begin{equation}
  \Delta G_{\text{sol},\varepsilon}^{x} =
  G\left(\left . \ce{{Asc}H^{x}}\right\vert {\mathbf{R_{\varepsilon}^x},\varepsilon}\right) -
   G\left(\left . \ce{{Asc}H^{x}}\right\vert {\mathbf{R_{\text{vacuum}}^x},\text{vacuum}}\right)
\end{equation}

\subsection{Global Chemical Reactivity Descriptors}
\label{sec:eta}
To characterize the intrinsic electron-donating and electron-accepting tendencies of ascorbic acid in different chemical environments, we employ a set of well-established global reactivity indices. These descriptors are derived from the ionization potential (IP) and electron affinity (EA) and provide complementary information about the molecule's stability, polarizability, and redox behavior \cite{Parr:89,Gazquez:07,Rajan:18,Baldea:2019f,Baldea:2019g}.

The complete set of indices investigated in this work is given by the following expressions, where $E_g = \text{IP} - \text{EA}$ is
fundamental (or transport or charge) ``HOMO-LUMO''
gap \cite{Parr:89,Burke:12,Baldea:2014c,Baldea:2008,Baldea:2012i}

\begin{equation}
\begin{aligned}
\eta &= \frac{E_g}{2} && \text{(chemical hardness)} \\
\sigma &= \frac{1}{\eta} = \frac{2}{E_g} && \text{(chemical softness)} \\
\chi &= \frac{\text{IP} + \text{EA}}{2} && \text{(Mulliken electronegativity)} \\
\omega &= \frac{\chi^2}{2\eta} && \text{(electrophilicity index)} \\
\omega^+ &= \frac{(\text{IP} + 3\,\text{EA})^2}{16  (\text{IP} - \text{EA}) } && \text{(electroaccepting power)} \\
\omega^- &= \frac{(3\,\text{IP} + \text{EA})^2}{16 (\text{IP} - \text{EA})} && \text{(electrodonating power)}
\end{aligned}
\label{eq:reactivity-indices}
\end{equation}

The rationale for including this rather comprehensive set is the following:

- Hardness ($\eta$) and softness ($\sigma$) reflect the molecule's resistance or propensity to undergo changes in electron density --- a property particularly relevant for antioxidants that must rapidly donate or accept electrons without large energetic cost.

- Electronegativity ($\chi$) indicates the thermodynamic driving force for electron transfer in interactions with other species, while the chemical potential ($\mu = -\chi$) represents a form of ``electronic pressure'': if two molecules come into contact, electrons will flow from the molecule with higher (less negative) $\mu$ (``nucleophiles'' with electrons held loosely) to the one with lower (more negative) $\mu$ (``electrophiles'' having ``suction'' or ``pull'' for electrons from other systems).

- The electrophilicity index ($\omega$) quantifies the overall capacity to accept electrons and is especially useful when comparing the relative strength of different antioxidants.

- The electroaccepting and electrodonating powers $\omega^+$ and $\omega^-$ allow a more nuanced description,
namely of a fractional charge donation and acceptance in weak interactions. This is important for understanding the behavior of ascorbic acid in biological media where it often participates in partial charge-transfer processes rather than complete redox events.

These indices collectively offer a robust framework for assessing the intrinsic stability and reactivity of a molecule, as well as its tendency to participate in charge-transfer processes.
Taken together, these indices provide a multidimensional picture of reactivity that cannot be captured by IP or EA alone.
In particular, they have proven useful for comparative studies of molecular species, enabling the prediction of electron flow direction and relative stability in chemical environments \cite{Parr:99,Gazquez:07,Gazquez:08}.
They are particularly valuable when solvent effects are included, as the stabilization of charged species (cation, anion) can dramatically alter the values and relative importance of donation vs. acceptance pathways. The full set is therefore essential for a rigorous assessment of ascorbic acid's antioxidant performance across environments, and for a meaningful comparison with other compounds reported in the literature.
\section{Results}
\label{sec:results}
\Cref{tab:reactivity_vacuum,tab:reactivity_water} collect the values of the global reactivity indices
computed using the adiabatic values of IP and EA for ascorbic acid in vacuo and in water, respectively.
Results for the other five solvents
are presented in \Cref{tab:reactivity_benzene,tab:reactivity_toluene,tab:reactivity_chlorobenzene,tab:reactivity_methanol,tab:reactivity_ethanol}.
\Cref{tab:solvation_water} provides data for Gibbs free energies of hydration; Gibbs free energy of solvation for
the other solvents are collected in \Cref{tab:solvation_benzene,tab:solvation_toluene,tab:solvation_chlorobenzene,tab:solvation_methanol,tab:solvation_ethanol}.
\begin{table*}[htbp]
\centering
\caption{Global reactivity descriptors for ascorbic acid in \textbf{vacuo}. Values in eV,
  Upper subrow: absolute values; lower subrow: signed deviation from CBS-QB3.}
\label{tab:reactivity_vacuum}
\setlength{\tabcolsep}{4.5pt}
\scriptsize
\begin{tabular}{lrrrrrrrrr}
\hline
Method & IP & EA & $E_g$ & $\eta$ & $\sigma$ & $\chi$ & $\omega$ & $\omega^+$ & $\omega^-$ \\
\hline
\textbf{CBS-QB3} & 8.524 & -0.179 & 8.703 & 4.351 & 0.115 & 4.172 & 2.000 & 0.458 & 4.630 \\
 & 0.000 & 0.000 & 0.000 & 0.000 & 0.000 & 0.000 & 0.000 & 0.000 & 0.000  \\
B3LYP/\allowbreak 6-311++G(3df,3pd) & 8.242 & 0.063 & 8.179 & 4.090 & 0.122 & 4.152 & 2.108 & 0.543 & 4.695 \\
 & -0.282 & +0.242 & -0.523 & -0.262 & +0.007 & -0.020 & +0.108 & +0.085 & +0.065  \\
B3LYP/\allowbreak 6-311+G(3df,3pd) & 8.242 & -0.080 & 8.322 & 4.161 & 0.120 & 4.081 & 2.001 & 0.481 & 4.561  \\
 & -0.282 & +0.099 & -0.380 & -0.190 & +0.005 & -0.091 & +0.001 & +0.023 & -0.069  \\
B3LYP/\allowbreak 6-311G(3df,3pd) & 8.087 & -0.420 & 8.508 & 4.254 & 0.118 & 3.833 & 1.727 & 0.342 & 4.176 \\
 & -0.436 & -0.241 & -0.195 & -0.098 & +0.003 & -0.339 & -0.273 & -0.116 & -0.455 \\
B3LYP/\allowbreak 6-31+G(d,p) & 8.245 & -0.016 & 8.261 & 4.130 & 0.121 & 4.115 & 2.049 & 0.508 & 4.623  \\
 & -0.278 & +0.163 & -0.442 & -0.221 & +0.006 & -0.058 & +0.049 & +0.050 & -0.007  \\
M062X/\allowbreak 6-311++G(3df,3pd) & 8.386 & -0.227 & 8.613 & 4.307 & 0.116 & 4.080 & 1.932 & 0.431 & 4.510 \\
 & -0.137 & -0.048 & -0.090 & -0.045 & +0.001 & -0.093 & -0.068 & -0.027 & -0.120  \\
M062X/\allowbreak 6-311+G(3df,3pd) & 8.386 & -0.232 & 8.618 & 4.309 & 0.116 & 4.077 & 1.929 & 0.429 & 4.506  \\
 & -0.137 & -0.052 & -0.085 & -0.042 & +0.001 & -0.095 & -0.071 & -0.029 & -0.124  \\
M062X/\allowbreak 6-311G(3df,3pd) & 8.261 & -0.495 & 8.757 & 4.378 & 0.114 & 3.883 & 1.722 & 0.328 & 4.211  \\
 & -0.262 & -0.316 & +0.054 & +0.027 & -0.001 & -0.289 & -0.278 & -0.130 & -0.419  \\
M062X/\allowbreak 6-31+G(d,p) & 8.368 & -0.172 & 8.540 & 4.270 & 0.117 & 4.098 & 1.967 & 0.451 & 4.549 \\
 & -0.156 & +0.007 & -0.163 & -0.081 & +0.002 & -0.074 & -0.034 & -0.007 & -0.081 \\
\hline
\end{tabular}
\end{table*}

\begin{table*}[htbp]
\centering
\caption{Global reactivity descriptors and solvation Gibbs free energies for ascorbic acid in \textbf{water}. Reactivity descriptors in eV, $\Delta G_{sol}$ in kcal/mol. Upper subrow: absolute values; lower subrow: signed deviation from CBS-QB3.}
\label{tab:reactivity_water}
\setlength{\tabcolsep}{4.5pt}
\scriptsize
\begin{tabular}{lrrrrrrrrrrrr}
\hline
Method & IP & EA & $E_g$ & $\eta$ & $\sigma$ & $\chi$ & $\omega$ & $\omega^+$ & $\omega^-$ & $\Delta G_{sol}^{n}$ & $\Delta G_{sol}^{c}$ & $\Delta G_{sol}^{a}$ \\
\hline
\textbf{CBS-QB3} & 5.586 & 0.826 & 4.760 & 2.380 & 0.210 & 3.206 & 2.159 & 0.854 & 4.059 & -10.928 & -54.496 & -58.306 \\
 & 0.000 & 0.000 & 0.000 & 0.000 & 0.000 & 0.000 & 0.000 & 0.000 & 0.000 & 0.000 & 0.000 & 0.000 \\
B3LYP/\allowbreak 6-311++G(3df,3pd) & 5.284 & 0.884 & 4.400 & 2.200 & 0.227 & 3.084 & 2.161 & 0.894 & 3.978 & -11.407 & -55.243 & -55.084 \\
 & -0.302 & +0.058 & -0.360 & -0.180 & +0.017 & -0.122 & +0.002 & +0.041 & -0.081 & -0.479 & -0.747 & +3.222 \\
B3LYP/\allowbreak 6-311+G(3df,3pd) & 5.284 & 0.883 & 4.401 & 2.200 & 0.227 & 3.083 & 2.160 & 0.894 & 3.977 & -11.409 & -55.245 & -57.653 \\
 & -0.302 & +0.057 & -0.359 & -0.180 & +0.017 & -0.122 & +0.001 & +0.040 & -0.082 & -0.481 & -0.749 & +0.653 \\
B3LYP/\allowbreak 6-311G(3df,3pd) & 5.136 & 0.598 & 4.538 & 2.269 & 0.220 & 2.867 & 1.812 & 0.662 & 3.529 & -10.144 & -53.853 & -57.731 \\
 & -0.450 & -0.227 & -0.222 & -0.111 & +0.010 & -0.339 & -0.347 & -0.192 & -0.531 & +0.784 & +0.643 & +0.575 \\
B3LYP/\allowbreak 6-31+G(d,p) & 5.302 & 0.952 & 4.349 & 2.175 & 0.230 & 3.127 & 2.248 & 0.957 & 4.084 & -12.725 & -56.148 & -59.115 \\
 & -0.284 & +0.127 & -0.411 & -0.205 & +0.020 & -0.079 & +0.089 & +0.103 & +0.024 & -1.797 & -1.652 & -0.809 \\
M062X/\allowbreak 6-311++G(3df,3pd) & 5.437 & 0.804 & 4.633 & 2.316 & 0.216 & 3.121 & 2.102 & 0.831 & 3.952 & -11.180 & -55.114 & -59.207 \\
 & -0.149 & -0.021 & -0.128 & -0.064 & +0.006 & -0.085 & -0.057 & -0.022 & -0.107 & -0.252 & -0.618 & -0.901 \\
M062X/\allowbreak 6-311+G(3df,3pd) & 5.437 & 0.804 & 4.633 & 2.317 & 0.216 & 3.120 & 2.102 & 0.831 & 3.951 & -11.180 & -55.116 & -59.305 \\
 & -0.149 & -0.022 & -0.127 & -0.063 & +0.006 & -0.085 & -0.057 & -0.023 & -0.108 & -0.252 & -0.620 & -0.999 \\
M062X/\allowbreak 6-311G(3df,3pd) & 5.319 & 0.573 & 4.746 & 2.373 & 0.211 & 2.946 & 1.828 & 0.652 & 3.598 & -10.188 & -53.965 & -59.017 \\
 & -0.267 & -0.253 & -0.014 & -0.007 & +0.001 & -0.260 & -0.331 & -0.202 & -0.462 & +0.740 & +0.531 & -0.711 \\
M062X/\allowbreak 6-31+G(d,p) & 5.432 & 0.868 & 4.564 & 2.282 & 0.219 & 3.150 & 2.174 & 0.885 & 4.035 & -12.472 & -56.168 & -60.741 \\
 & -0.154 & +0.042 & -0.196 & -0.098 & +0.009 & -0.056 & +0.015 & +0.031 & -0.025 & -1.544 & -1.672 & -2.435 \\
\hline
\end{tabular}
\end{table*}

DFT-based values of IP$_{\text{vert}}$ and EA$_{\text{vert}}$
are reported in
\Cref{tab:ip_ea_vacuum,tab:ip_ea_water,tab:ip_ea_benzene,tab:ip_ea_toluene,tab:ip_ea_chlorobenzene,tab:ip_ea_methanol,tab:ip_ea_ethanol}.

\begin{table*}[htbp]
\centering
\caption{Adiabatic and vertical IP/EA (eV) in vacuo ($\varepsilon$ = 1.00).
  Notice the substantial differences between the adiabatic and vertical values
  (equal to the cation's and anion's reorganization energies
  $\lambda_{c} = \text{IP}_{\text{vert}} - \text{IP}_{\text{adiab}}$,
  $\lambda_{a} = \text{EA}_{\text{adiab}} - \text{EA}_{\text{vert}}$, cf.~\Cref{fig:lambdas}) correlating with the
  significant bond lengths' change upon electron removal/attachment.}
\label{tab:ip_ea_vacuum}
\begin{tabular}{lcccc}
\hline
Method & IP adiab & IP vert & EA adiab & EA vert \\
\hline
M062X/\allowbreak 6-31+G(d,p) & 8.368 & 8.874 & -0.172 & -0.621 \\
B3LYP/\allowbreak 6-31+G(d,p) & 8.245 & 8.673 & -0.016 & -0.243 \\
B3LYP/\allowbreak 6-311++G(3df,3pd) & 8.242 & 8.673 & 0.063 & -0.005 \\
M062X/\allowbreak 6-311++G(3df,3pd) & 8.386 & 8.900 & -0.227 & -0.297 \\
\hline
\end{tabular}
\end{table*}

\begin{table*}[htbp]
\centering
\caption{Adiabatic and vertical IP/EA (eV) in water ($\varepsilon$ = 78.36).
  Notice the large differences between the IP and adiabatic and vertical values
  (equal to the cation's and anion's reorganization energies, cf.~\Cref{fig:lambdas})
  correlating with the significant bond lengths' change upon electron removal/attachment.}
\label{tab:ip_ea_water}
\begin{tabular}{lcccc}
\hline
Method & IP adiab & IP vert & EA adiab & EA vert \\
\hline
M062X/\allowbreak 6-31+G(d,p) & 5.432 & 5.832 & 0.868 & 0.224 \\
B3LYP/\allowbreak 6-31+G(d,p) & 5.302 & 5.638 & 0.952 & 0.359 \\
B3LYP/\allowbreak 6-311++G(3df,3pd) & 5.284 & 5.629 & 0.884 & 0.307 \\
M062X/\allowbreak 6-311++G(3df,3pd) & 5.437 & 5.848 & 0.804 & 0.172 \\
\hline
\end{tabular}
\end{table*}

Although there are many studies on ascorbic acid, comparing the present results with experiment
is problematic.\cite{Niki:91,Du:12,Matsui:15,Ricci:16,Abyar:16,Cimino:18,Levine:20,Liu:20b,Dosedel:21,Marshall:21}
Direct experimental determination of gas-phase ionization potentials and electron affinities for ascorbic acid remains challenging due to its thermal instability and propensity for decomposition upon vaporization. Consequently, no direct gas-phase IP or EA measurements are available in standard databases (e.g., NIST).

While electrochemical studies in aqueous solution provide valuable redox potentials \cite{Niki:91,Buettner:93}, converting these to precise gas-phase values through thermochemical cycles requires accurate solvation energies for both the neutral and ionic species, introducing substantial uncertainties. Such conversions, while sometimes attempted in the literature, must be treated with caution due to the cumulative errors in the various terms involved.

Rather than attempting precise quantitative comparisons with indirect experimental data, we emphasize the qualitative trends: our computed gas-phase adiabatic IP ($\sim$8.5\,eV) and EA ($\sim -0.2$\,eV) fall within the expected range for organic molecules of this size and functionalization. More importantly, the dramatic solvent-induced shifts---most notably the $\sim$3\,eV reduction in IP when moving from vacuum to water---are physically intuitive and align with the well-known strong stabilization of charged species by polar solvents. These trends, which are the primary focus of this work, are robust and method-independent, as confirmed by the consistency across our CBS-QB3 and DFT calculations.

\section{Discussion}
\subsection{Computational Aspects}
\label{sec:computations}
The tested DFT functionals exhibit markedly different performance when benchmarked against the most elaborate CBS-QB3 reference (\Cref{tab:ip_dev,tab:ea_dev,tab:eta_dev,tab:sigma_dev,tab:devs_chi,tab:omega_dev,tab:omegap_dev,tab:omegam_dev,tab:solvation_dev,tab:sol_neutral_dev,tab:sol_cation_dev,tab:sol_anion_dev}).

Among the global chemical reactivity indices (IP, EA, $E_g$, $\eta$, $\sigma$, $\chi$, $\omega$, $\omega^+$, $\omega^-$), the M062X functionals demonstrate superior agreement with CBS-QB3. In particular, M062X/6-31+G(d,p) achieves the lowest mean absolute deviation (MAD $\approx$ 0.10 eV) and maximum absolute deviation (MaxAD $\approx$ 0.21 eV) when averaged over all descriptors and solvents, closely followed by M062X/6-311+G(3df,3pd). By contrast, B3LYP/6-311G(3df,3pd) shows the poorest performance, with an MAD of approximately 0.32 eV and MaxAD approaching 0.68 eV---errors that are most pronounced in polar solvents.

Overall, the M062X family, particularly with moderate basis sets such as 6-31+G(d,p), provides the most consistent and
accurate description of both electronic reactivity indices and thermodynamic solvation properties for ascorbic acid
in the studied continuum solvent models when benchmarked against CBS-QB3.

\subsection{HOMO and LUMO Spatial Distributions}
\label{sec:homo-lumo}

\begin{figure*}
  \centerline{
        \includegraphics[width=0.3\textwidth,angle=0]{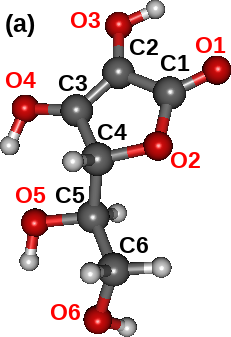}
    \includegraphics[width=0.3\textwidth,angle=0]{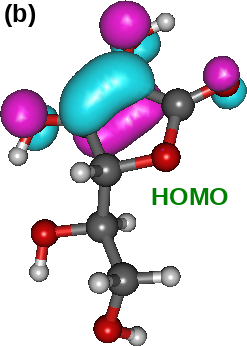}}
  \centerline{
    \includegraphics[width=0.3\textwidth,angle=0]{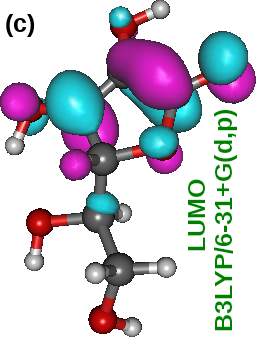}
    \includegraphics[width=0.3\textwidth,angle=0]{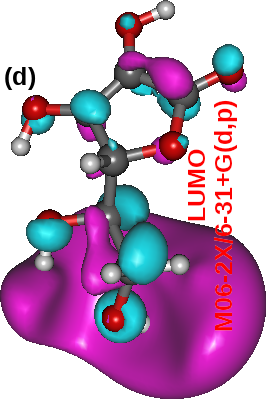}
    \includegraphics[width=0.3\textwidth,angle=0]{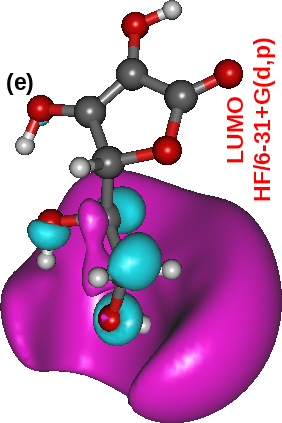}
  }
  \caption{(a) Optimized geometry of ascorbic acid with IUPAC atom numbering (indistinguishable within drawing accuracy across all methods employed).
(b) HOMO spatial distribution (also indistinguishable within drawing accuracy across all methods), concentrated along the \ce{C2=C3} bond, with significant contributions from the adjacent oxygen atoms \ce{O3} and \ce{O4}, and to a lesser extent from \ce{O1} (bonded to \ce{C1}).
LUMO spatial distributions computed at (c) B3LYP/6-31+G(d,p), (d) M062X/6-31+G(d,p), \textcolor{black}{and (e) HF/6-31+G(d,p).}
Note the marked contrast: the B3LYP LUMO shows substantial delocalization over the lactone-enediol conjugated system (major contributions from \ce{C1-C2} and \ce{C3}), whereas the M062X LUMO is nonphysically dominated by the side chain, \textcolor{black}{as is also the case for the Hartree-Fock LUMO.}}
      \label{fig:vitC}
\end{figure*}

Ascorbic acid (vitamin C, \ce{C6H8O6}) is a water-soluble organic compound whose core structure is a five-membered dihydrofuran-2-one ring (a $\gamma$-lactone), formed by the atoms \ce{C1}, \ce{O2}, \ce{C4}, \ce{C3}, and \ce{C2} in cyclic order
(IUPAC numbering, \Cref{fig:vitC}a).
Within this ring, the enediol moiety is located at \ce{C2=C3},
a carbon double bond flanked by two hydroxy groups on adjacent carbons
(\ce{C2(O3H)=C3(O4H)}), conferring the molecule's strong reducing properties. Attached to \ce{C4} is a chiral diol side chain (-\ce{CH(OH)CH2OH}) at \ce{C5} and \ce{C6}.
The molecule contains two chiral centers at \ce{C4} and \ce{C5}, with the biologically active form being the L-enantiomer. This combination of the conjugated enediol and lactone system enables reversible redox cycling, underpinning ascorbic acid's role as a biological antioxidant and electron donor.

The computed HOMO and LUMO distributions (\Cref{fig:vitC}b,c) provide insight into the electronic structure of ascorbic acid,
highlighting the role of the enediol moiety (a carbon double bond flanked by two hydroxy groups on adjacent carbons)
in its antioxidant activity. The HOMO is predominantly a $\pi$-bonding orbital localized along the enediol \ce{C=C} bond (C2--C3 in standard numbering),
with additional contributions from adjacent oxygen atoms (O1, O3, and O4). This distribution indicates that electron donation (oxidation) is facilitated by the conjugated enediol system, where the $\pi$ electrons are readily available for transfer.

When interpreting the LUMO, a cautionary note is warranted. In Kohn--Sham (KS) density functional theory---as in Hartree--Fock theory---the virtual orbitals are mathematical constructs determined solely by orthogonality to the occupied manifold and lack direct physical significance.\cite{Parr:89,Gunnarson:89,Kohn:96,HaugJauho,Baldea:2014c} (The sole exception is the KS HOMO, whose eigenvalue would equal the negative ionization energy if the exact exchange--correlation functional were used.\cite{Kohn:96})
Nevertheless, with basis sets not too large while still including diffuse functions---essential for describing the spatial extension of an added electron---and when guided by chemical intuition, the LUMO can offer a physically meaningful picture of electron-accepting tendencies, particularly for anionic states.

\Cref{fig:vitC}c--e, where LUMO densities computed in several ways are depicted,
illustrate how critical the method utilized influences the LUMO representation.
While the M06-2X/6-31+G(d,p) (\Cref{fig:vitC}d) and HF/6-31+G(d,p) (\Cref{fig:vitC}e) densities appear nonphysically extended over the aliphatic side chain, the B3LYP/6-31+G(d,p) LUMO (\Cref{fig:vitC}c) provides a chemically reasonable description. It is delocalized over the carbonyl--enediol conjugated framework, with moderate contributions from the enediol oxygen atoms (O4 and O3). This picture is consistent with a $\pi^*$ antibonding orbital that spans the conjugated system, reflecting the molecule's capacity to stabilize additional electron density via delocalization.

The orbital picture aligns with the computed thermodynamic and solvation data. The anion (AscH$^-$) exhibits the most negative $\Delta G_{\text{sol}}$ (ca. $-58$ kcal/mol in water, cf.~\Cref{tab:solvation_water}), indicating strong stabilization of the delocalized negative charge by the polar solvent. The cation (AscH$^+$) also gains substantial solvation energy (ca. $-54$ kcal/mol), consistent with the HOMO's delocalized nature facilitating charge separation upon oxidation. These features are reflected in the aqueous ionization potential and electron affinity: the low IP ($\sim$5.6~eV, equivalent to $\sim$129~kcal/mol) and positive EA ($\sim$0.83~eV, equivalent to $\sim$19~kcal/mol) (cf.~\Cref{tab:reactivity_water}) confirm the ease of electron loss from the HOMO and the stability of the anion, respectively. The solvent-induced reduction of the IP by $\sim$3.0~eV ($\sim$69~kcal/mol) when moving from vacuum to water (cf.~\Cref{tab:reactivity_vacuum,tab:reactivity_water}) underscores how aqueous solvation markedly enhances the molecule's electron-donating capacity, a key factor in its biological antioxidant function.

\subsection{Bond Lengths' Change upon Electron Removal/Attachment}
\label{sec:metrics}

The orbital distributions have direct structural consequences, which are quantified by the bond length changes upon electron addition or removal. \Cref{tab:bond_changes_vacuo,tab:bond_changes_water} present selected bond lengths for neutral, cationic, and anionic ascorbic acid in both vacuum and aqueous solution. The most pronounced changes occur in bonds that are integral to the conjugated $\pi$-system identified in the HOMO and LUMO, while bonds peripheral to this network---such as \ce{C4-C5} in the side chain---exhibit minimal variation, confirming the localization of redox activity on the lactone-enediol framework.

\begin{table}[htbp]
\centering
\caption{Selected bond lengths ({\AA}) in ascorbic acid in vacuo (neutral, cation, anion) highlighting structural changes upon electron removal or addition (M062X/6-31+G(d,p) level).}
\label{tab:bond_changes_vacuo}
\begin{tabular}{lccc}
\hline
Bond (IUPAC)               &  Neutral  & Cation   &  Anion   \\
\hline
\ce{C1=O1} (carbonyl)      &   1.2038  &  1.1909  &  1.2448  \\
\ce{C1-C2}                 &   1.4645  &  1.4978  &  1.4004  \\
\ce{C2=C3} (enediol)       &   1.3403  &  1.4052  &  1.4006  \\
\ce{C2-O3} (OH on \ce{C2}) &   1.3469  &  1.2865  &  1.3787  \\
\ce{C3-O4} (OH on \ce{C3}) &   1.3365  &  1.2767  &  1.4041  \\
\ce{C3-C4}                 &   1.5101  &  1.5099  &  1.5131  \\
\ce{C4-O2} (ring O)        &   1.4249  &  1.4279  &  1.4262  \\
\ce{C1-O2} (ring O)        &   1.3684  &  1.3527  &  1.4140  \\
\ce{C4-C5}                 &   1.5229  &  1.5288  &  1.5222  \\
\hline
\end{tabular}
\end{table}

\begin{table}[htbp]
\centering
\caption{Selected bond lengths ({\AA}) in ascorbic acid in water (neutral, cation, anion) highlighting structural changes upon electron removal or addition (M062X/6-31+G(d,p) level).}
\label{tab:bond_changes_water}
\begin{tabular}{lccc}
\hline
Bond (IUPAC)               &  Neutral  & Cation   &  Anion   \\
\hline
\ce{C1=O1} (carbonyl)      &  1.2107   &  1.1960  &  1.2548  \\
\ce{C1-C2}                 &  1.4591   &  1.4951  &  1.3942  \\
\ce{C2=C3} (enediol)       &  1.3412   &  1.4049  &  1.4031  \\
\ce{C2-O3} (OH on \ce{C2}) &  1.3511   &  1.2854  &  1.3791  \\
\ce{C3-O4} (OH on \ce{C3}) &  1.3350   &  1.2758  &  1.3978  \\
\ce{C3-C4}                 &  1.5084   &  1.5048  &  1.5127  \\
\ce{C4-O2} (ring O)        &  1.4297   &  1.4319  &  1.4308  \\
\ce{C1-O2} (ring O)        &  1.3605   &  1.3474  &  1.4075  \\
\ce{C4-C5}                 &  1.5243   &  1.5282  &  1.5206  \\
\hline
\end{tabular}
\end{table}

\begin{itemize}
\item \textbf{Electron Removal (Cation vs Neutral)}: Oxidation (electron removal) primarily affects the conjugated $\pi$-system. The C2=C3 enediol double bond lengthens by 0.065\,\AA\ in vacuo (0.064\,\AA\ in water), indicating significant $\pi$-bond weakening as electrons are removed from the HOMO. Concurrently, the C2--O3 and C3--O4 hydroxyl bonds shorten by $\sim$0.06\,\AA, suggesting increased bond strength due to reduced electron-electron repulsion and enhanced electrostatic attraction.
  The C1--C2 bond lengthens by 0.033\,\AA, reflecting decreased $\pi$-conjugation between the carbonyl and enediol groups.

\item \textbf{Electron Attachment (Anion vs Neutral)}: Reduction (electron addition) substantially alters bond lengths throughout the conjugated system. The C1=O1 carbonyl bond lengthens by 0.041\,\AA\ in vacuo (0.044\,\AA\ in water), indicating reduced double-bond character as the LUMO (primarily $\pi^*_{\text{C1=O}}$) becomes populated. Conversely, the C1--C2 bond shortens by 0.064\,\AA\ in vacuo (0.065\,\AA\ in water), gaining double-bond character due to increased $\pi$-delocalization.
  The C2=C3 bond also lengthens by 0.060\,\AA, while both hydroxyl bonds (C2--O3, C3--O4) elongate by 0.03--0.07\,\AA, consistent with increased electron density on oxygen lone pairs.

\item \textbf{Solvent Effects (Water vs Vacuum)}: Solvation induces subtle but systematic changes.
  For all redox states, water lengthens the C1=O1 carbonyl bond by 0.007--0.010\,{\AA} due to enhanced polarization. The neutral and cation show slightly shortened C1--C2 bonds in water (by 0.005--0.003\,\AA), while the anion's C1--C2 bond is further shortened (by 0.006\,\AA), suggesting water enhances charge delocalization. Hydroxyl bonds are generally longer in water by 0.004--0.008\,\AA due to hydrogen bonding.

    \item \textbf{Connection to Orbital Picture}: These structural changes are fully consistent with the HOMO and LUMO distributions (Section~\ref{sec:homo-lumo}). The HOMO, a $\pi$-bonding orbital centered on the C2=C3 bond, explains its sensitivity to electron removal (oxidation) and addition (reduction). The LUMO, a $\pi^*$ orbital delocalized over the carbonyl-enediol system, accounts for the opposite behavior of the C1=O1 and C1--C2 bonds upon electron attachment. Bonds outside the conjugated framework (e.g., C4--C5) remain nearly invariant, underscoring the localization of redox activity on the lactone-enediol moiety.

    \item \textbf{Chemical Interpretation}: The bond length changes reflect the redistribution of electron density upon redox changes. Oxidation removes electron density primarily from the enediol $\pi$-system, weakening C2=C3 while strengthening polar O--H bonds. Reduction adds electron density to the $\pi^*$-system, weakening both C=O and C=C bonds while strengthening the connecting C1--C2 bond---consistent with increased quinoid-like character. Solvent effects modestly amplify these trends through dielectric stabilization and hydrogen bonding.
\end{itemize}

\textbf{Highlighted Changes (most affected bonds):}
\begin{itemize}
    \item \textcolor{black}{C2=C3: +0.065\,\AA\ (cation)} and \textcolor{black}{+0.060\,\AA\ (anion)} --- largest changes, indicating this $\pi$-bond is most sensitive to redox state
    \item \textcolor{black}{C2--O3: -0.060\,\AA\ (cation)} and \textcolor{black}{+0.032\,\AA\ (anion)} --- hydroxyl bonds show opposite trends upon oxidation vs reduction
    \item \textcolor{black}{C1--C2: -0.064\,\AA\ (anion)} --- largest bond shortening, reflecting increased double-bond character upon electron addition
    \item \textcolor{black}{C1=O1: +0.044\,\AA\ (anion in water)} --- carbonyl bond weakening is amplified by solvation
\end{itemize}

In \Cref{tab:bond_changes_vacuo,tab:bond_changes_water} we also included the \ce{C4-C5} bond length.
Its minimal change ($\pm 0.006$\,{\AA}) across all redox states and negligible solvent effects confirm the orbital interpretation:
the HOMO/LUMO are concentrated on the lactone ring, not on the aliphatic side chain.

\subsection{Vertical versus Adiabatic Ionization Potentials and Electron Affinities}
\label{sec:vertical}

Since CBS-QB3 calculations rely on their own geometry optimizations, they cannot directly provide vertical IP and EA values.
Vertical IP$_{\text{vert}}$ and EA$_{\text{vert}}$ are therefore obtained from DFT calculations and reported for vacuum and water in \Cref{tab:ip_ea_vacuum,tab:ip_ea_water}, and for the other five solvents in \Cref{tab:ip_ea_benzene,tab:ip_ea_toluene,tab:ip_ea_chlorobenzene,tab:ip_ea_methanol,tab:ip_ea_ethanol}.

Vertical values correspond to Franck-Condon transitions where the geometry remains frozen at the neutral molecule's equilibrium structure during electron removal or addition. The adiabatic values, in contrast, account for full nuclear relaxation in the resulting ion.

The functional and basis-set dependence of these values is moderate. The meta-hybrid M06-2X yields slightly higher IPs and lower (more negative) EAs in vacuo compared to B3LYP, consistent with its higher exact-exchange content and improved treatment of medium-range correlation. The largest Pople basis set 6-311++G(3df,3pd) generally lowers the EA by a few tenths of an eV relative to 6-31+G(d,p), but does not qualitatively alter the trends to be analyzed in the next section.

The key observation from the data is the consistent $0.4-0.5$\,eV difference between vertical and adiabatic IPs, and the
$0.4-0.6$\,eV difference between vertical and adiabatic EAs in vacuo. These differences represent the \textit{reorganization energies} of the cation and anion, respectively (\Cref{fig:lambdas}). These values, which are substantially larger than for molecules with aromatic rings,\cite{Baldea:2013a,Baldea:2014a} correlate directly with the substantial bond-length changes shown in \Cref{tab:bond_changes_vacuo,tab:bond_changes_water}. For example, the cation reorganization energy (IP$_{\text{vert}} -$ IP$_{\text{adiab}}$) of $\sim 0.5$\,eV reflects the structural cost of accommodating the hole in the conjugated $\pi$-system, primarily the lengthening of the enediol \ce{C2=C3} bond and the shortening of the hydroxyl C-O bonds. Similarly, the anion reorganization energy ($\text{EA}_{\text{adiab}} - \text{EA}_{\text{vert}}$) of
$\sim 0.5 - 0.6$\,eV arises from the geometric adjustments required to delocalize the added electron, notably the elongation of the carbonyl \ce{C1=O1} and enediol \ce{C2=C3} bonds and the contraction of the \ce{C1-C2} linkage.
\begin{table*}[htbp]
\centering
\caption{Global reactivity descriptors for ascorbic acid computed using vertical IP and EA.}
\label{tab:vertical_reactivities}
\setlength{\tabcolsep}{4.5pt}
\scriptsize
\begin{tabular}{lrrrrrrr}
\hline
Method & Vacuum & Benzene & Toluene & Chlorobenzene & Methanol & Ethanol & Water \\
\hline
\multicolumn{8}{l}{\textbf{B3LYP/6-31+G(d,p)}} \\
\hline
IP (eV) & 8.673 & 7.370 & 7.278 & 5.961 & 5.890 & 5.973 & 5.638 \\
EA (eV) & -0.243 & 0.398 & 0.373 & -0.025 & 0.532 & 0.573 & 0.359 \\
$E_g$ (eV) & 8.916 & 6.972 & 6.905 & 5.986 & 5.358 & 5.400 & 5.279 \\
$\eta$ (eV) & 4.458 & 3.486 & 3.452 & 2.993 & 2.679 & 2.700 & 2.639 \\
$\sigma$ (eV$^{-1}$) & 0.112 & 0.143 & 0.145 & 0.167 & 0.187 & 0.185 & 0.189 \\
$\chi = - \mu$ (eV) & 4.215 & 3.884 & 3.825 & 2.968 & 3.211 & 3.273 & 2.998 \\
$\omega$ (eV) & 1.993 & 2.164 & 2.119 & 1.472 & 1.924 & 1.984 & 1.703 \\
$\omega^{-}$ (eV) & 4.657 & 4.541 & 4.464 & 3.330 & 3.865 & 3.958 & 3.532 \\
$\omega^{+}$ (eV) & 0.442 & 0.657 & 0.638 & 0.362 & 0.654 & 0.685 & 0.534 \\
\hline
\multicolumn{8}{l}{\textbf{B3LYP/6-311++G(3df,3pd)}} \\
\hline
IP (eV) & 8.673 & 7.370 & 7.278 & 5.954 & 5.882 & 5.965 & 5.629 \\
EA (eV) & -0.005 & 0.518 & 0.488 & -0.038 & 0.480 & 0.522 & 0.307 \\
$E_g$ (eV) & 8.678 & 6.852 & 6.790 & 5.992 & 5.402 & 5.443 & 5.322 \\
$\eta$ (eV) & 4.339 & 3.426 & 3.395 & 2.996 & 2.701 & 2.721 & 2.661 \\
$\sigma$ (eV$^{-1}$) & 0.115 & 0.146 & 0.147 & 0.167 & 0.185 & 0.184 & 0.188 \\
$\chi = - \mu$ (eV) & 4.334 & 3.944 & 3.883 & 2.958 & 3.181 & 3.244 & 2.968 \\
$\omega$ (eV) & 2.165 & 2.270 & 2.221 & 1.460 & 1.873 & 1.933 & 1.655 \\
$\omega^{-}$ (eV) & 4.874 & 4.670 & 4.586 & 3.314 & 3.801 & 3.895 & 3.472 \\
$\omega^{+}$ (eV) & 0.540 & 0.726 & 0.703 & 0.356 & 0.620 & 0.651 & 0.504 \\
\hline
\multicolumn{8}{l}{\textbf{M062X/6-31+G(d,p)}} \\
\hline
IP (eV) & 8.874 & 7.574 & 7.481 & 6.158 & 6.083 & 6.166 & 5.832 \\
EA (eV) & -0.621 & 0.125 & 0.112 & -0.178 & 0.395 & 0.436 & 0.224 \\
$E_g$ (eV) & 9.495 & 7.449 & 7.369 & 6.336 & 5.688 & 5.730 & 5.608 \\
$\eta$ (eV) & 4.748 & 3.724 & 3.684 & 3.168 & 2.844 & 2.865 & 2.804 \\
$\sigma$ (eV$^{-1}$) & 0.105 & 0.134 & 0.136 & 0.158 & 0.176 & 0.175 & 0.178 \\
$\chi = - \mu$ (eV) & 4.127 & 3.849 & 3.796 & 2.990 & 3.239 & 3.301 & 3.028 \\
$\omega$ (eV) & 1.793 & 1.989 & 1.956 & 1.411 & 1.844 & 1.902 & 1.635 \\
$\omega^{-}$ (eV) & 4.450 & 4.380 & 4.315 & 3.302 & 3.819 & 3.910 & 3.499 \\
$\omega^{+}$ (eV) & 0.324 & 0.530 & 0.518 & 0.312 & 0.580 & 0.609 & 0.471 \\
\hline
\multicolumn{8}{l}{\textbf{M062X/6-311++G(3df,3pd)}} \\
\hline
IP (eV) & 8.900 & 7.595 & 7.502 & 6.175 & 6.100 & 6.183 & 5.848 \\
EA (eV) & -0.297 & 0.089 & 0.074 & -0.230 & 0.343 & 0.383 & 0.172 \\
$E_g$ (eV) & 9.197 & 7.506 & 7.428 & 6.405 & 5.757 & 5.800 & 5.676 \\
$\eta$ (eV) & 4.599 & 3.753 & 3.714 & 3.203 & 2.878 & 2.900 & 2.838 \\
$\sigma$ (eV$^{-1}$) & 0.109 & 0.133 & 0.135 & 0.156 & 0.174 & 0.172 & 0.176 \\
$\chi = - \mu$ (eV) & 4.301 & 3.842 & 3.788 & 2.972 & 3.221 & 3.283 & 3.010 \\
$\omega$ (eV) & 2.012 & 1.967 & 1.932 & 1.380 & 1.803 & 1.858 & 1.596 \\
$\omega^{-}$ (eV) & 4.737 & 4.357 & 4.290 & 3.266 & 3.773 & 3.862 & 3.456 \\
$\omega^{+}$ (eV) & 0.436 & 0.515 & 0.502 & 0.294 & 0.552 & 0.579 & 0.446 \\
\hline
\end{tabular}
\end{table*}

In summary, the substantial reorganization energies mirror the bond-length changes documented in the previous section, confirming that the redox activity of ascorbic acid is intimately linked to the structural flexibility of its enediol-carbonyl $\pi$-system.

\subsection{Solvent Effects on Reactivity Indices: Solvent is Paramount}
\label{sec:solvent}
The chemical reactivity of ascorbic acid --- and of antioxidants in general --- cannot be meaningfully assessed in vacuo. These molecules exert their protective action in highly polar, aqueous biological environments, where solvent effects dominate the electronic structure, energetic stability, and global reactivity indices. Solvent effects are therefore paramount: they stabilize charged species (cations and anions) far more strongly than the neutral form, leading to dramatic changes in global reactivity indices.

\begin{figure*}
  \centerline{
    \includegraphics[width=0.23\textwidth,angle=0]{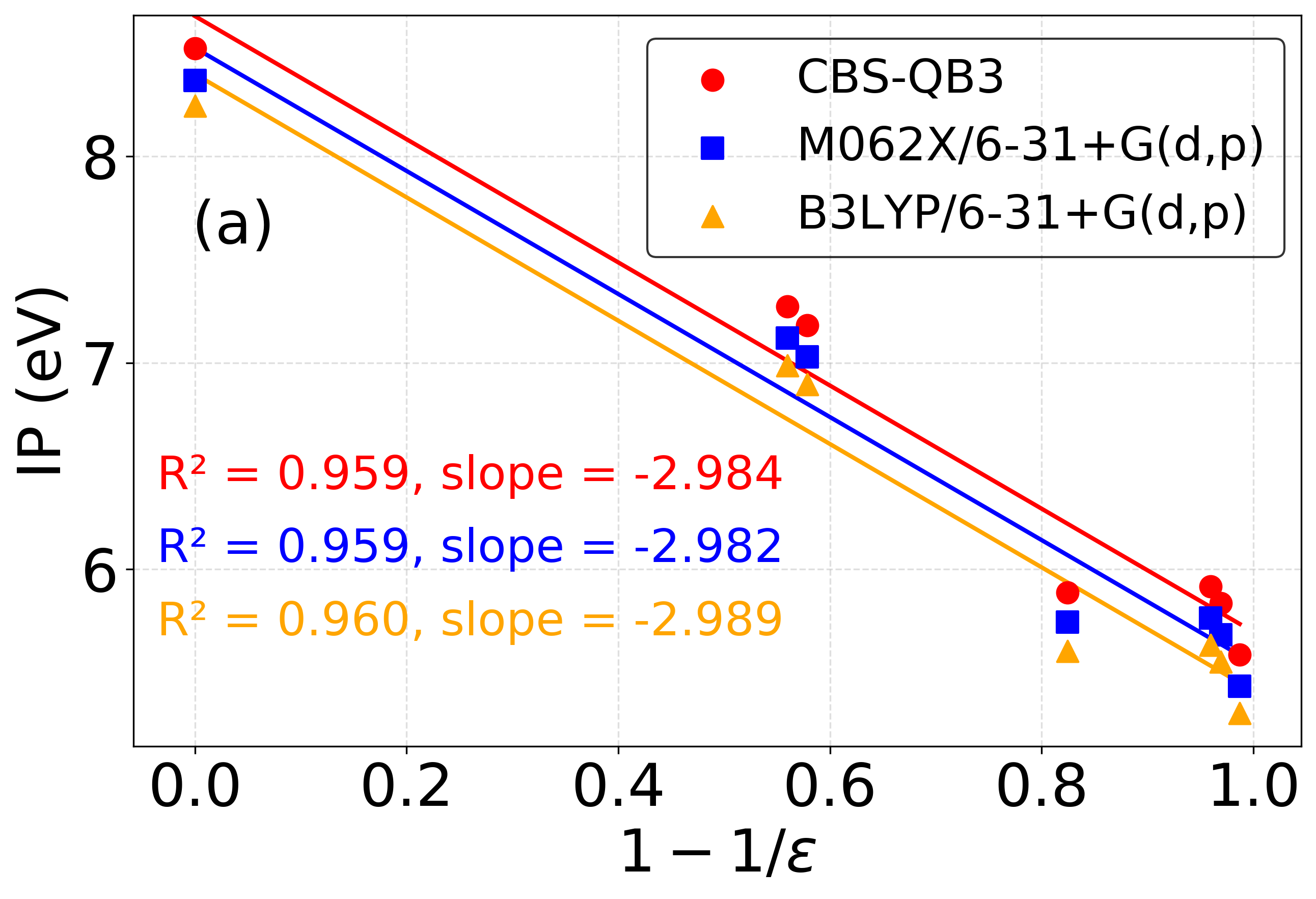}
    \includegraphics[width=0.23\textwidth,angle=0]{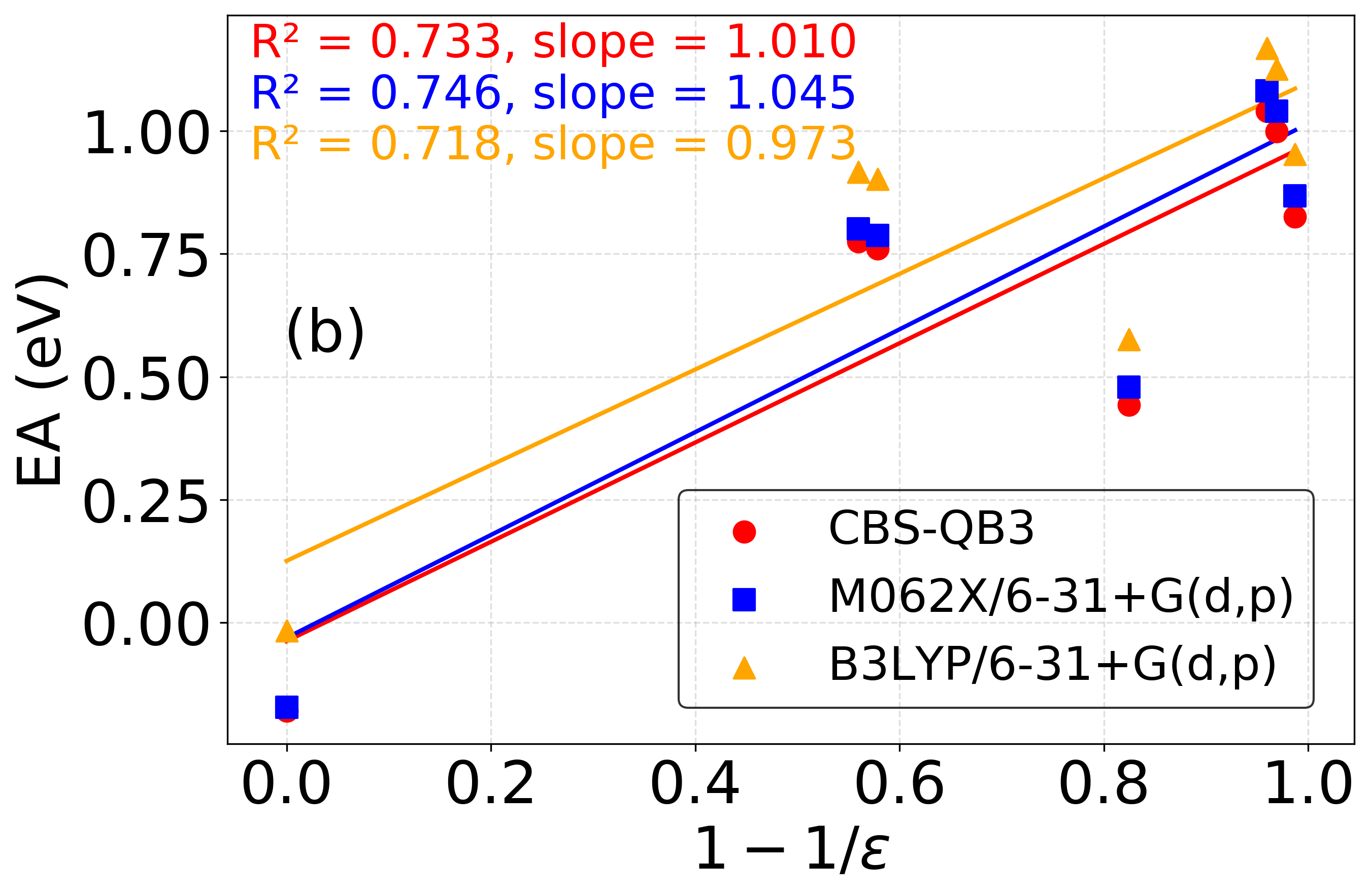}
    \includegraphics[width=0.23\textwidth,angle=0]{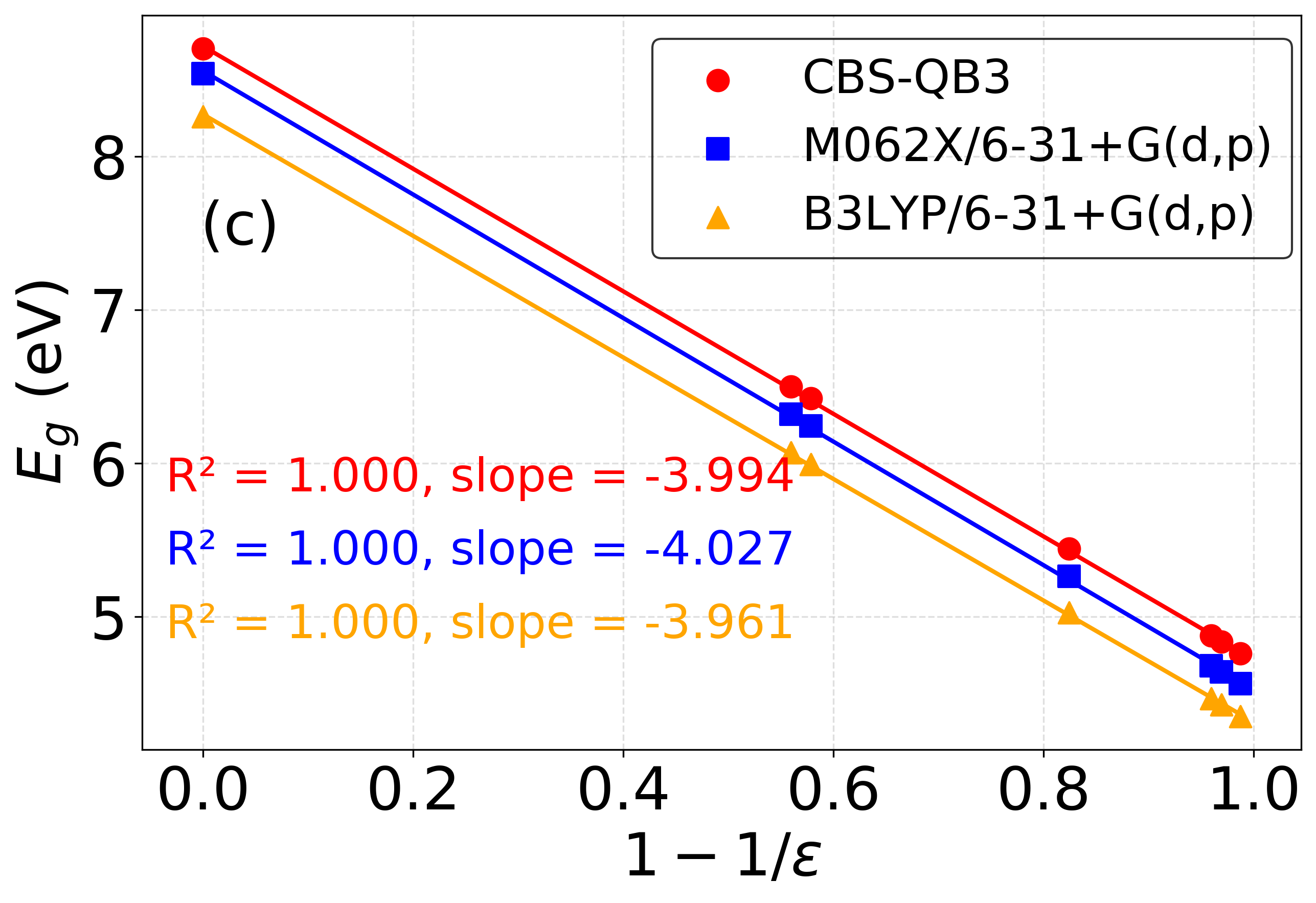}
    \includegraphics[width=0.23\textwidth,angle=0]{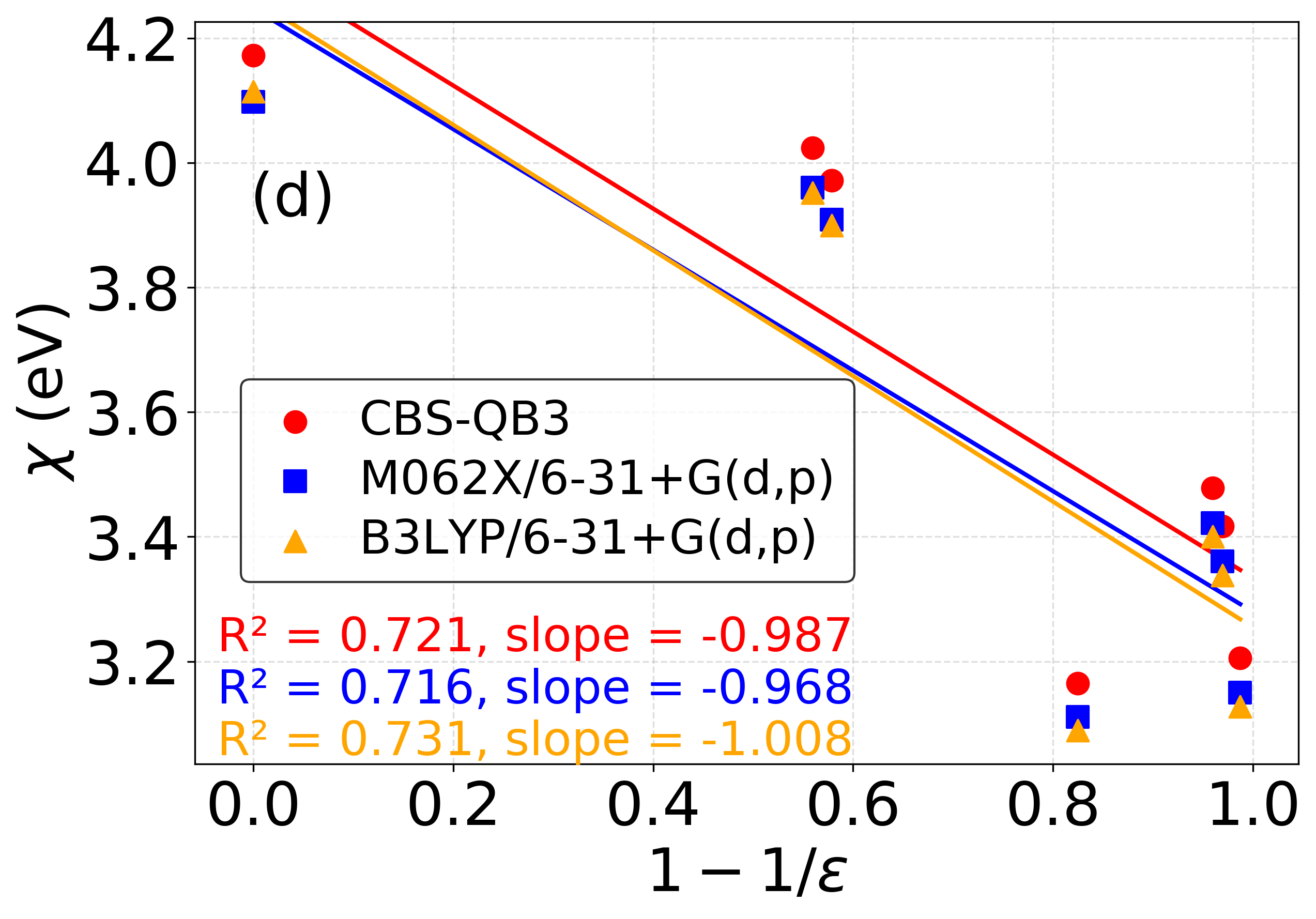}
  }
  \centerline{
    \includegraphics[width=0.23\textwidth,angle=0]{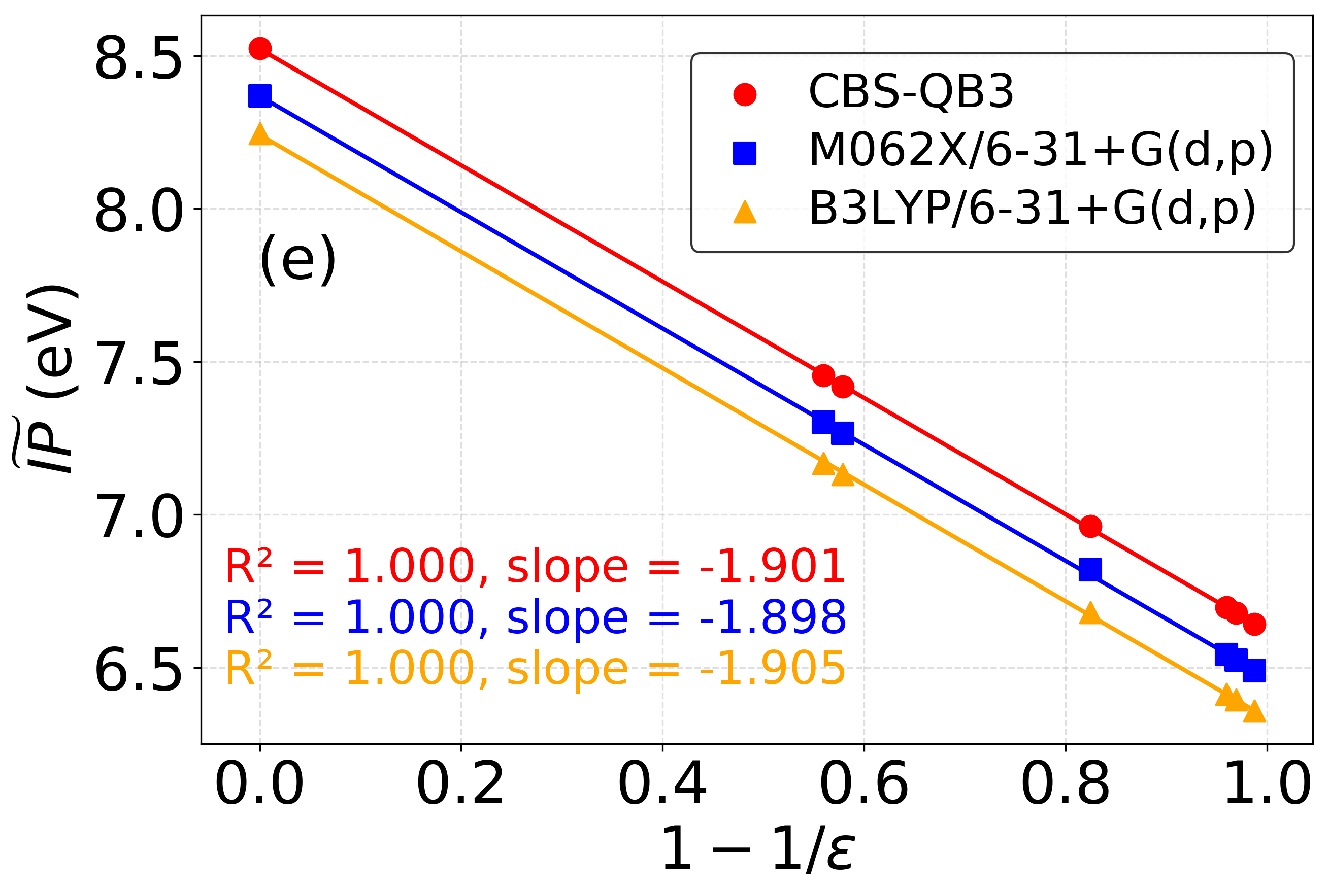}
    \includegraphics[width=0.23\textwidth,angle=0]{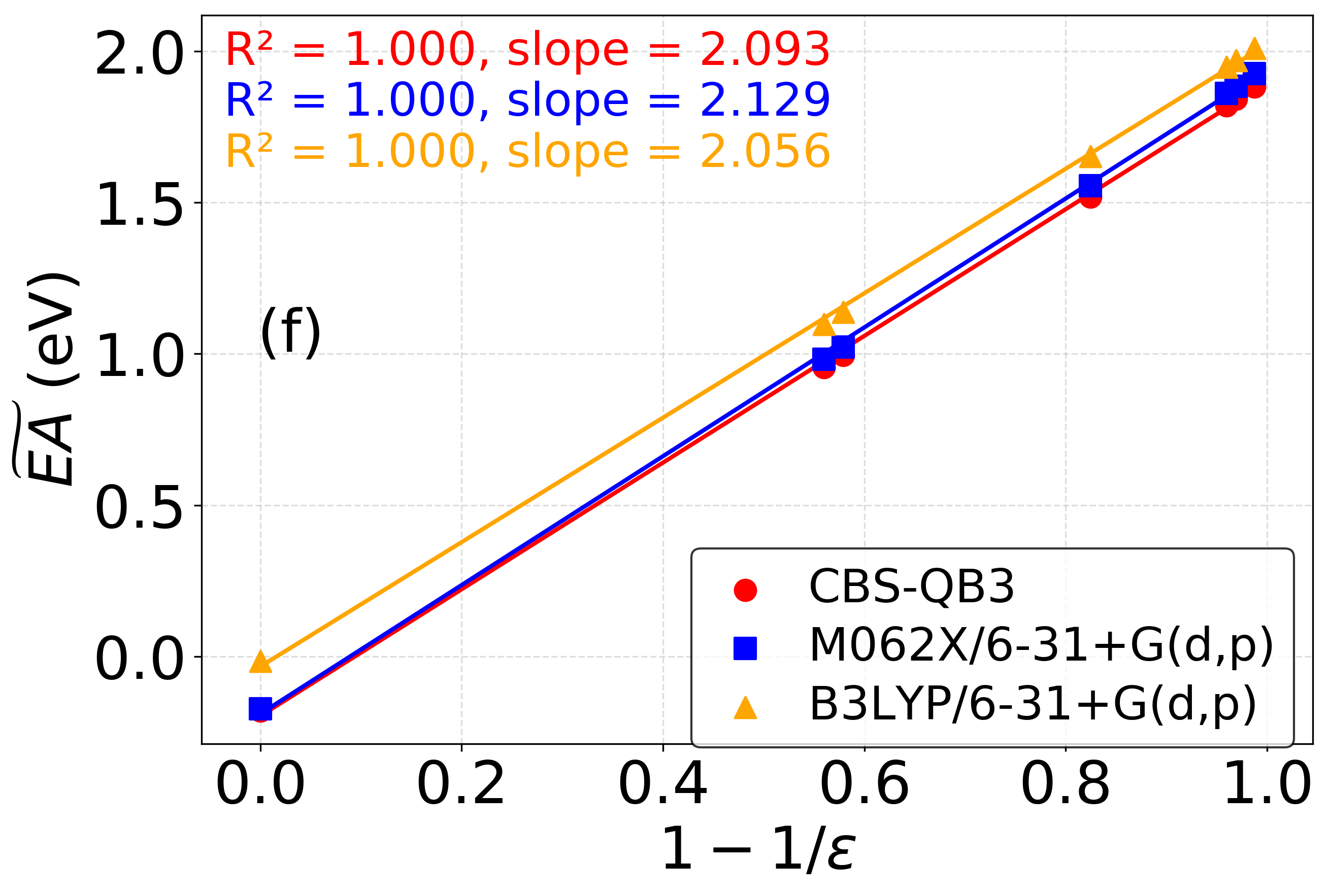}
    \includegraphics[width=0.23\textwidth,angle=0]{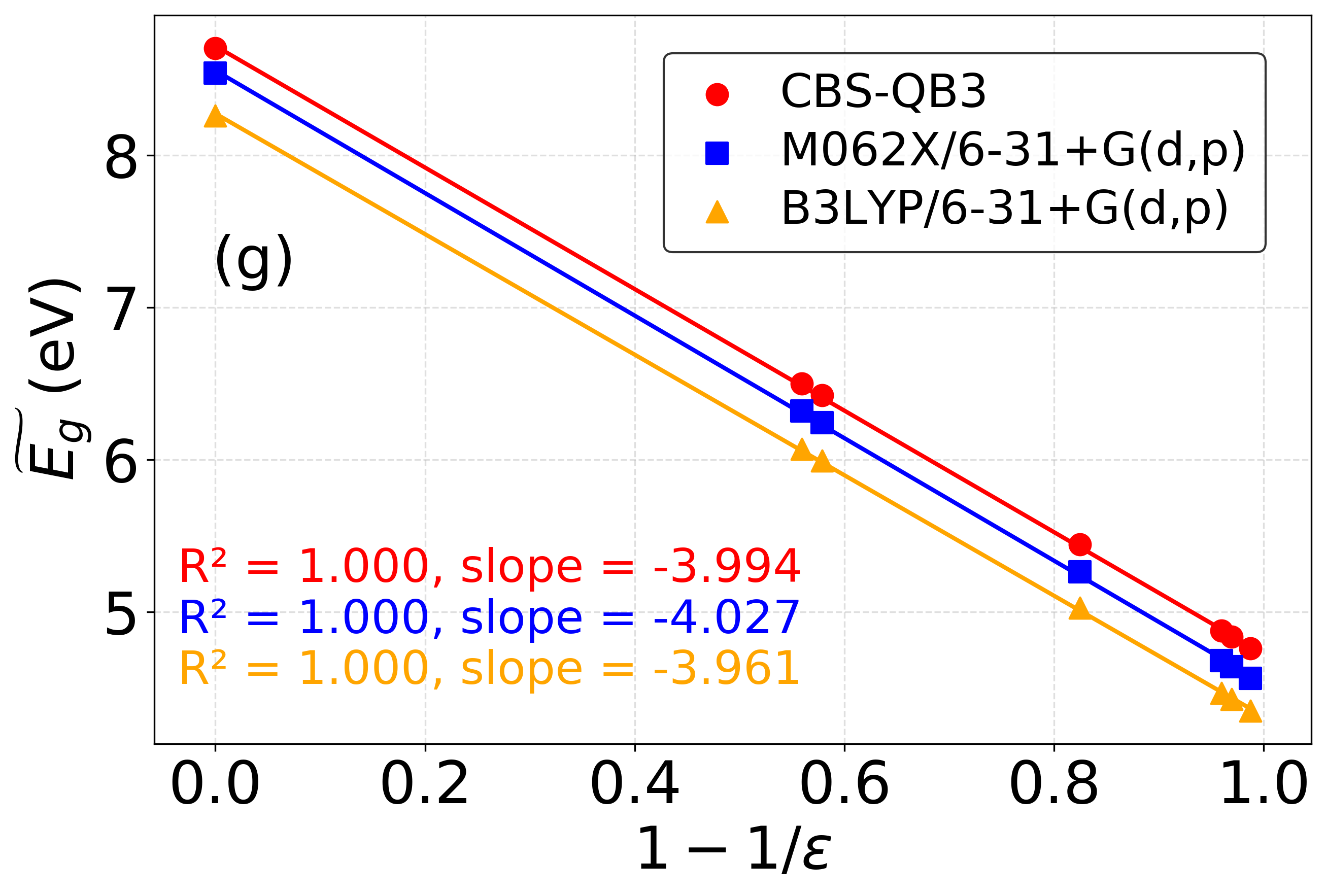}
    \includegraphics[width=0.23\textwidth,angle=0]{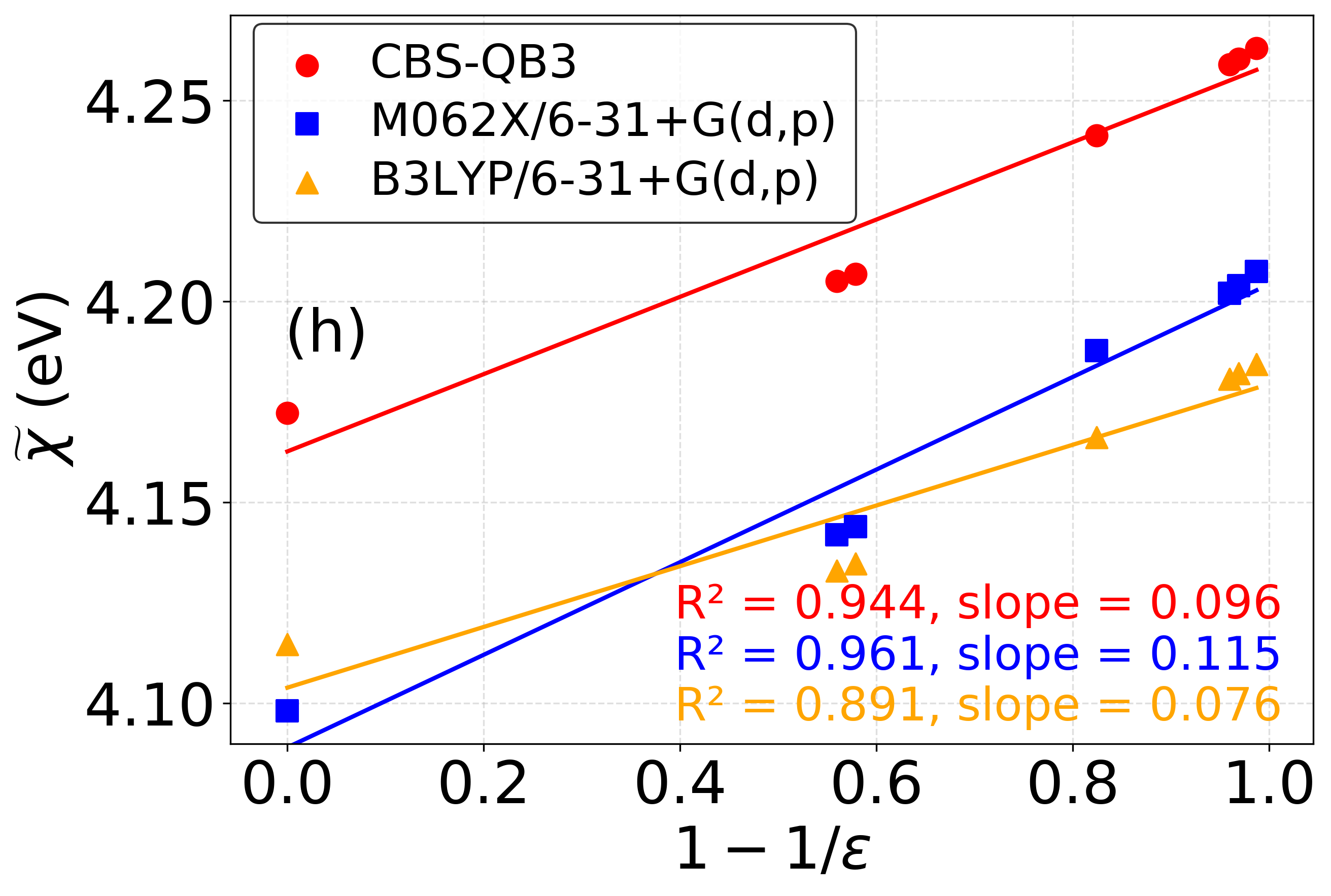}
  }
  \caption{Adiabatic quantities
    plotted against $1 - 1/\varepsilon$:
    (a) ionization potential IP, (b) electron affinity EA, (c) HOMO-LUMO gap, and (d) electronegativity $\chi$
    using data for vacuum and six solvents spanning a broad spectrum which  confirm by and large the Born-like
    \cite{Bard:01,AtkinsBook} effect of the solvents. (e--h) Their counterparts obtained by subtracting the electron solvation enthalpy
    from IP and EA calculated according to the standard prescription (eqs~(\ref{eq-IP}) and (\ref{eq-EA})) exhibiting a nearly perfect linearity suggest either
    an observable deviation from Born picture or the need to re-consider electron solvation data
    available in the literature \cite{Markovic:16} and used in the present calculations.
    Shown here are results of CBS-QB3 calculations and two extreme DFT flavors,
    M06-2X/6-31+G(d,p) and B3LYP/6-31+G(d,p), which exhibit the smallest and the largest from
    CBS-QB3 (cf.~\Cref{tab:ip_dev,tab:ea_dev,tab:eta_dev,tab:sigma_dev,tab:omega_dev,tab:omegap_dev,tab:omegam_dev,tab:solvation_dev,tab:sol_neutral_dev,tab:sol_cation_dev,tab:sol_anion_dev}). See the main text for details.
    }
      \label{fig:ip-ea-eg-chi}
\end{figure*}

\begin{figure*}
  \centerline{
    \includegraphics[width=0.23\textwidth,angle=0]{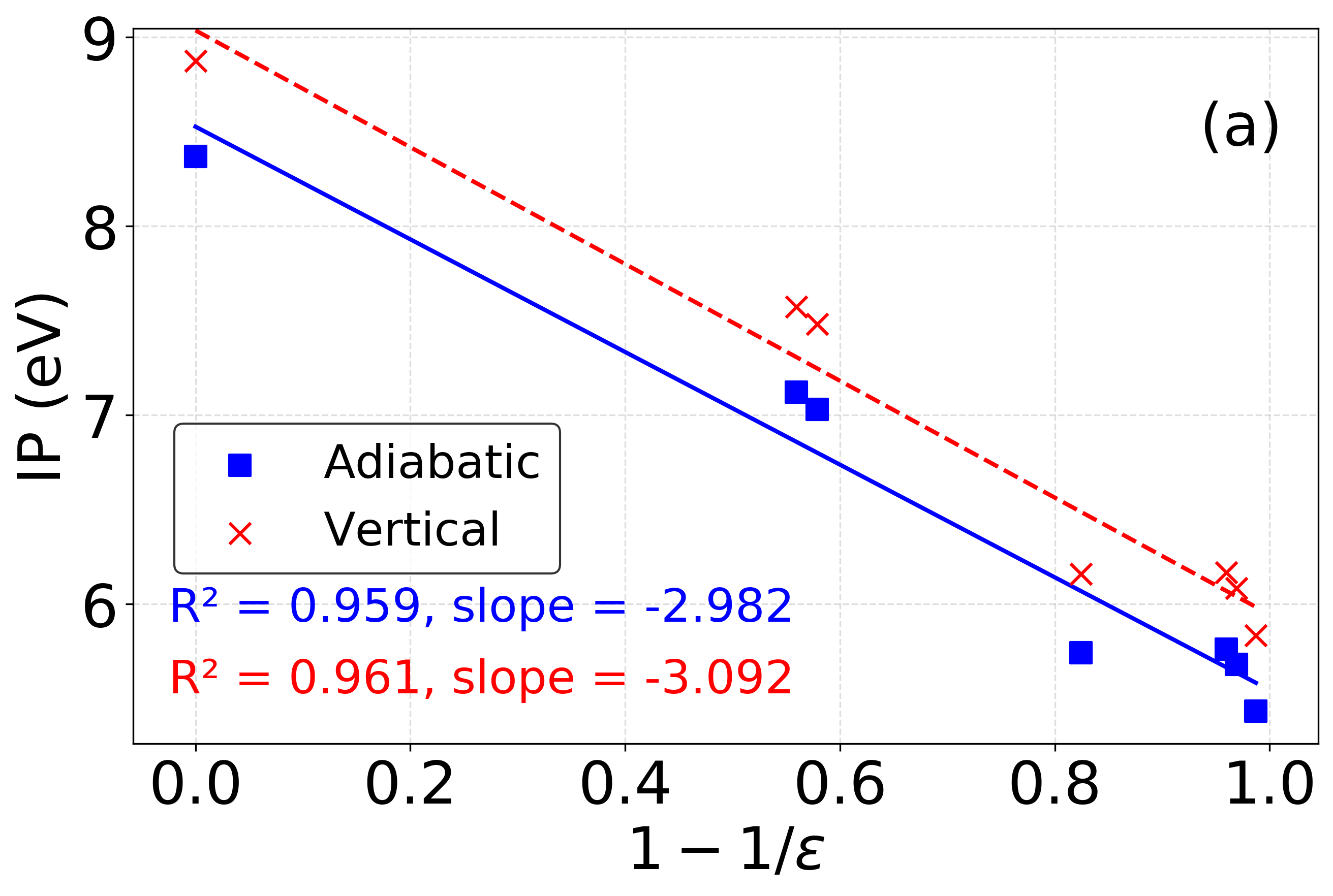}
    \includegraphics[width=0.23\textwidth,angle=0]{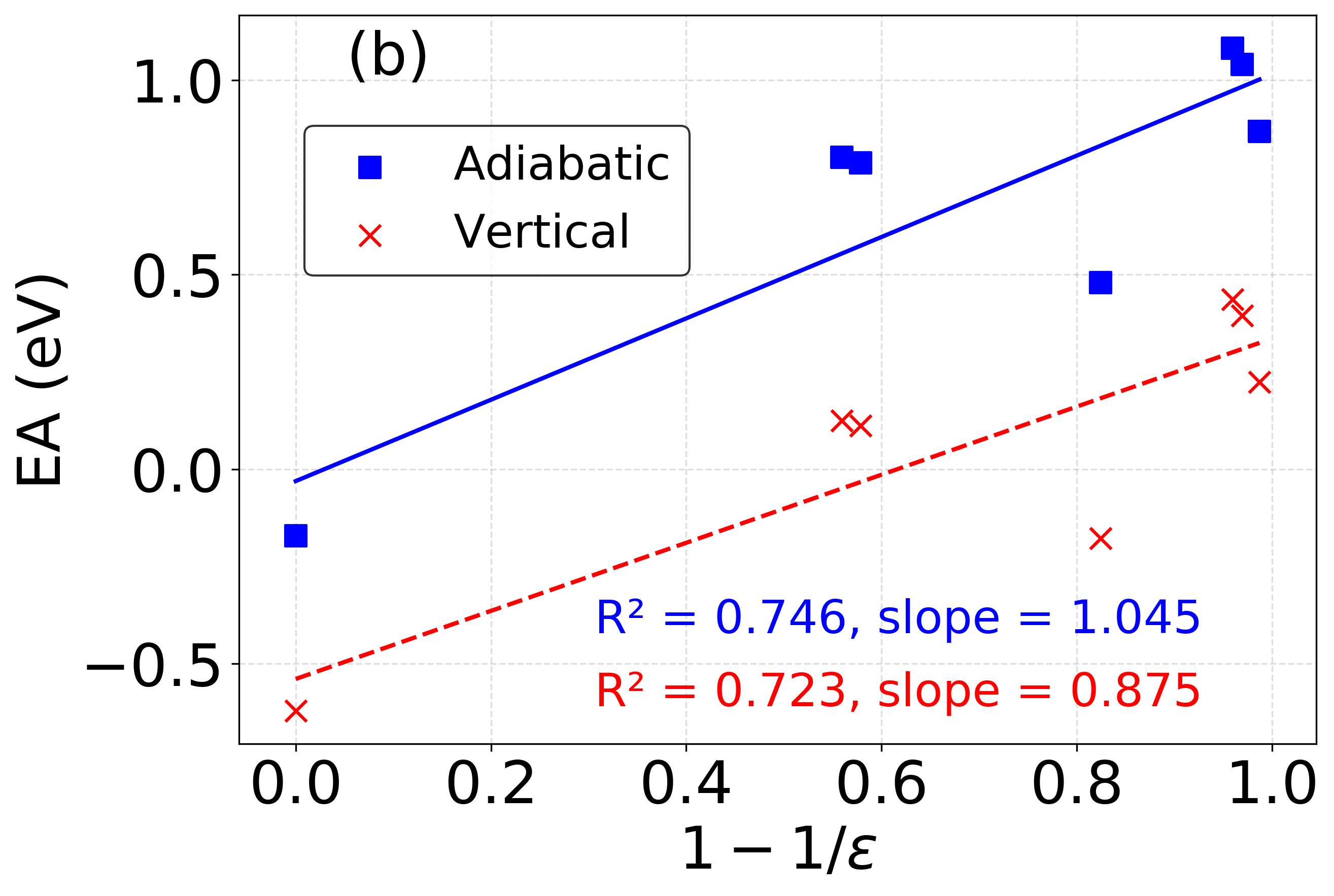}
    \includegraphics[width=0.23\textwidth,angle=0]{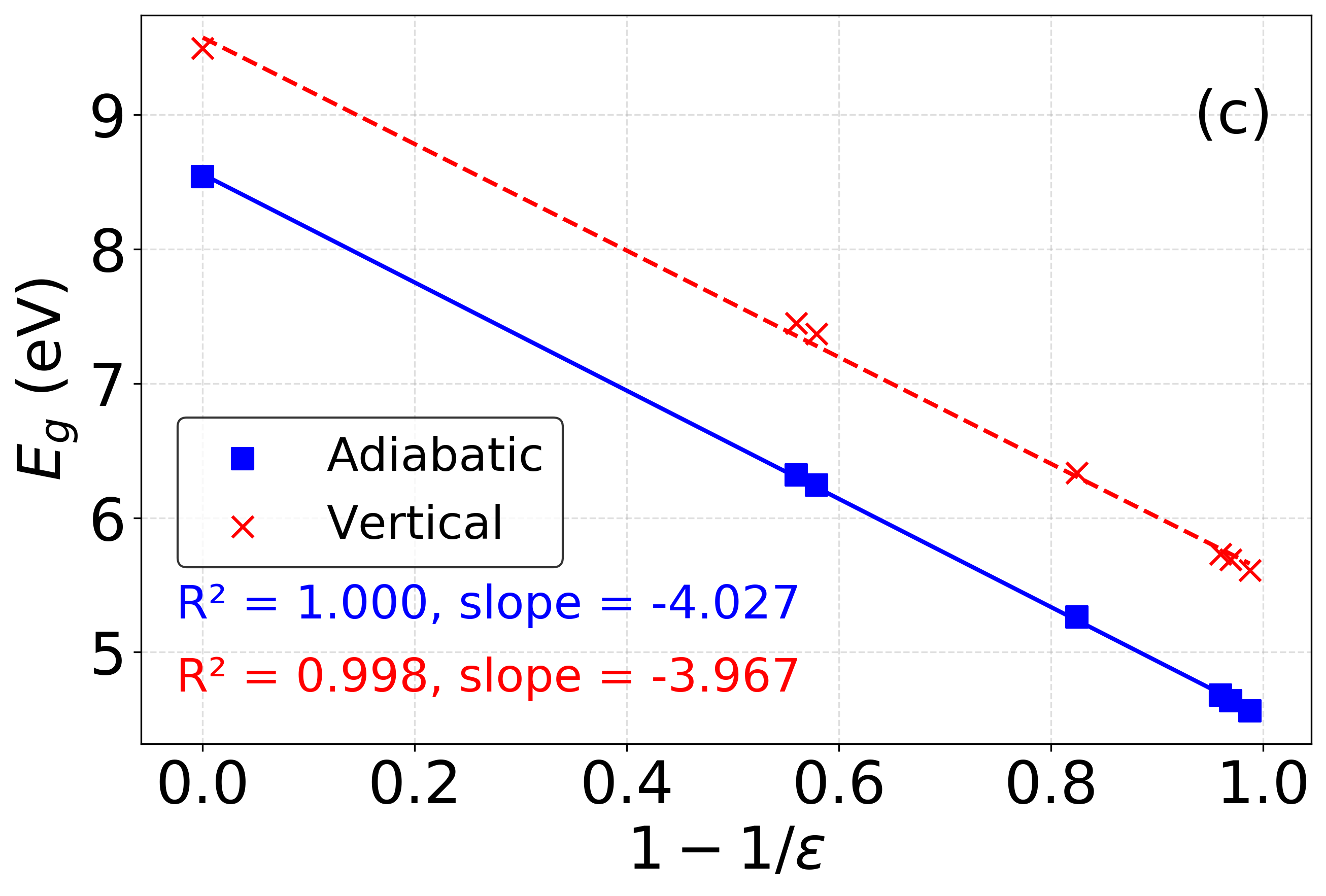}
    \includegraphics[width=0.23\textwidth,angle=0]{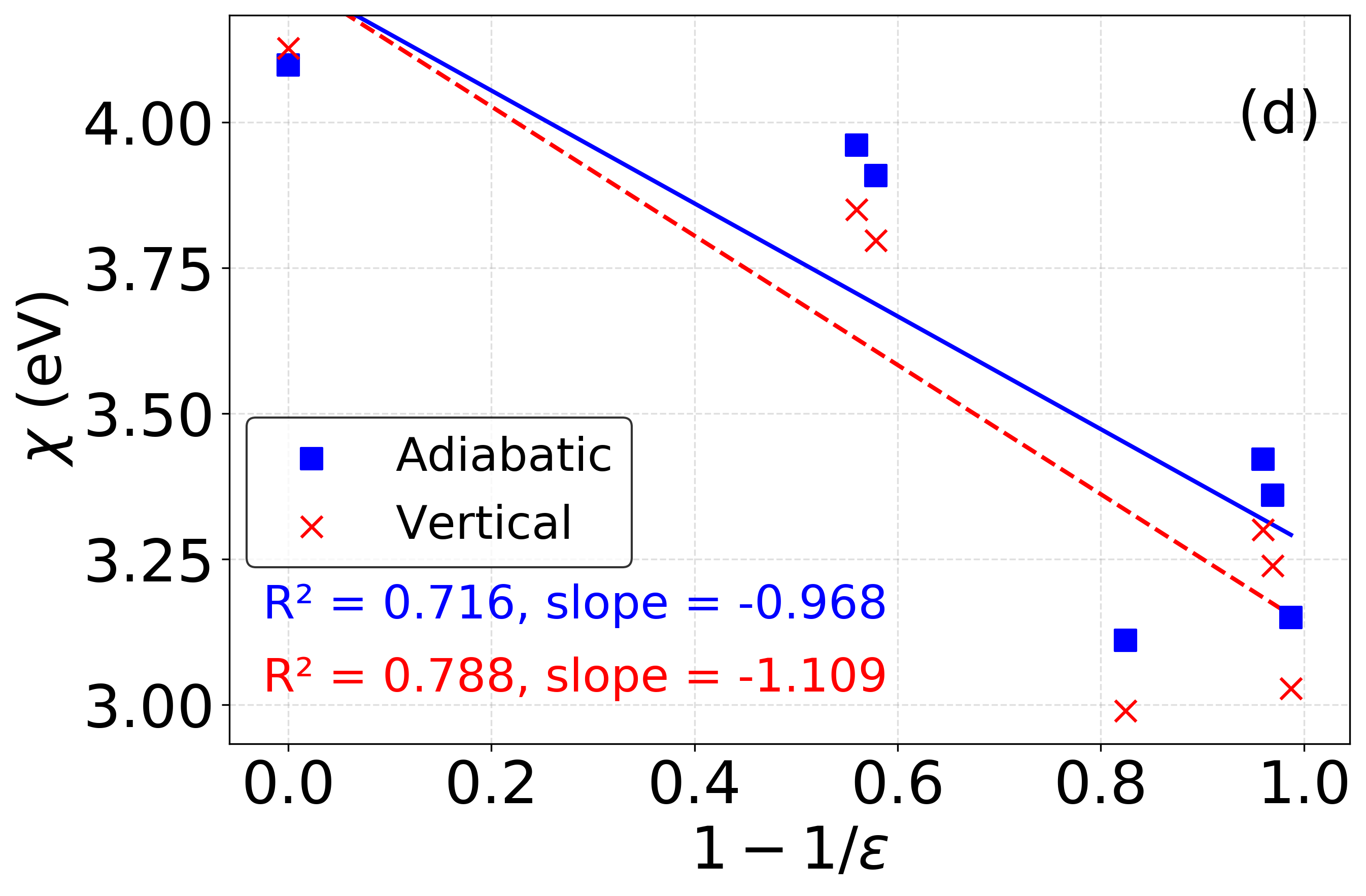}
    }
  \caption{Vertical and adiabatic
    (a) ionization potential IP and (b) electron affinity EA and the quantities (c) $E_g$ and (d) $\chi$
    depending linearly on them plotted versus $1 - 1/\varepsilon$. Notice that the approximately linear
    vertical IP and EA are less inclined than the adiabatic ones, reflecting a similar dependence of
    the reorganization energies, which are equal to the difference between them.
    (e--h) Counterpart of the quantities shown in panels (a--d) but adding the electron's solvation enthalpy
  to the Koopmans' estimates.}
      \label{fig:ip-ea-eg-chi-vert}
\end{figure*}

The redox behavior of ascorbic acid is profoundly modulated by the solvent environment, as evidenced by the dramatic shifts in both adiabatic (\Cref{fig:ip-ea-eg-chi}) and vertical (\Cref{fig:ip-ea-eg-chi-vert}) IP and EA values across media of varying polarity. To quantify this dependence, we analyze the computed ionization potentials, electron affinities, and the fundamental gap ($E_g$) as functions of the solvent's response, parameterized in terms of Born polarity function ($1 - 1/\varepsilon$).
This Born-like functional form \cite{Bard:01,AtkinsBook}
emerges from continuum solvation models and captures the electrostatic stabilization of a charge in a polarizable medium.

\Cref{fig:ip-ea-eg-chi-vert} plots the adiabatic and vertical IP, EA, $E_g$, and $\chi$ against $1 - 1/\varepsilon$ for all solvents studied (benzene, toluene, chlorobenzene, methanol, ethanol, water) and vacuum ($\varepsilon=1$). The data are extracted from \Cref{tab:ip_ea_vacuum,tab:ip_ea_water} and the analogous tables for the other five solvents
(\Cref{tab:ip_ea_benzene,tab:ip_ea_toluene,tab:ip_ea_chlorobenzene,tab:ip_ea_methanol,tab:ip_ea_ethanol}).
Several robust trends emerge.

\begin{figure*}
  \centerline{
    \includegraphics[width=0.23\textwidth,angle=0]{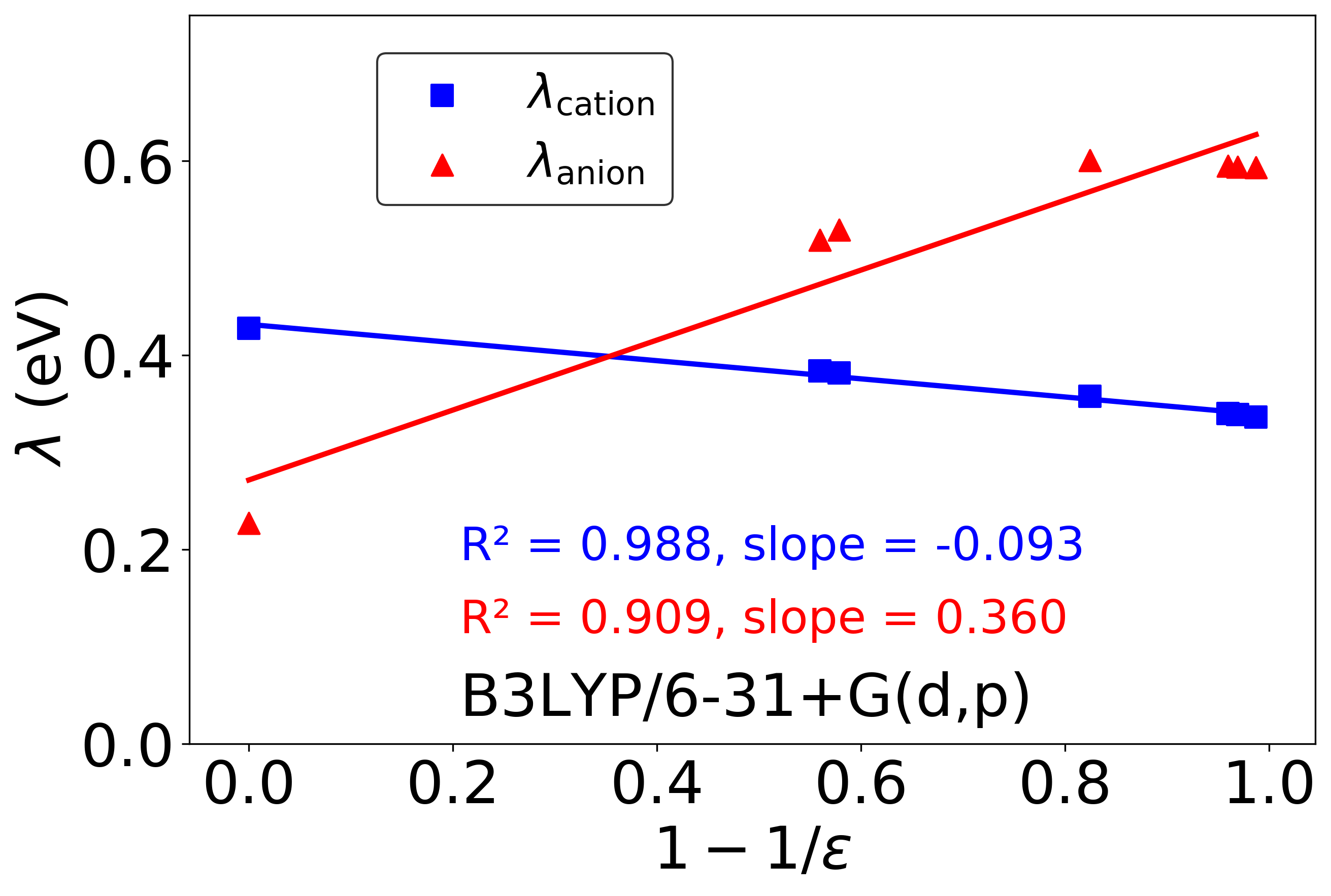}
    \includegraphics[width=0.23\textwidth,angle=0]{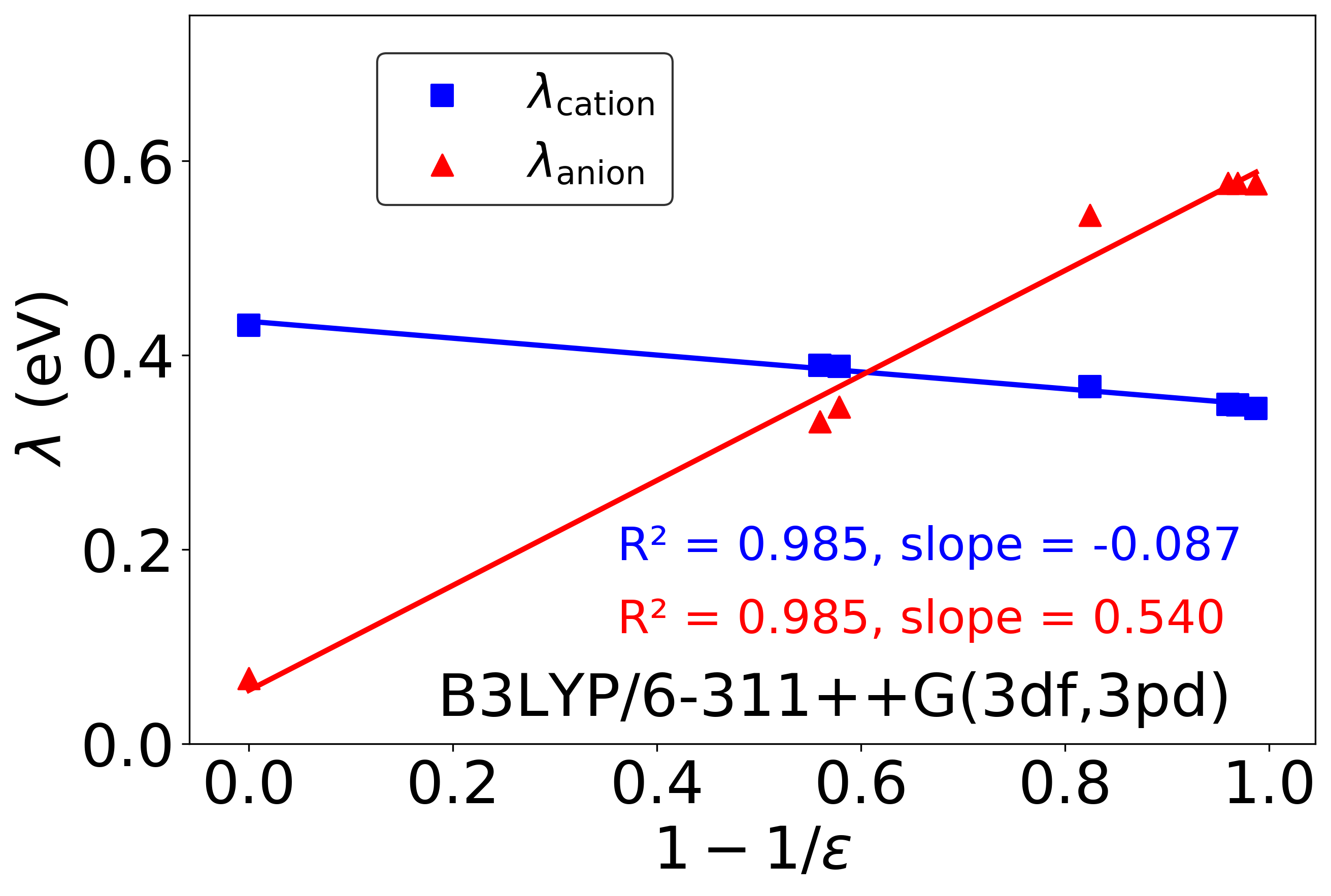}
    \includegraphics[width=0.23\textwidth,angle=0]{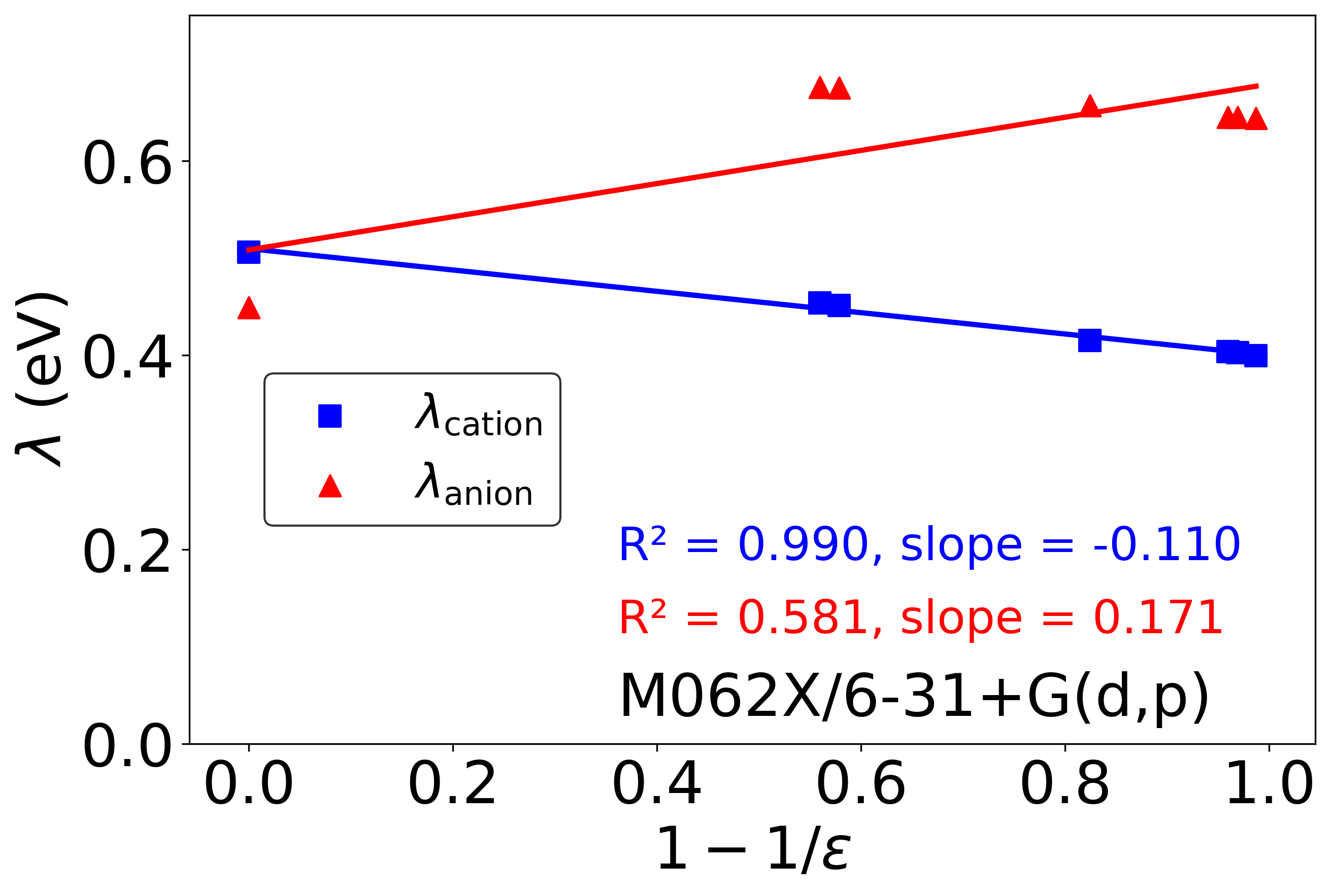}
    \includegraphics[width=0.23\textwidth,angle=0]{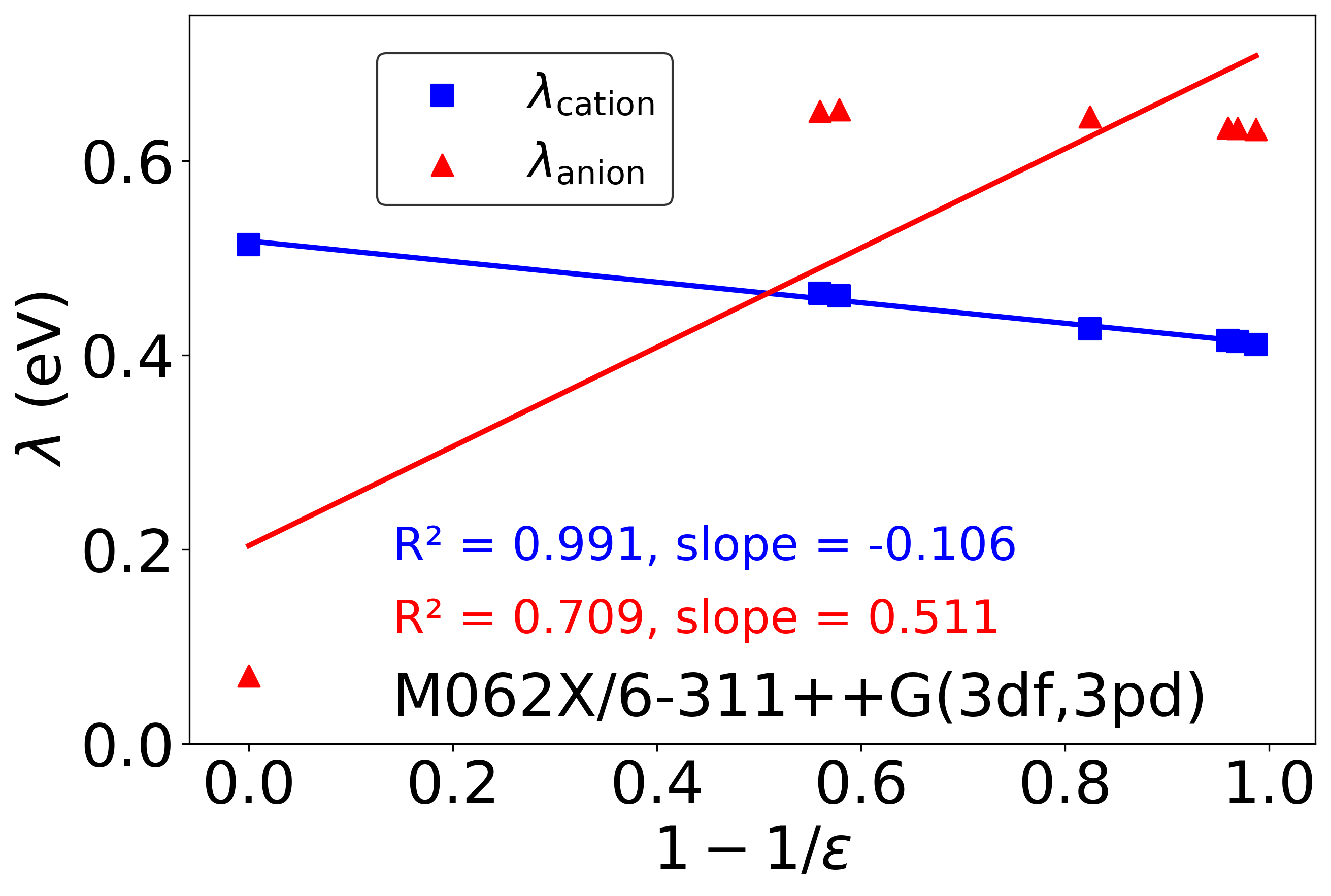}
  }
  \caption{Cation and anion reorganization energies $\Lambda$ plotted versus $1 - 1/\varepsilon$ computed by using (left to right):
    B3LYP/6-31+G(d,p), B3LYP/6-311++G(3df,3pd), M06-2X/6-31+G(d,p), and M06-2X/6-311++G(3df,3pd).
    These results show a series of contrasts (cation versus anion, B3LYP versus M06-2X, smaller basis set 6-31+G(d,p) versus larger basis set 6-311++G(3df,3pd))
    discussed in the main text.
  }
      \label{fig:lambdas}
\end{figure*}

Both IP$_{\text{adiab}}$ and IP$_{\text{vert}}$ decrease nearly linearly with increasing $1 - 1/\varepsilon$, reflecting the enhanced stabilization of the cation by polar solvents. The slope is different from
  that for the adiabatic values because they incorporate the full solvation of the relaxed cation, whereas the vertical values correspond to the instantaneous charge distribution of the neutral geometry (\Cref{fig:lambdas}).
  The total solvent-induced reduction in IP amounts to $\approx$3\,eV when moving from vacuum to water.

Similarly, EA$_{\text{adiab}}$ and EA$_{\text{vert}}$ increase with solvent polarity, shifting from negative (unbound) values in vacuo to positive (bound) values in polar solvents. This roughly linear versus $1 - 1/\varepsilon$ trend confirms that the anion becomes progressively more stable as the dielectric constant grows. The slope for EA$_{\text{adiab}}$ is again different from that for EA$_{\text{vert}}$, consistent with the additional relaxation of the anion's structure in the adiabatic process
  (\Cref{fig:lambdas}). Similar anion's stabilization upon solvation in water \cite{Baldea:2012i,Baldea:2013a}
  was previous found to be relevant for
  understanding charge transport in molecular junctions immersed in solvents.\cite{Baldea:2013b,Baldea:2013c}

  The narrowing of the HOMO-LUMO gap $E_g = \mbox{IP} - \mbox{EA}$ is perhaps the most striking solvent effect.
In vacuo, ascorbic acid exhibits a large $E_g$($\approx$8.7\,eV, cf.~\Cref{tab:reactivity_vacuum}), indicating a hard, electronically stable molecule with low propensity for electron transfer.
Upon solvation, $E_g$ decreases sharply. In benzene ($\varepsilon = 2.27$), $E_g$ narrows by $\approx$2.2\,eV relative to vacuum
(\Cref{tab:reactivity_benzene}).
As the solvent becomes more polar (\Cref{tab:ip_ea_toluene,tab:ip_ea_chlorobenzene,tab:ip_ea_methanol,tab:ip_ea_ethanol}), this reduction
  increases, reaching $\approx$4\,eV for the most polar solvent considered (water, $\varepsilon \approx 78.4$, \Cref{tab:ip_ea_water};
see \Cref{fig:ip-ea-eg-chi}).

\begin{figure*}
  \centerline{
    \includegraphics[width=0.23\textwidth,angle=0]{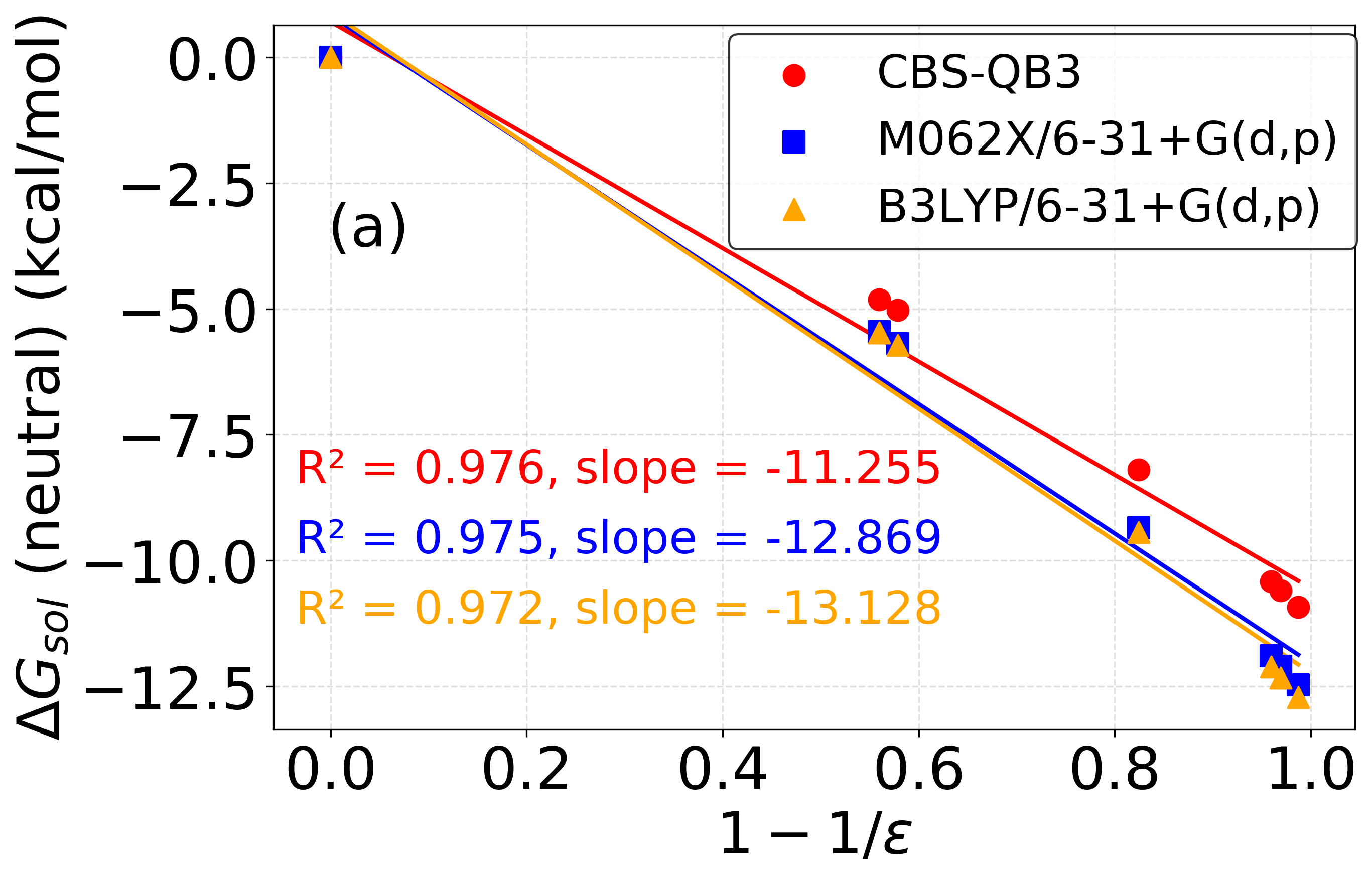}
    \includegraphics[width=0.23\textwidth,angle=0]{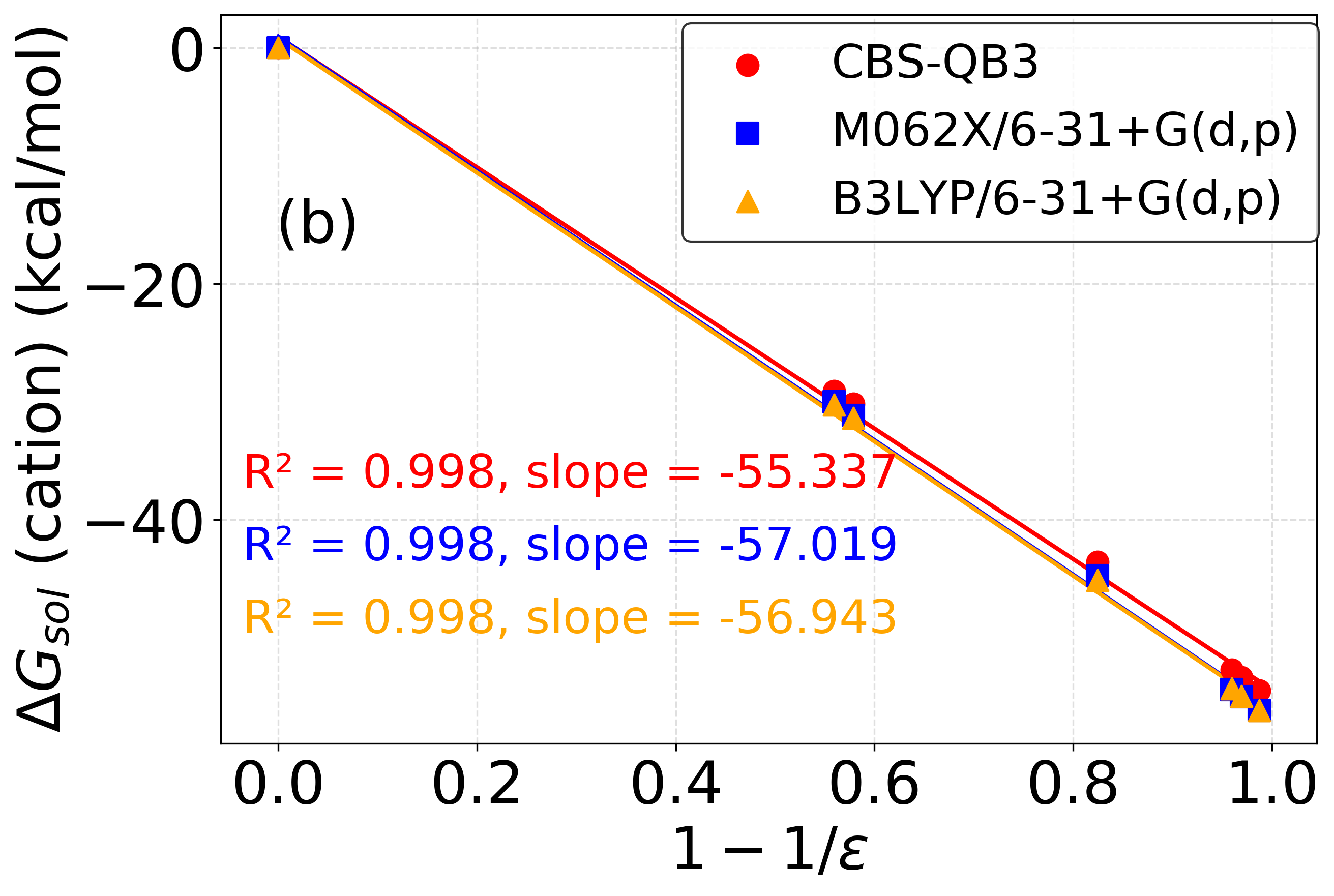}
    \includegraphics[width=0.23\textwidth,angle=0]{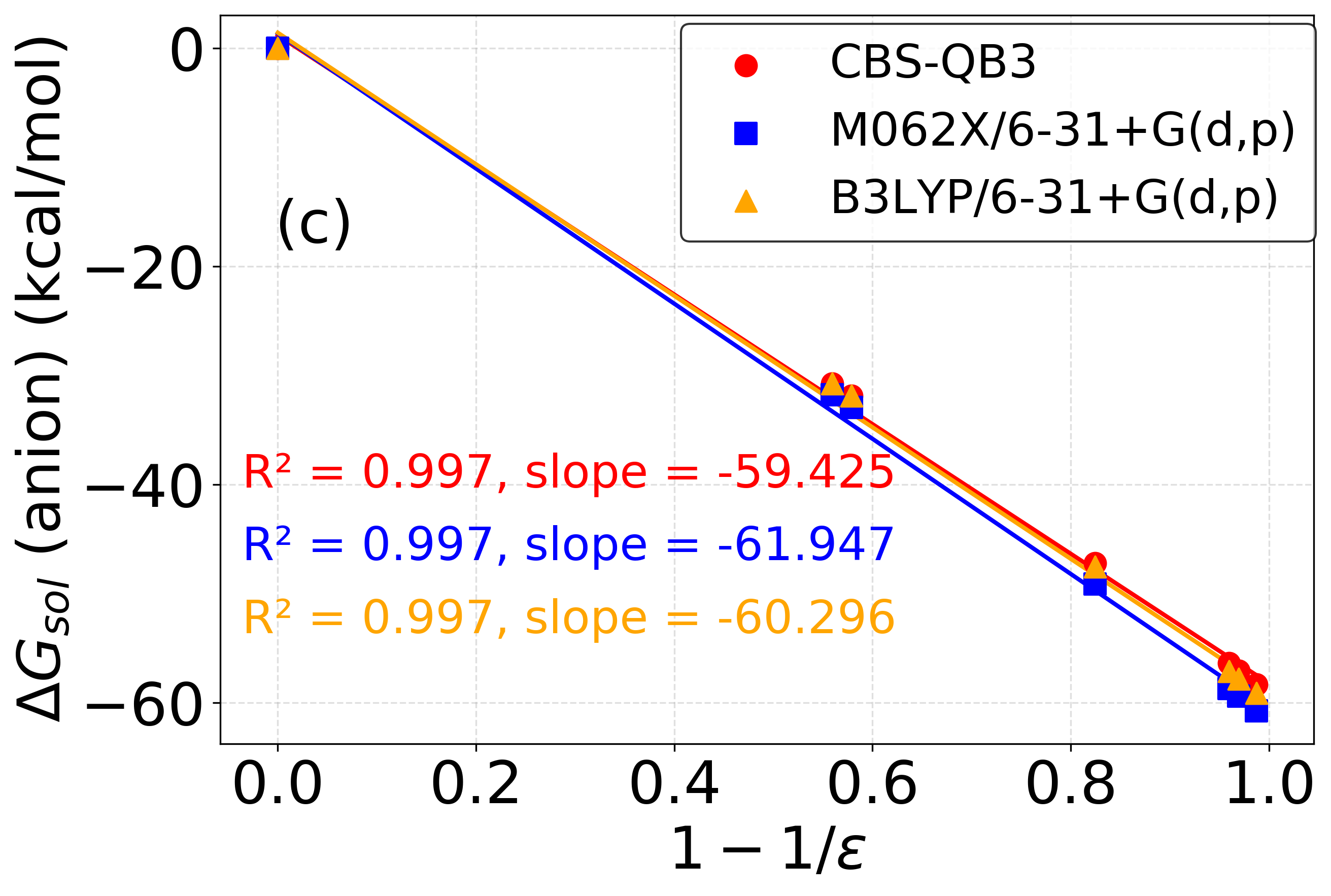}
  }
  \centerline{
    \includegraphics[width=0.23\textwidth,angle=0]{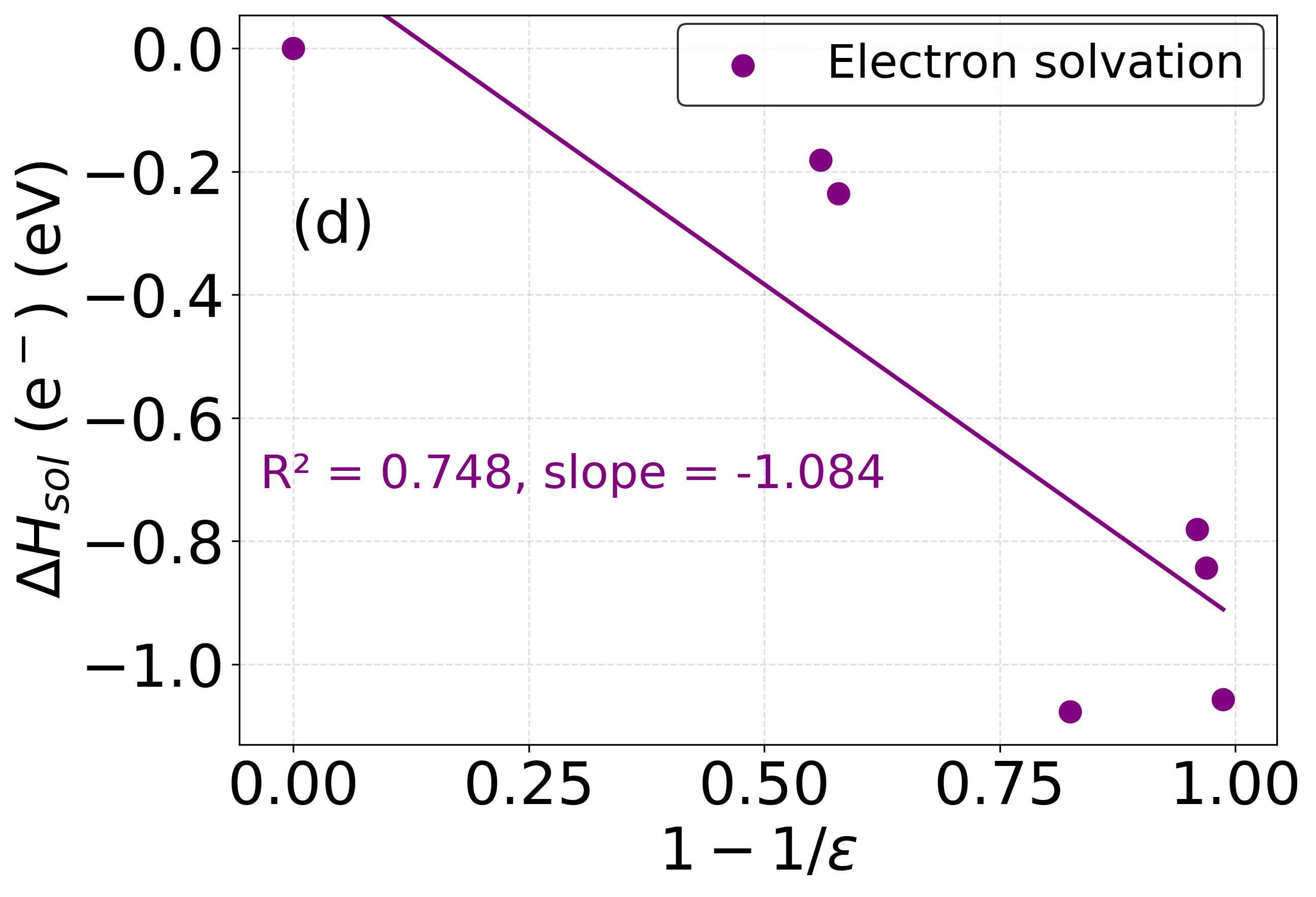}
  }
  \caption{Gibbs free energy of solvation $\Delta G_{\text{sol}}$ plotted against $1 - 1/\varepsilon$ for (a) neutral, (c) cationic, and (c) anionic species. Notice the nearly perfect linear dependence for the charged species which contrasts
    to the neutral molecules, a behavior which might be related to the deviation from linearity visible both in \Cref{fig:ip-ea-eg-chi}a,b,d and in the present panel (d), which depicts the electron enthalpy of solvation taken from literature.\cite{Markovic:16}
  }
      \label{fig:Delta_g}
\end{figure*}

Basically, this gap closure arises from differential solvation: both the cation and the anion are much stronger stabilized than the neutral (\Cref{fig:Delta_g}).
The net effect is a softer molecule (hardness $\eta = E_g/2$ drops from $\sim 4.35$\,eV in vacuo to $\sim 2.38$\,eV in water) and increased softness $\sigma = 1/\eta = 2 /E_g$ (from $\sim 0.12$ to $\sim 0.21$\,eV$^{-1}$, cf.~\Cref{tab:reactivity_vacuum,tab:reactivity_water})).

Interestingly, inspection of \Cref{fig:ip-ea-eg-chi} reveals a notable contrast.
$E_g$-data plotted versus Born polarity function lie virtually perfect on a straight line while,
although not dramatic, deviation from linearity of the
IP and especially EA data is visible ($R^2 = 0.96$ for IP and $R^2 \approx 0.90$ for EA).

The fact that, unlike the only approximately linear IP and EA (eqs.~(\ref{eq-IP}) and (\ref{eq-EA})),
the electron solvation enthalpy ($ \Delta H_{\text{sol}} \left(\left . \ce{e-}\right\vert {\varepsilon}\right) $)
does not contribute to ``perfectly'' linear $E_g$ (eq.~(\ref{eq-Eg})) may suggest
the departure from linearity visible for IP and EA arises because the
electron solvation enthalpy ($ H\left(\left . \ce{e-}\right\vert {\varepsilon}\right) $)
departs from linearity (\Cref{fig:Delta_g}d).
Indeed, upon subtracting literature values for electron's solvation enthalpy \cite{Markovic:16}, the resulting quantities
[EA $- \Delta H_{\text{sol}}\left(\left . \ce{e-}\right\vert {\varepsilon}\right)$ and
 IP $- \Delta H_{\text{sol}}\left(\left . \ce{e-}\right\vert {\varepsilon}\right)$]
display near-perfect linear dependence on $1 - 1/\epsilon$, achieving $R^2 = 1.0000$ (corrected IP) and $R^2 = 0.9996$ (corrected EA) for the reference CBS-QB3 method. Similar improvements in linearity are observed across all tested DFT functionals. This fact suggests that either the solvation enthalpies for the excess electron reported in Ref.~\citenum{Markovic:16} may warrant re-consideration or, alternatively,
that specific solvent-electron interactions beyond the simple Born model play a non-negligible role.
The latter possibility is
supported by the similar deviation from linearity characterizing cation's and anion's solvation
(\Cref{fig:Delta_g}a,b).

Chemically, the narrowed gap and increased softness facilitate electron donation to free radicals---the primary antioxidant mechanism.
The electroaccepting power $\omega^+$ rises sharply (from $\sim 0.46$\,eV in vacuo to $\sim 0.85$\,eV in water), while electrodonating power
$\omega^-$ decreases from 4.63\,eV to 4.06\,eV, consistent with ascorbic acid's role as a reducing agent.
The electrophilicity index $\omega$ also increases from 2.0\,eV to 2.2\,eV, reflecting greater reactivity toward electron-deficient species.

These trends are not merely quantitative; they explain why ascorbic acid is highly effective in aqueous physiological environments but far less so in non-polar media.

The foregoing results underscore that any assessment of ascorbic acid's antioxidant potency must account for the solvent environment. The large, systematic variations in IP, EA, and $E_g$ with dielectric constant demonstrate that gas-phase calculations alone are insufficient to describe redox activity in solution. The following section will critically examine the consequences of this solvent dependence for reactivity descriptors and for the common practice of estimating redox potentials from orbital energies (Koopmans theorem).

\subsection{Why Koopmans Theorem Fails to Correctly Capture the Chemical Reactivity in Solvents}
\label{sec:koopmans}

Koopmans' theorem (KT) approximates the vertical ionization potential and electron affinity as
\(\text{IP}_{\text{KT}} = -\varepsilon_{\text{HOMO}}\) and
\(\text{EA}_{\text{KT}} = -\varepsilon_{\text{LUMO}}\) of the neutral molecule at its equilibrium geometry. While computationally inexpensive, this approximation rests on severe assumptions: frozen orbitals, no nuclear relaxation, and a vacuum reference for the added/removed electron. For antioxidants operating in solution, these assumptions break down catastrophically, as demonstrated below.

\subsubsection{The Arbitrariness of Orbital Energies in Koopmans' Theorem}
\label{sec:koopmans-arbitrary}

Before discussing the physical shortcomings, it is essential to recognize that the orbital energies themselves---particularly the LUMO---are not well-defined physical observables in Kohn--Sham (KS) or Hartree--Fock (HF) theory. The KS/HF LUMO is a mathematical construct determined solely by orthogonality to the occupied manifold and lacks direct physical meaning \cite{Parr:89,Gunnarson:89,Kohn:96,Baldea:2014c}. Consequently, its numerical value behaves essentially as a random number generator: it varies wildly with the choice of functional, exact-exchange admixture, and basis set.

This arbitrariness is glaringly evident in our computed data. For instance, in vacuo, the KT-derived EA (\(-E_{\text{LUMO}}\)) for ascorbic acid ranges from \(+1.167\)\,eV (B3LYP/6-31+G(d,p)) to \(-0.347\)\,eV (M062X/6-311G(3df,3pd))---a spread exceeding \(1.5\)\,eV (Table~\ref{tab:koopmans}).
Similarly, the HOMO energy (and hence KT-IP) varies by more than \(2\)\,eV across the
the set of methods considered, with the Hartree-Fock yielding the highest values.
In water, the situation is even more extreme: the KT-based EA shifts from \(+1.265\)\,eV (B3LYP/6-31+G(d,p)) to \(-0.220\)\,eV (M062X/6-311G(3df,3pd)), while the reference adiabatic EA from CBS-QB3 is \(+0.826\)\,eV. Such wild, method-dependent fluctuations render any single KT estimate physically meaningless and highlight that KT's failure begins with the ill-defined nature of the orbital energies on which it relies.

The ensuing subsections demonstrate that, even when fixing a method, KT fails to capture the essential physics of solvation and nuclear relaxation. However, this inherent arbitrariness of the orbital energies provides an additional, fundamental reason to avoid KT for quantitative predictions.

\subsubsection{In Vacuo: Large Deviations Even in the Gas Phase}
In vacuo, KT already yields substantial errors.

For instance, at the M06-2X/6-31+G(d,p) level, \(\text{IP}_{\text{KT}} \approx 7.90\)\,eV while the vertical IP (without zero-point energy) is 8.88\,eV---a deviation of nearly 1.0\,eV. The corresponding KT-EA estimate is \(+0.195\)\,eV (Table~\ref{tab:koopmans}), whereas the vertical EA is \(-0.621\)\,eV---an error exceeding 0.8\,eV. Hartree--Fock (HF) yields an even more erratic LUMO energy, giving \(\text{EA}_{\text{KT}} = -1.543\)\,eV in vacuo (Table~\ref{tab:koopmans}), i.e., a bound anion is predicted to be unbound, which is qualitatively wrong. These discrepancies translate into unreliable reactivity indices. For example, the KT-derived hardness \(\eta_{\text{KT}} = (-\varepsilon_{\text{LUMO}} + \varepsilon_{\text{HOMO}})/2\) in vacuo is 3.85\,eV (M06-2X/6-31+G(d,p)), whereas the adiabatic hardness from CBS-QB3 is 4.35\,eV---a 0.5\,eV underestimation that would artificially enhance the predicted softness and reactivity.

\subsubsection{In Solution: A Complete Breakdown}
The failure becomes dramatic when moving to solvents. KT predicts IP and EA values that are virtually independent of the dielectric environment (see \Cref{tab:koopmans} and \Cref{fig:koopmans}).
For water, \(\text{IP}_{\text{KT}}\) remains around 8.02\,eV (M06-2X/6-31+G(d,p)) and \(\text{EA}_{\text{KT}} \approx 0.085\)\,eV (Table~\ref{tab:koopmans}), whereas the adiabatic IP drops to 5.44\,eV and the adiabatic EA rises to 0.80\,eV (CBS-QB3).
The KT-derived HOMO-LUMO gap (\(\sim\)7.93\,eV in water) is more than double the actual adiabatic gap (4.76\,eV), leading to a hardness overestimation of about 1.6\,eV. Consequently, all global reactivity indices computed via KT are qualitatively wrong in solution: they suggest a hard, poorly reactive molecule, whereas the true solvent-stabilized species is soft and an excellent electron donor.

\subsubsection{Making the Comparison ``Fair'': Vertical Values and Electron Solvation}
\begin{figure*}
  \centerline{
    \includegraphics[width=0.23\textwidth,angle=0]{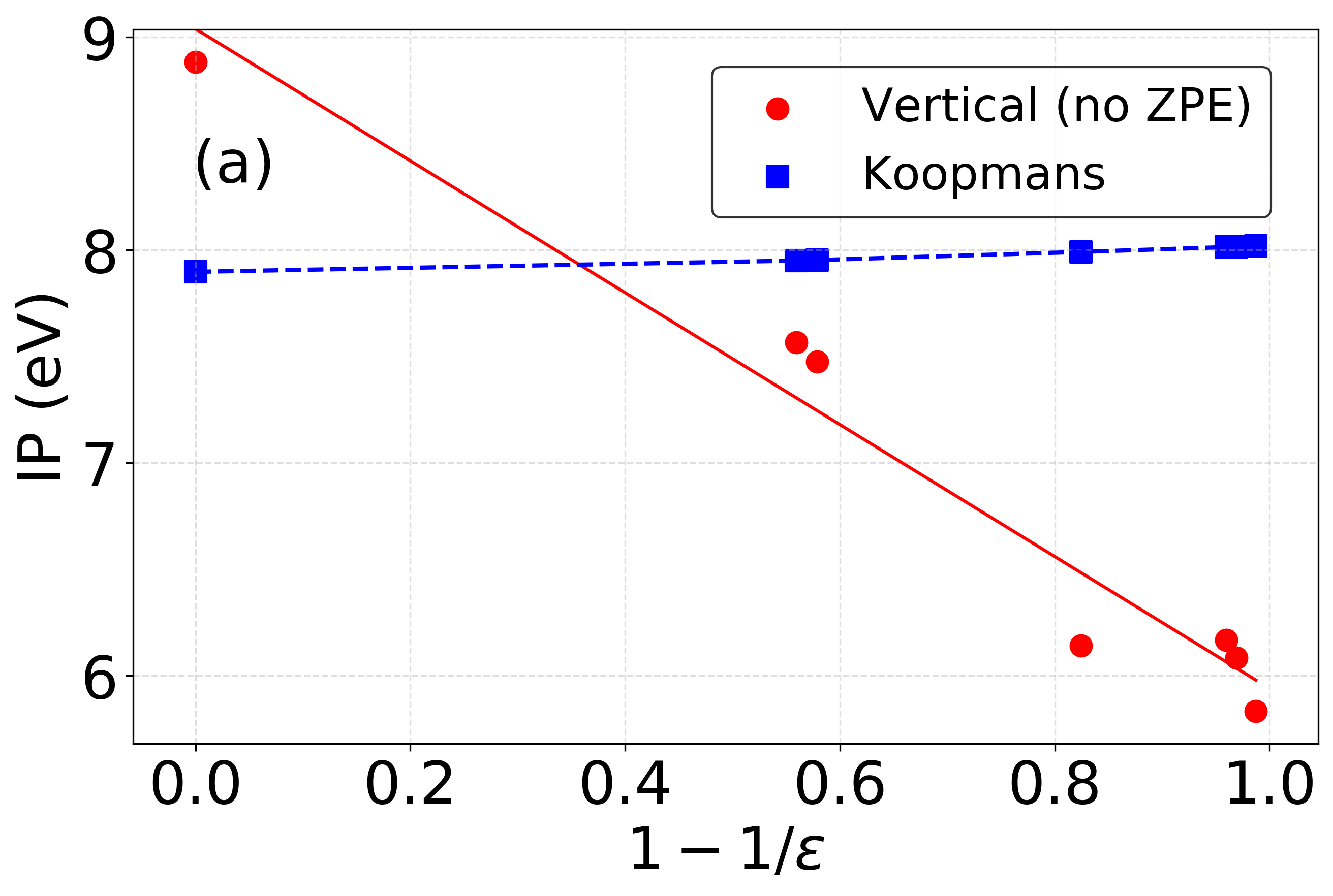}
    \includegraphics[width=0.23\textwidth,angle=0]{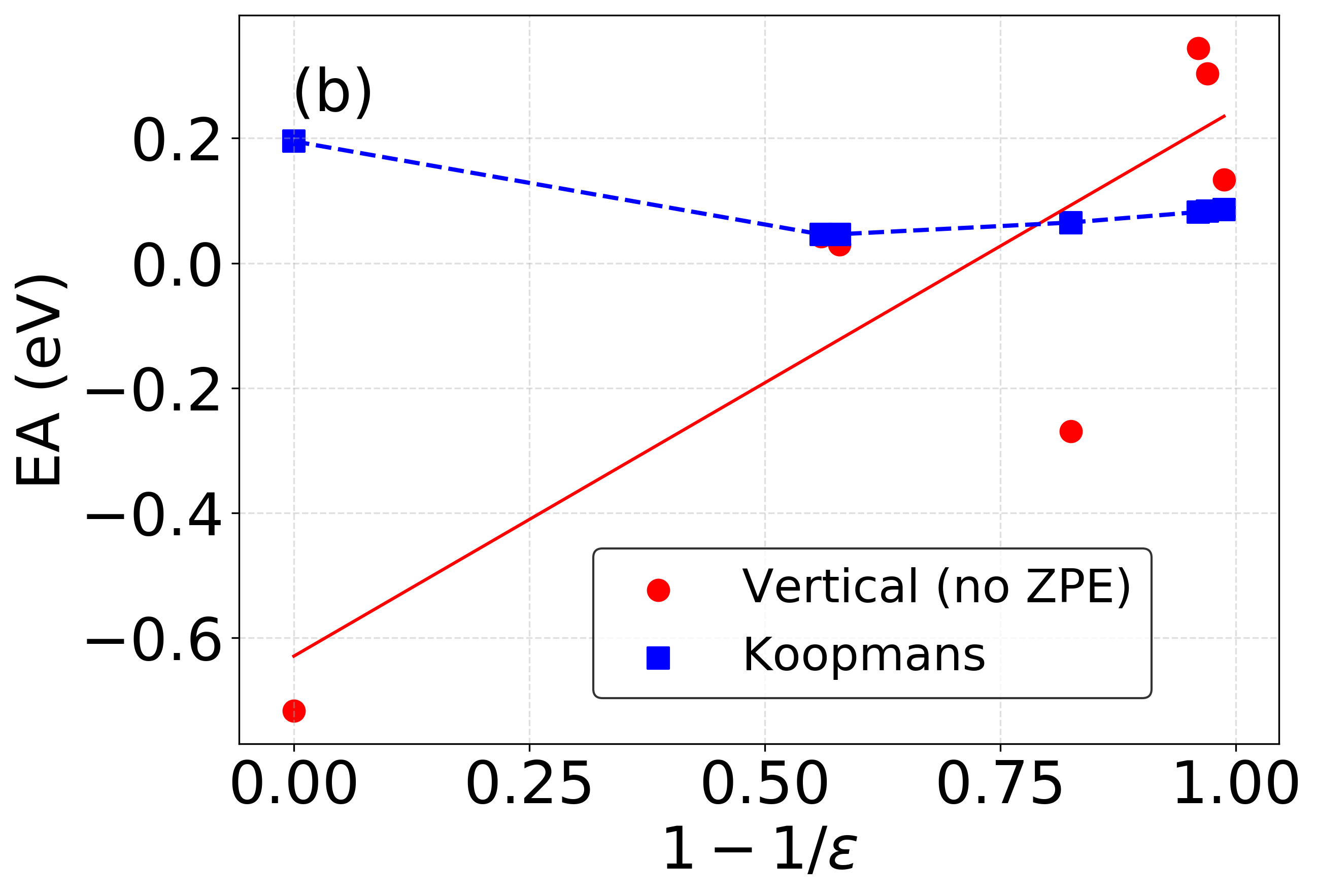}
    \includegraphics[width=0.23\textwidth,angle=0]{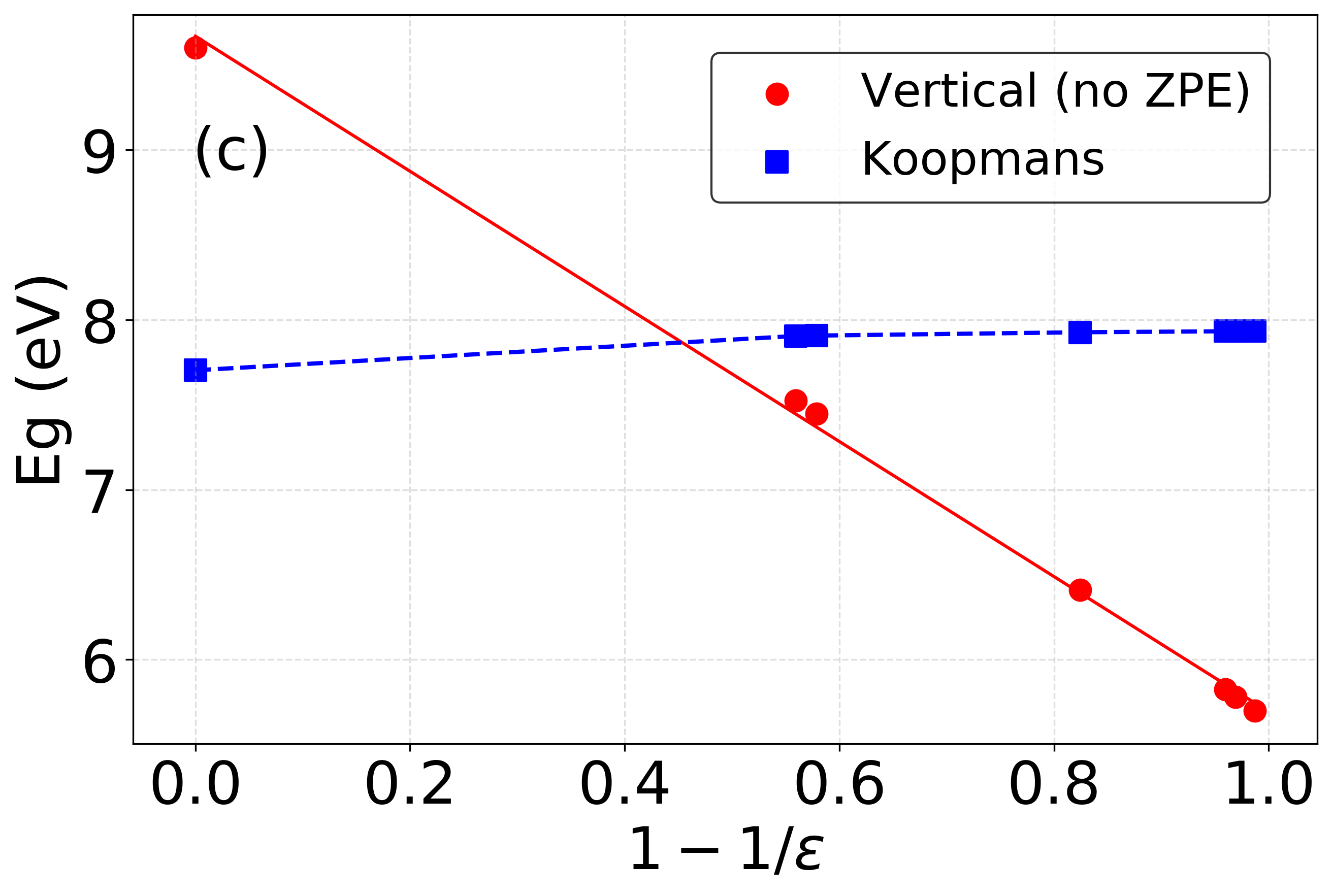}
    \includegraphics[width=0.23\textwidth,angle=0]{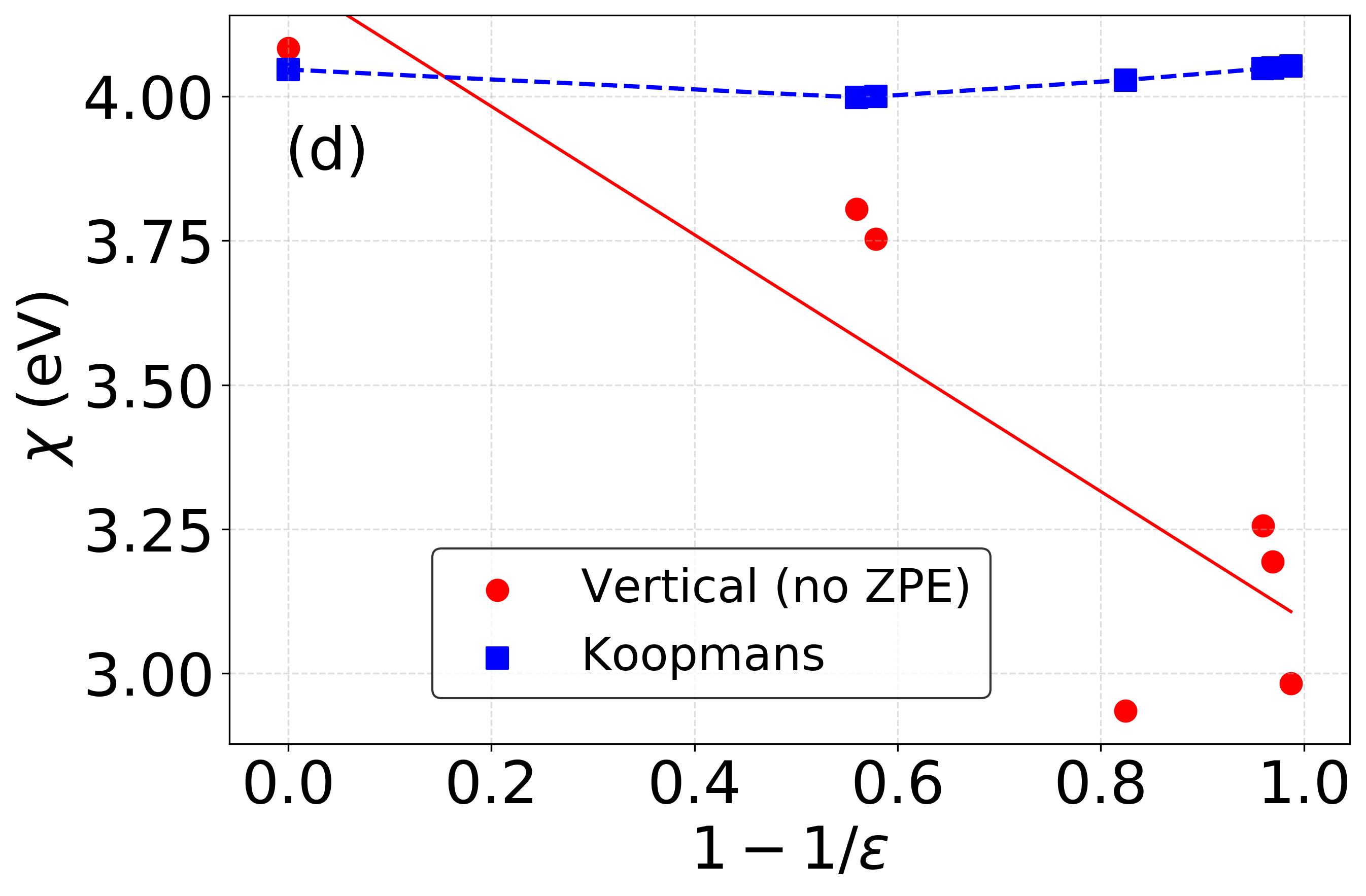}
  }
  \centerline{
    \includegraphics[width=0.23\textwidth,angle=0]{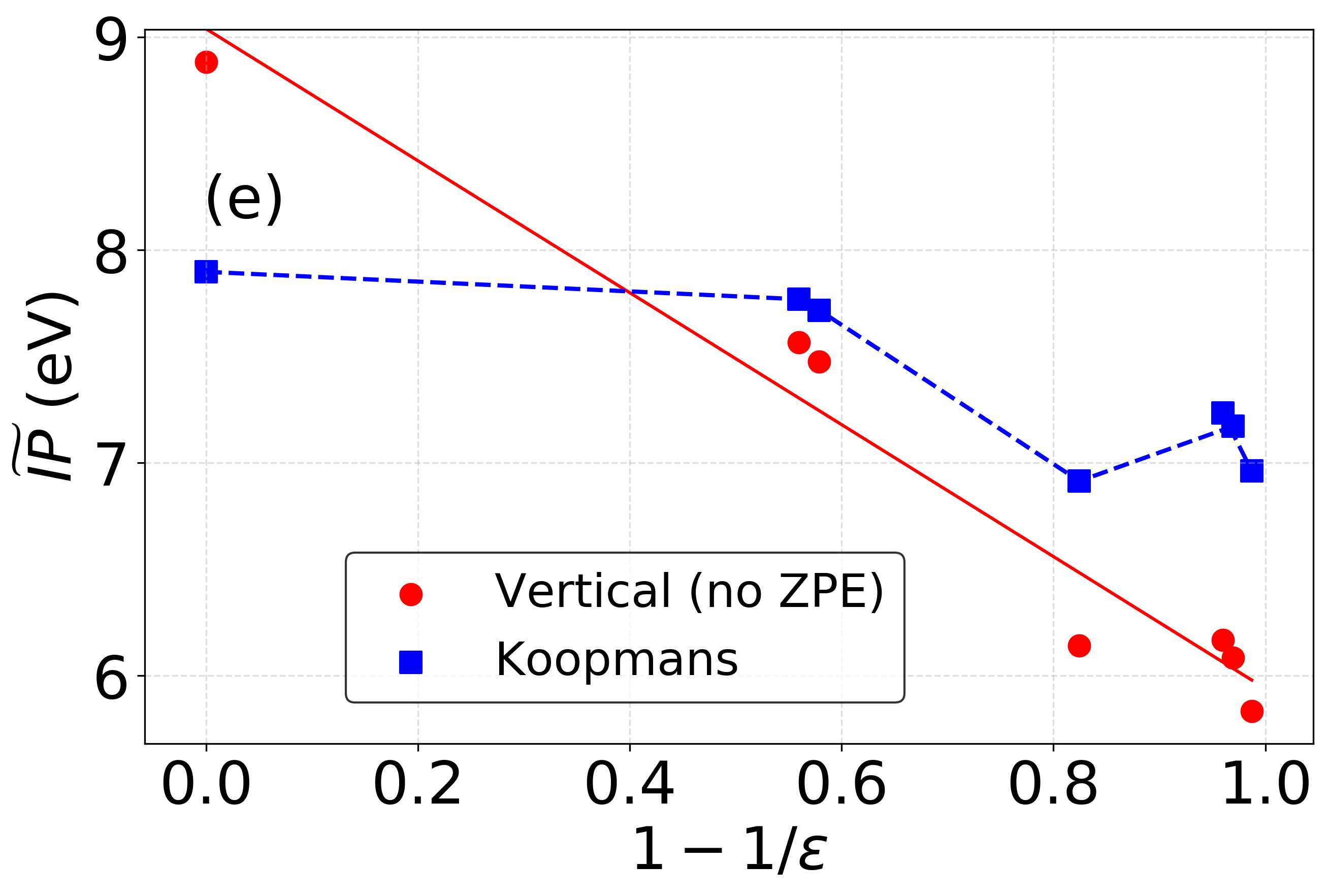}
    \includegraphics[width=0.23\textwidth,angle=0]{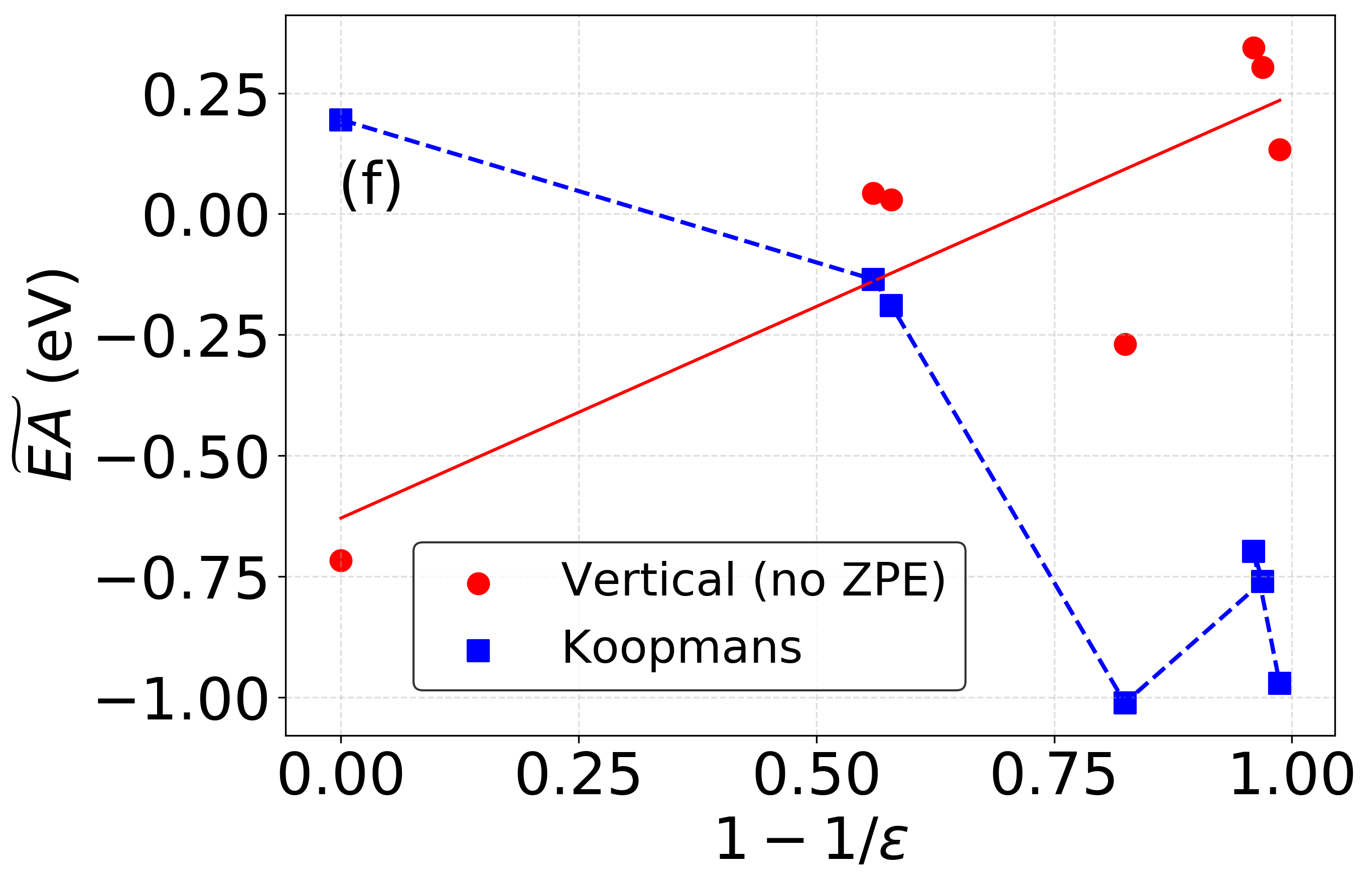}
    \includegraphics[width=0.23\textwidth,angle=0]{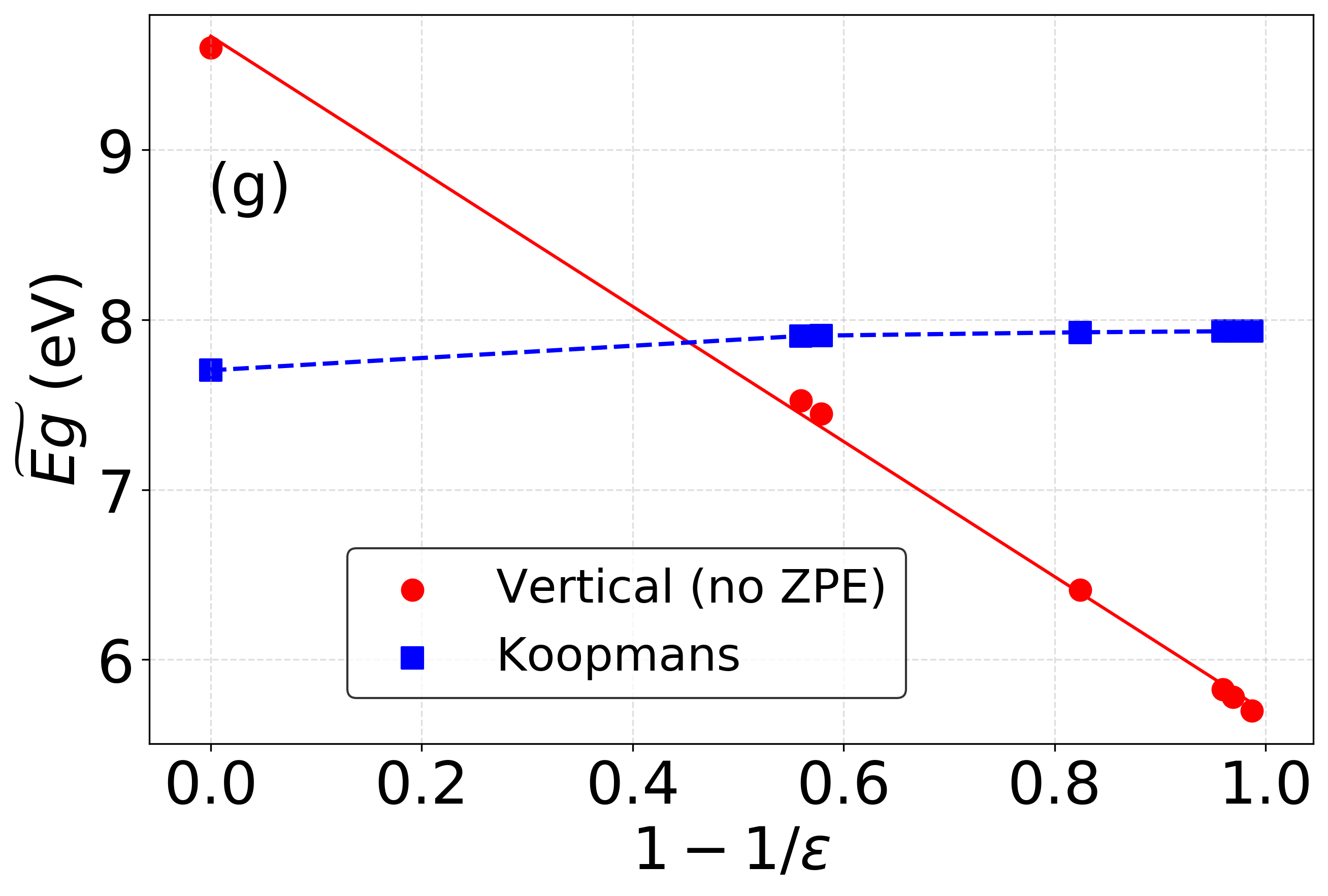}
    \includegraphics[width=0.23\textwidth,angle=0]{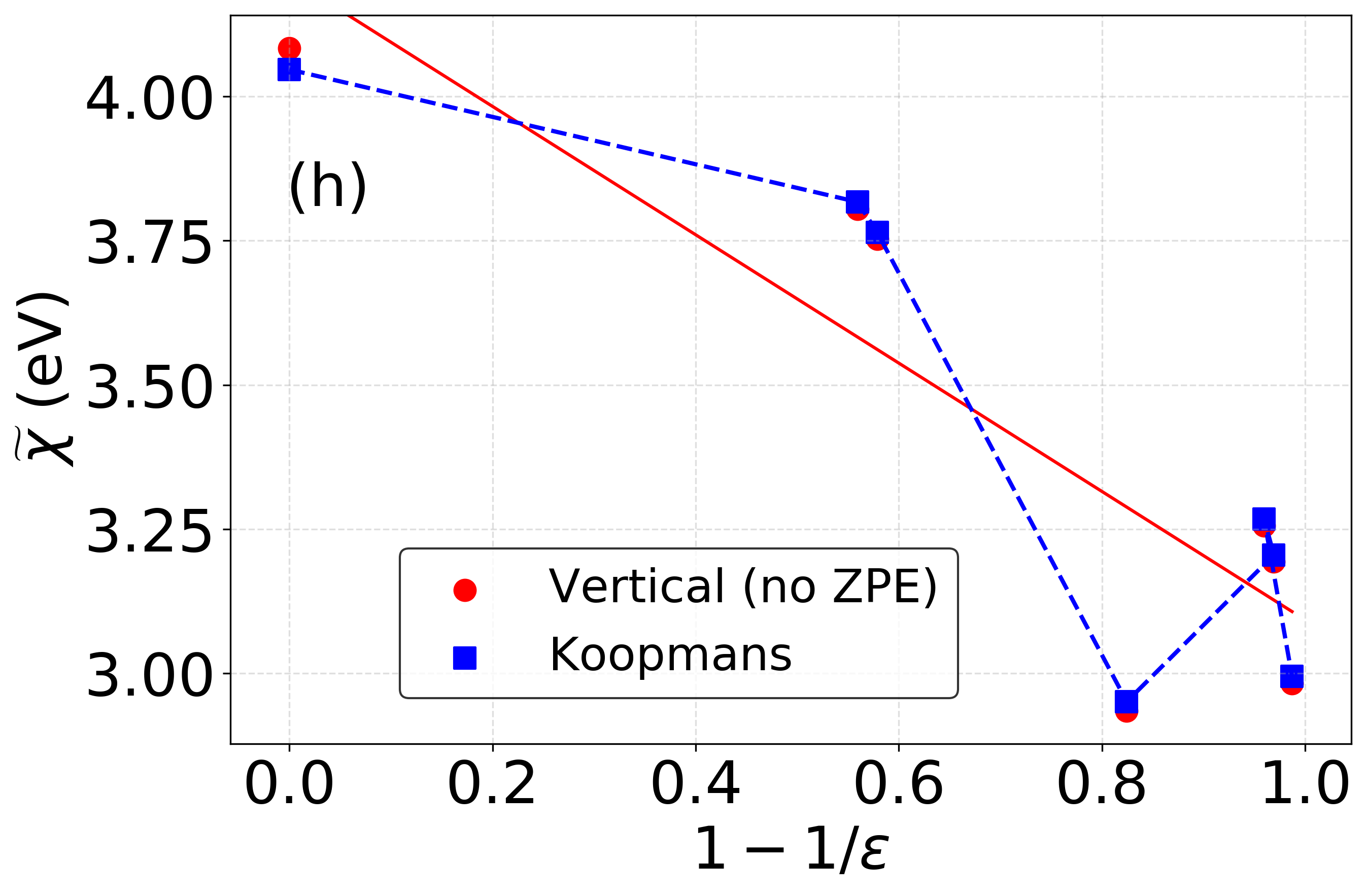}
  }
  \caption{(a) Ionization potential ($-E_{\text{HOMO}}$) and (b) electron affinity ($-E_{\text{LUMO}}$)
    and related quantities---(c) HOMO-LUMO gap $E_g$ and (d) electronegativity $\chi$---
    estimated via Koopmans theorem plotted versus $1 - 1/\varepsilon$.
    To better demonstrate the failure of the Koopmans theorem, vertical IP and EA values which (as inherently the case
    for the Koopmans-based estimates) do not include corrections due to zero point motion.
    The quantitative and qualitative disagreement is evident. Letting alone substantial differences,
    the Koopmans-based estimates are virtually solvent independent (same value as in vacuo),
    in marked contrast to the vertical ones computed via eqs.~(\ref{eq-IP_vert}) and (\ref{eq-EA_vert}),
    whose strong, roughly linear dependence on $1 - 1/\varepsilon$ is similar both to that of their adiabatic
    counterparts (\Cref{fig:ip-ea-eg-chi}) and to the vertical properties including zero-point corrections
    (\Cref{fig:ip-ea-eg-chi-vert}).
  }
      \label{fig:koopmans}
\end{figure*}

One might argue that KT provides \emph{vertical} estimates, and a fairer comparison should be with vertical IP/EA (without zero-point corrections). \Cref{tab:vertical_reactivities} and \Cref{fig:ip-ea-eg-chi-vert}a--d show that vertical IP and EA decrease/increase roughly linearly with the Born polarity function \(1-1/\varepsilon\), reflecting the differential solvation of the instantaneously created charge. In contrast, KT-based values are flat lines (\Cref{fig:koopmans}a--d). Even after adding the experimental solvation enthalpy of the electron
\(\Delta H_{\text{sol}}\left(\left . \ce{e-}\right\vert {\varepsilon}\right)\)
to the KT estimates (to account for the fact that the electron is released into the solvent, not vacuum), the resulting ``corrected'' KT values remain poor and substantially deviate from the true vertical trends (\Cref{fig:koopmans}e--h). This confirms that the failure of KT is not merely a neglect of electron solvation; it stems from its inability to capture the orbital relaxation and the dielectric response of the solvent to the newly created charge.

\subsubsection{Why Koopmans Fails So Badly in Solution}
The physical reasons are twofold. First, KT ignores nuclear relaxation. The reorganization energies for ascorbic acid are substantial (\(\lambda_c \approx 0.5\)\,eV, \(\lambda_a \approx 0.6\)\,eV in water, see \Cref{fig:lambdas}), so adiabatic IP/EA---which determine thermodynamic redox potentials---differ significantly from vertical ones. Antioxidant reactions are not Franck--Condon processes; they occur on timescales that allow full geometric relaxation of the oxidized/reduced species. Second, and more critically, KT treats the added/removed electron as a vacuum particle. In reality, the electron is solvated, and the solvent dielectric strongly stabilizes the cation and anion. This differential stabilization, which scales roughly as \(1-1/\varepsilon\), is entirely absent from the orbital energies of the neutral molecule. Hence, KT cannot reproduce the dramatic solvent-induced reductions in IP and increases in EA that are essential for understanding antioxidant activity in biological media.\\

The present analysis, combining high-level reference calculations (CBS-QB3) with a broad range of DFT functionals and solvents, unequivocally demonstrates that Koopmans' theorem is both quantitatively and qualitatively inadequate for estimating chemical reactivity indices of antioxidants. Beyond the well-known physical approximations (frozen orbitals, no nuclear relaxation, vacuum reference), KT relies on orbital energies that are themselves arbitrary---varying by more than 1.5\,eV with functional and basis set---and bear no consistent relation to true vertical or adiabatic electron affinities. In solution, KT's failure is catastrophic: it predicts solvent-independent IP and EA values, completely missing the several-eV shifts induced by dielectric screening. Even after empirical corrections for electron solvation, KT cannot capture the essential physics of charge stabilization in polar media. Therefore, we strongly advise against using Koopmans' theorem for any study of antioxidants in solution. Adiabatic IP and EA from explicit \(\Delta\)-SCF calculations---including proper treatment of solvation and full geometry relaxation---are mandatory for meaningful predictions.
\begin{table*}[htbp]
\centering
\caption{Global reactivity descriptors for ascorbic acid computed using Koopmans theorem computed with selected methods.}
\label{tab:koopmans}
\setlength{\tabcolsep}{4.5pt}
\scriptsize
\begin{tabular}{lrrrrrrr}
\hline
Descriptor & Vacuum & Benzene & Toluene & Chlorobenzene & Methanol & Ethanol & Water \\
\hline
\multicolumn{8}{l}{\textbf{M062X/6-31+G(d,p)}} \\
\hline
IP (eV) & 7.898 & 7.950 & 7.953 & 7.991 & 8.015 & 8.014 & 8.019 \\
EA (eV) & 0.195 & 0.046 & 0.046 & 0.065 & 0.083 & 0.082 & 0.085 \\
$E_g$ (eV) & 7.702 & 7.904 & 7.907 & 7.926 & 7.932 & 7.932 & 7.933 \\
$\eta$ (eV) & 3.851 & 3.952 & 3.953 & 3.963 & 3.966 & 3.966 & 3.967 \\
$\sigma$ (eV$^{-1}$) & 0.130 & 0.127 & 0.126 & 0.126 & 0.126 & 0.126 & 0.126 \\
$\chi$ (eV) & 4.046 & 3.998 & 4.000 & 4.028 & 4.049 & 4.048 & 4.052 \\
$\omega$ (eV) & 2.126 & 2.022 & 2.023 & 2.047 & 2.067 & 2.066 & 2.070 \\
$\omega^{-}$ (eV) & 4.630 & 4.515 & 4.517 & 4.557 & 4.587 & 4.585 & 4.592 \\
$\omega^{+}$ (eV) & 0.584 & 0.517 & 0.518 & 0.528 & 0.538 & 0.537 & 0.539 \\
\hline
\multicolumn{8}{l}{\textbf{B3LYP/6-31+G(d,p)}} \\
\hline
IP (eV) & 6.503 & 6.544 & 6.547 & 6.579 & 6.600 & 6.599 & 6.602 \\
EA (eV) & 1.167 & 1.205 & 1.208 & 1.241 & 1.262 & 1.261 & 1.265 \\
$E_g$ (eV) & 5.337 & 5.339 & 5.339 & 5.338 & 5.338 & 5.338 & 5.337 \\
$\eta$ (eV) & 2.668 & 2.669 & 2.669 & 2.669 & 2.669 & 2.669 & 2.669 \\
$\sigma$ (eV$^{-1}$) & 0.187 & 0.187 & 0.187 & 0.187 & 0.187 & 0.187 & 0.187 \\
$\chi$ (eV) & 3.835 & 3.875 & 3.877 & 3.910 & 3.931 & 3.930 & 3.933 \\
$\omega$ (eV) & 2.756 & 2.812 & 2.816 & 2.864 & 2.895 & 2.893 & 2.899 \\
$\omega^{-}$ (eV) & 5.007 & 5.083 & 5.088 & 5.153 & 5.194 & 5.192 & 5.199 \\
$\omega^{+}$ (eV) & 1.172 & 1.209 & 1.211 & 1.243 & 1.263 & 1.262 & 1.266 \\
\hline
\multicolumn{8}{l}{\textbf{HF/6-31+G(d,p)}} \\
\hline
IP (eV) & 9.472 & 9.519 & 9.522 & --- & 9.569 & 9.569 & 9.571 \\
EA (eV) & -1.543 & -1.775 & -1.781 & --- & -1.849 & -1.849 & -1.848 \\
$E_g$ (eV) & 11.014 & 11.294 & 11.303 & --- & 11.418 & 11.418 & 11.420 \\
$\eta$ (eV) & 5.507 & 5.647 & 5.652 & --- & 5.709 & 5.709 & 5.710 \\
$\sigma$ (eV$^{-1}$) & 0.091 & 0.089 & 0.088 & --- & 0.088 & 0.088 & 0.088 \\
$\chi$ (eV) & 3.965 & 3.872 & 3.870 & --- & 3.860 & 3.860 & 3.861 \\
$\omega$ (eV) & 1.427 & 1.328 & 1.325 & --- & 1.305 & 1.305 & 1.306 \\
$\omega^{-}$ (eV) & 4.098 & 3.970 & 3.967 & --- & 3.949 & 3.948 & 3.950 \\
$\omega^{+}$ (eV) & 0.133 & 0.097 & 0.097 & --- & 0.089 & 0.088 & 0.089 \\
\hline
\end{tabular}
\end{table*}

\ib{

}

\section{Conclusion}
\label{sec:conclusion}

This comprehensive quantum chemical study of ascorbic acid---a prototypical biological antioxidant---delivers two clear, interconnected messages of broad relevance to the field of antioxidant research.

First, \textit{solvent effects are paramount} and must be explicitly included when assessing antioxidant reactivity. Ascorbic acid in vacuo is a hard, high-gap molecule
(IP\,$\approx$\,8.5\,eV, EA\,$\sim -0.2$\,eV, \(E_g \approx \) 8.7\,eV), exhibiting modest electron-donating propensity. In aqueous solution, however, the stabilization of the cation and anion by the polar environment dramatically lowers the ionization potential (to\,$\approx$\,5.6\,eV), raises the electron affinity (to\,$\approx$\,0.8\,eV), and narrows the fundamental gap to\,$\approx$\,4.8\,eV. This solvent-induced softening (hardness drops from $\sim$4.4\,eV to $\sim$2.4\,eV) and the concomitant increase in electrophilicity and electroaccepting power are precisely what enable ascorbic acid to function as an efficient electron donor in biological media. The trends across solvents of varying polarity follow a roughly Born-like dependence, underscoring the dominant role of electrostatic stabilization. Any meaningful prediction of antioxidant potency must therefore be based on calculations that properly account for solvation, ideally using adiabatic IP and EA obtained from geometry-optimized ions in the relevant dielectric environment.

Second, and more critically, \textit{Koopmans' theorem must never be used to estimate reactivity indices of antioxidants in solution}. Our analysis demonstrates that KT fails both quantitatively and qualitatively. In vacuo, it already deviates by $0.5-1.0$\,eV from vertical IP/EA due to its neglect of orbital and nuclear relaxation. In solution, the failure becomes catastrophic: KT predicts IP and EA values that are virtually independent of solvent polarity, completely missing the several-eV shifts that arise from differential solvation of the charged species. Consequently, KT-derived hardness, softness, electrophilicity, and donating/accepting powers are not merely inaccurate---they are physically misleading, suggesting a nearly solvent-insensitive, hard molecule when the true system is soft and highly responsive to its environment. This breakdown stems from KT's inherent assumptions (frozen orbitals, no geometry relaxation, vacuum reference for the electron), which are fundamentally incompatible with the physics of redox processes in condensed phases.

From a computational standpoint, the meta-hybrid functional M06-2X with a moderate basis set such as 6-31+G(d,p) provides a reliable and cost-effective alternative to high-level composite methods like CBS-QB3 for evaluating adiabatic reactivity indices in solution, showing mean absolute deviations below 0.1\,eV for most descriptors. However, the choice of functional is secondary to the necessity of performing explicit $\Delta$-SCF calculations (including full geometry optimization of the neutral and ionic species) coupled with a continuum solvation model. Vertical estimates, while computationally simpler, still capture the essential solvent trends but neglect the substantial reorganization energies ($\approx$\,0.5\,eV for both cation and anion) that are relevant for thermodynamic redox potentials.

In summary, the chemical reactivity of antioxidants cannot be understood in isolation from their solvation environment, and approximations that ignore this environment---most notably Koopmans' theorem---produce results that are not only quantitatively wrong but qualitatively inverted. We therefore urge researchers to abandon the use of Koopmans' theorem for antioxidant studies in solution and instead adopt adiabatic calculations that explicitly account for solvent stabilization of the charged species. This approach is essential for developing accurate, predictive models of antioxidant activity in biological systems.

\begin{figure*}
  \centerline{
    \includegraphics[width=0.45\textwidth,angle=0]{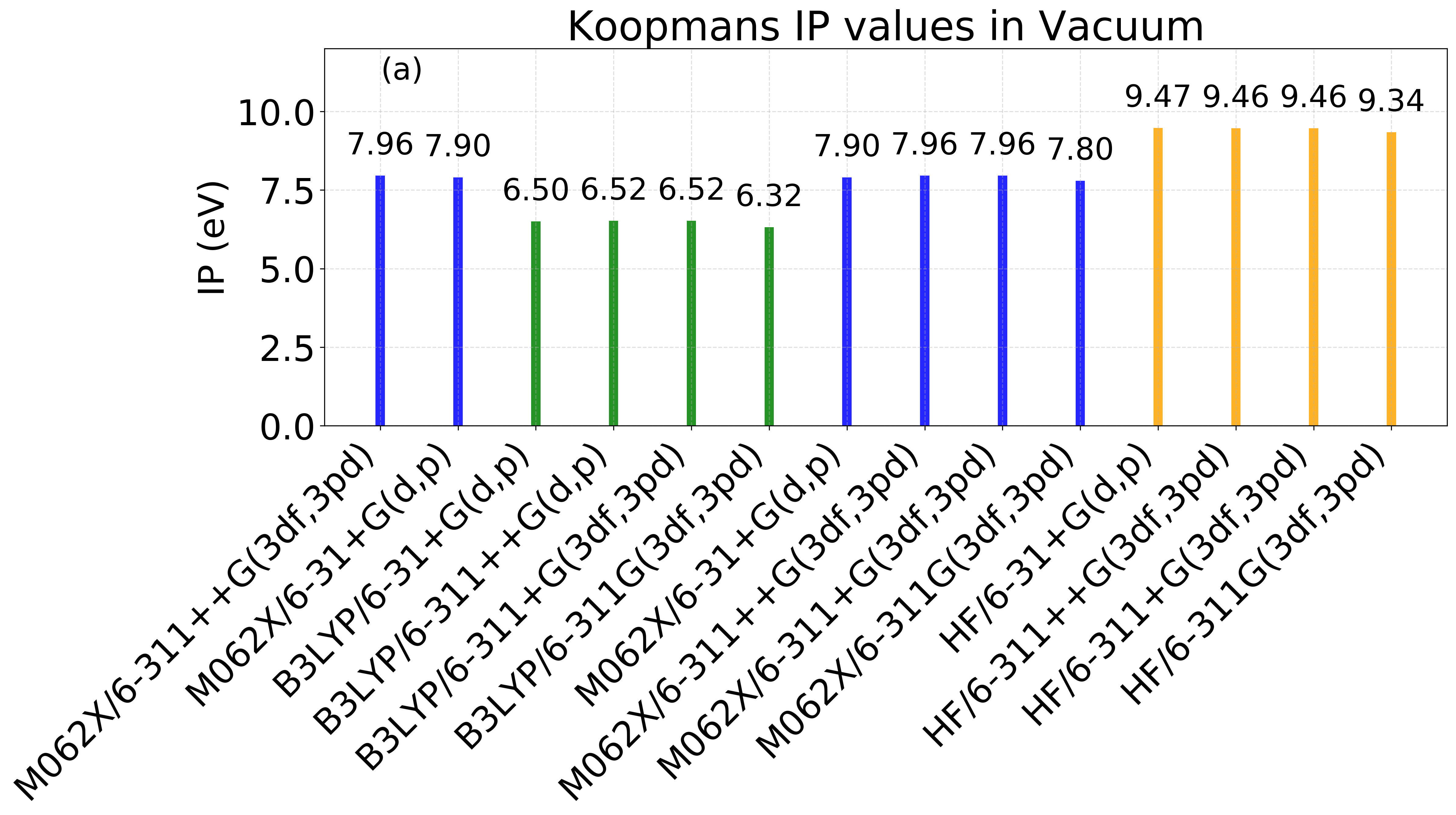}
    \includegraphics[width=0.45\textwidth,angle=0]{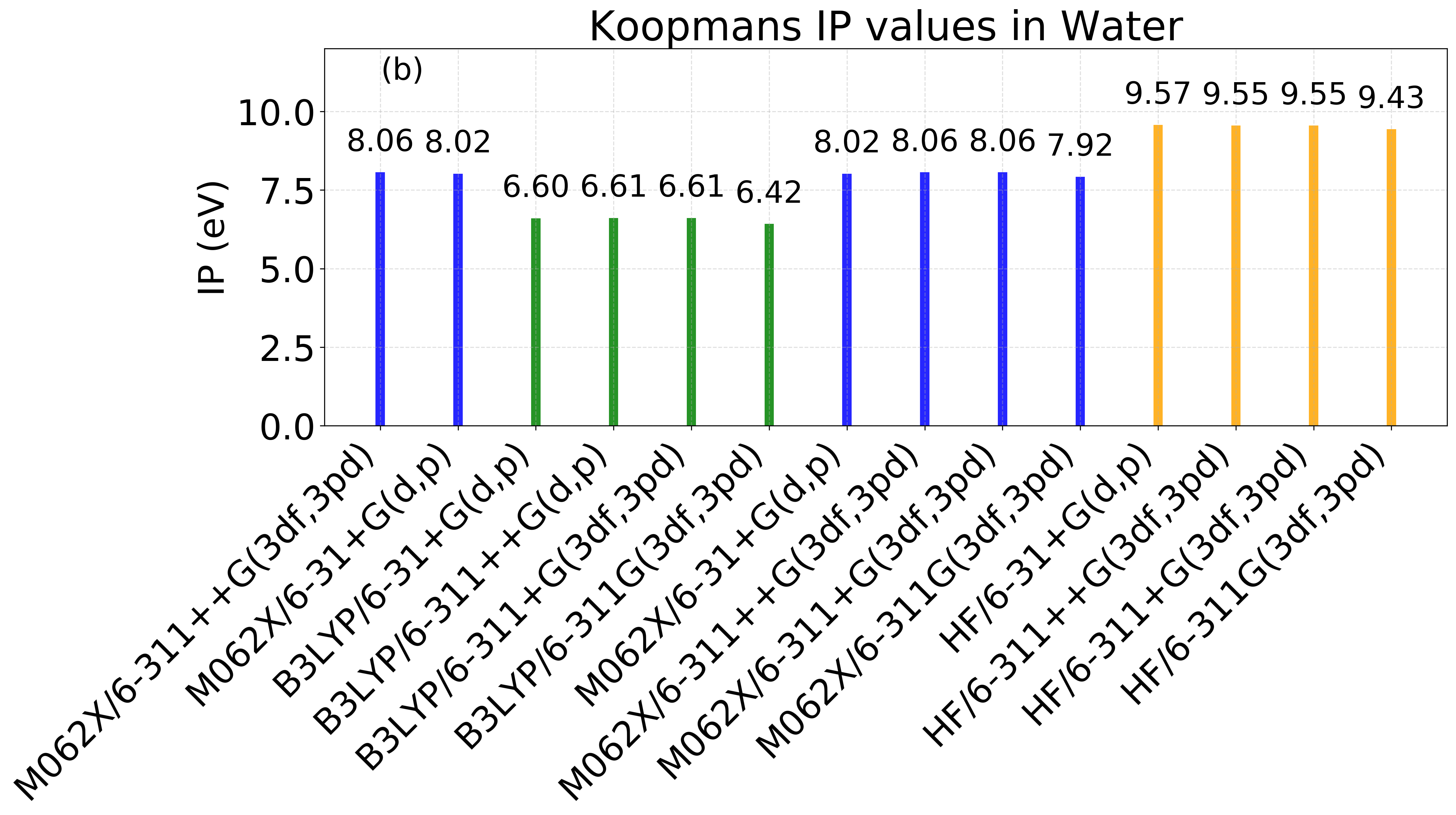}
  }
  \centerline{
    \includegraphics[width=0.45\textwidth,angle=0]{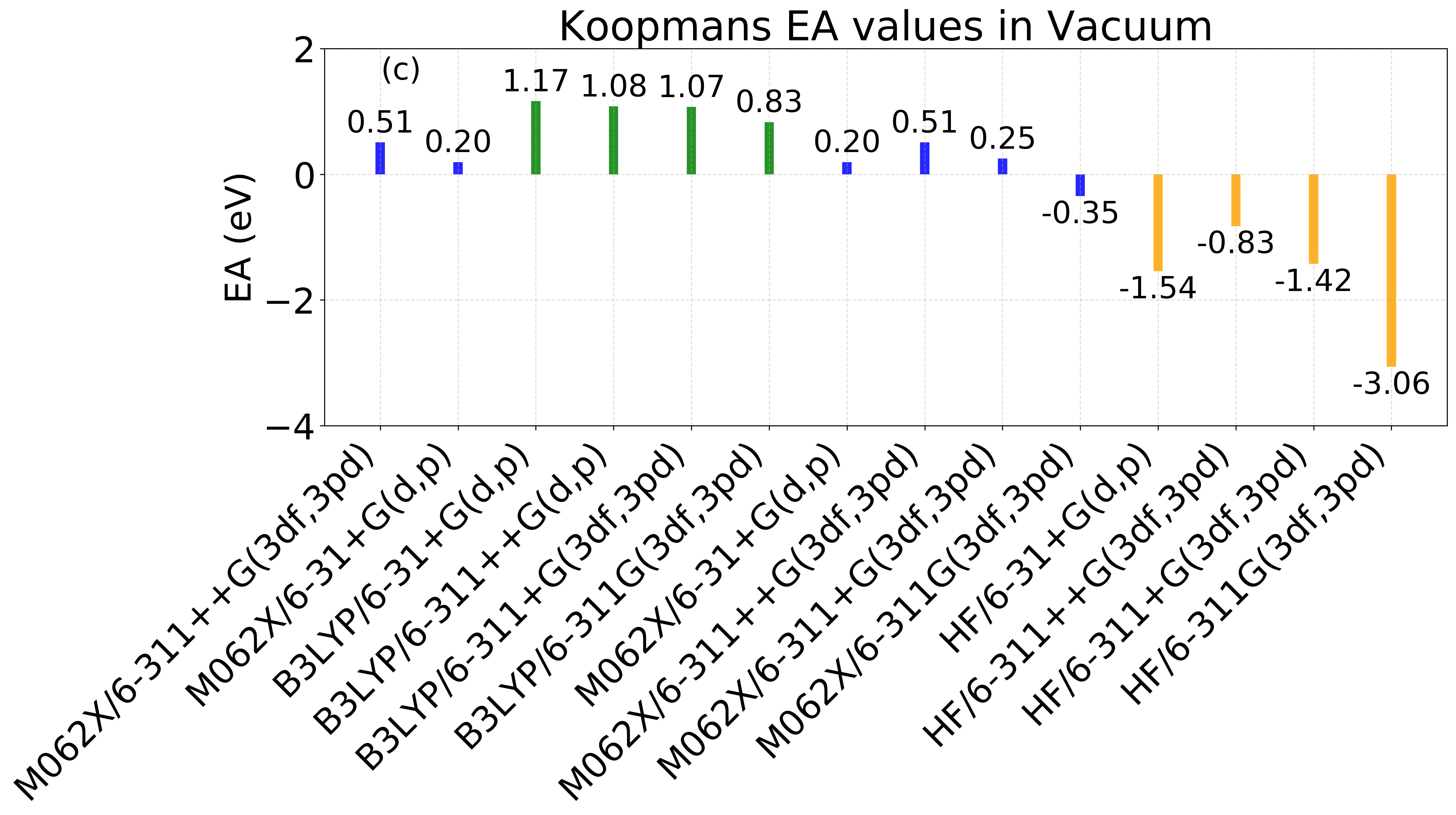}
    \includegraphics[width=0.45\textwidth,angle=0]{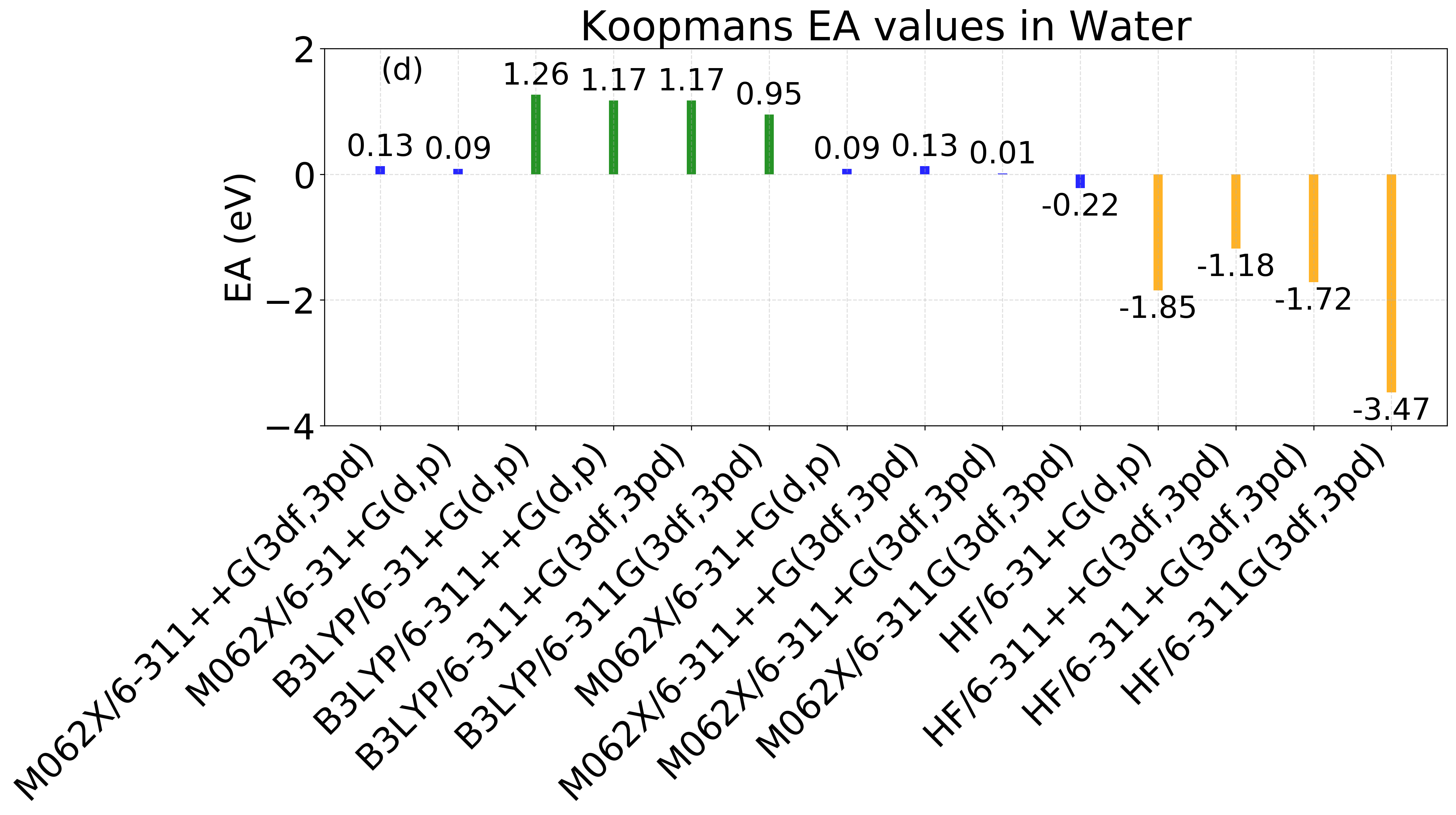}
  }
  \centerline{
    \includegraphics[width=0.45\textwidth,angle=0]{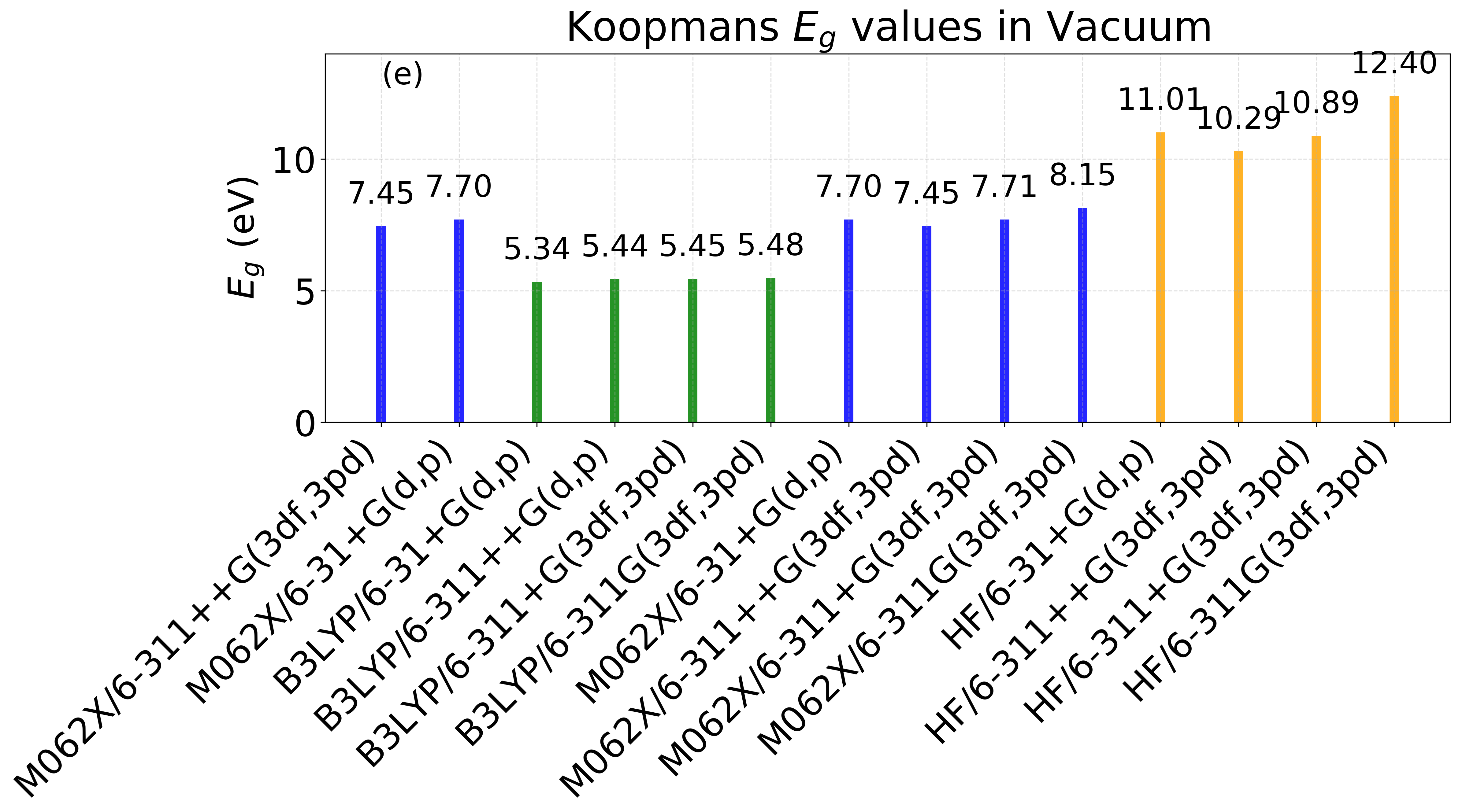}
    \includegraphics[width=0.45\textwidth,angle=0]{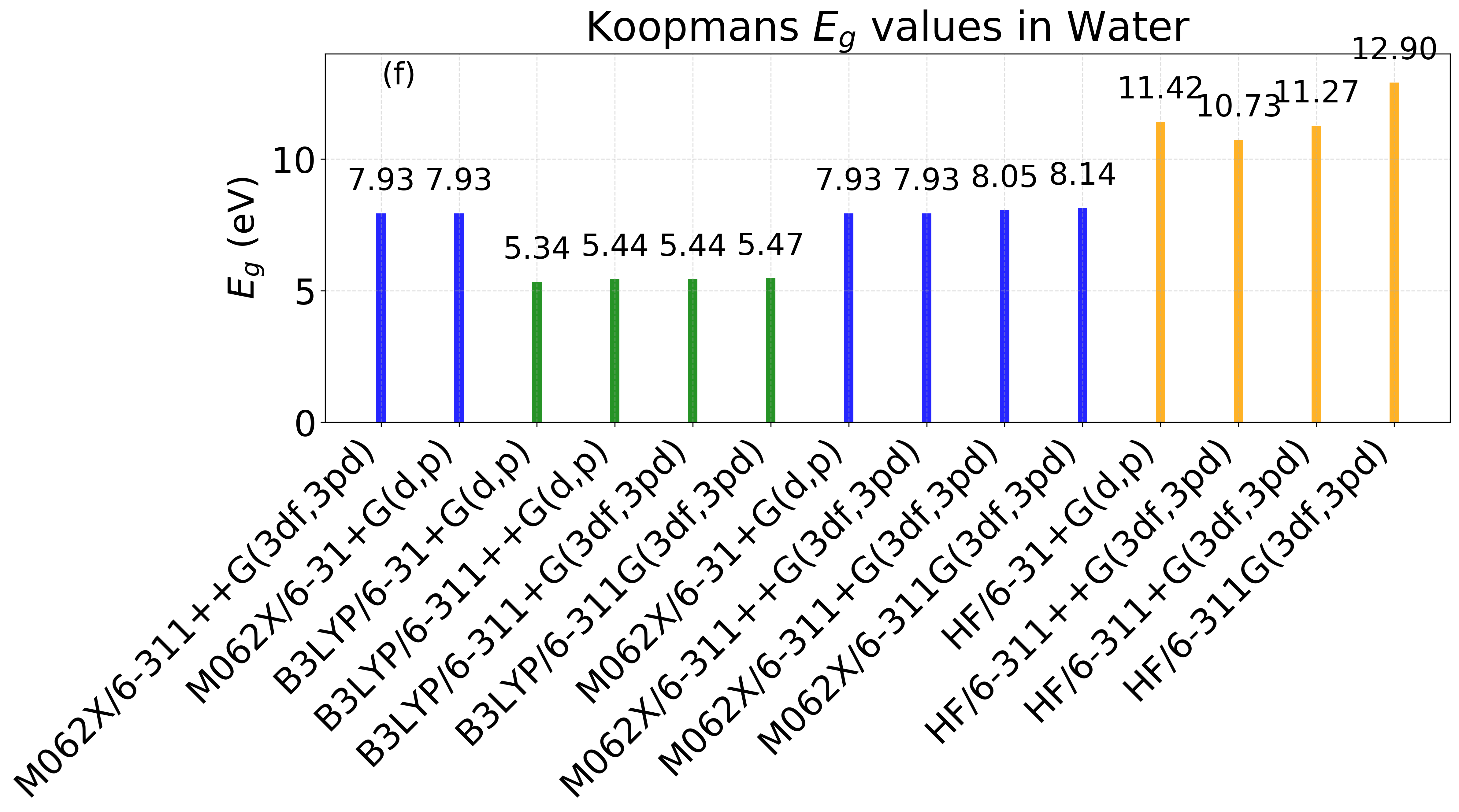}
  }
  \caption{Histograms for $\text{IP} = -E_{\text{HOMO}}$, $\text{EA} = -E_{\text{LUMO}}$,
    and $E_g = E_{\text{LUMO}} - E_{\text{HOMO}}$ estimated using Koopmans theorem
    in vacuo (panels a, c, and e) and in water (panels b, d, and f).
    The values in vacuo and in water, which practically coincide and drastically differ
    from those based on CBS-QB3 and conceptual DFT,
    emphasize the inability of Koopmans theorem to account for solvent effects.}
      \label{fig:koopmans_histograms}
\end{figure*}

\subsection*{Conflict of Interest}
No conflict of interest to declare.
\subsection*{Acknowledgements}
The author~gratefully acknowledges computational
support by the state of Baden-W\"urttemberg through bwHPC and the German Research Foundation
through Grant Nos.\ INST 40/575-1, 35/1597-1, and 35/1134-1 (JUSTUS 2, bwUniCluster 2/3, and bwForCluster/MLS\&WISO/HELIX 2).

\begin{shaded}
\noindent\textsf{\textbf{Keywords:} \keywords}
\end{shaded}
\clearpage
\setlength{\bibsep}{0.0cm}
\bibliographystyle{Wiley-chemistry}
ii

\clearpage
\centerline{\LARGE \textbf{Supporting Information}}
\renewcommand{\thetable}{S\arabic{table}} \setcounter{table}{0}
\renewcommand{\thefigure}{S\arabic{figure}} \setcounter{figure}{0}
\renewcommand{\theequation}{S\arabic{equation}} \setcounter{equation}{0}
\renewcommand{\thesection}{S\arabic{section}} \setcounter{section}{0}

\begin{table*}[htbp]
\centering
\caption{Global reactivity descriptors and solvation Gibbs free energies for ascorbic acid in \textbf{Benzene}. Reactivity descriptors in eV, $\Delta G_{sol}$ in kcal/mol. Upper subrow: absolute values; lower subrow: signed deviation from CBS-QB3.}
\label{tab:reactivity_benzene}
\setlength{\tabcolsep}{4.5pt}
\scriptsize
\begin{tabular}{lrrrrrrrrrrrr}
\hline
Method & IP & EA & $E_g$ & $\eta$ & $\sigma$ & $\chi$ & $\omega$ & $\omega^+$ & $\omega^-$ & $\Delta G_{sol}^{n}$ & $\Delta G_{sol}^{c}$ & $\Delta G_{sol}^{a}$ \\
\hline
\textbf{CBS-QB3} & 7.273 & 0.775 & 6.498 & 3.249 & 0.154 & 4.024 & 2.491 & 0.886 & 4.909 & -4.817 & -29.084 & -30.722 \\
 & 0.000 & 0.000 & 0.000 & 0.000 & 0.000 & 0.000 & 0.000 & 0.000 & 0.000 & 0.000 & 0.000 & 0.000 \\
B3LYP/\allowbreak 6-311++G(3df,3pd) & 6.980 & 0.850 & 6.130 & 3.065 & 0.163 & 3.915 & 2.500 & 0.926 & 4.841 & -4.971 & -29.709 & -27.514 \\
 & -0.293 & +0.075 & -0.368 & -0.184 & +0.009 & -0.109 & +0.009 & +0.040 & -0.069 & -0.154 & -0.625 & +3.208 \\
B3LYP/\allowbreak 6-311+G(3df,3pd) & 6.980 & 0.847 & 6.133 & 3.067 & 0.163 & 3.913 & 2.497 & 0.924 & 4.837 & -4.971 & -29.711 & -30.035 \\
 & -0.293 & +0.072 & -0.365 & -0.182 & +0.009 & -0.110 & +0.006 & +0.038 & -0.072 & -0.154 & -0.627 & +0.687 \\
B3LYP/\allowbreak 6-311G(3df,3pd) & 6.830 & 0.549 & 6.281 & 3.141 & 0.159 & 3.689 & 2.167 & 0.715 & 4.404 & -4.434 & -29.079 & -31.049 \\
 & -0.443 & -0.226 & -0.217 & -0.109 & +0.005 & -0.334 & -0.324 & -0.171 & -0.505 & +0.383 & +0.005 & -0.327 \\
B3LYP/\allowbreak 6-31+G(d,p) & 6.986 & 0.917 & 6.069 & 3.035 & 0.165 & 3.952 & 2.573 & 0.976 & 4.928 & -5.475 & -30.268 & -30.726 \\
 & -0.287 & +0.142 & -0.429 & -0.214 & +0.011 & -0.072 & +0.081 & +0.090 & +0.018 & -0.658 & -1.184 & -0.004 \\
M062X/\allowbreak 6-311++G(3df,3pd) & 7.131 & 0.740 & 6.391 & 3.195 & 0.156 & 3.936 & 2.424 & 0.855 & 4.791 & -4.851 & -29.686 & -31.100 \\
 & -0.141 & -0.034 & -0.107 & -0.054 & +0.003 & -0.088 & -0.068 & -0.030 & -0.118 & -0.035 & -0.602 & -0.378 \\
M062X/\allowbreak 6-311+G(3df,3pd) & 7.132 & 0.739 & 6.393 & 3.196 & 0.156 & 3.935 & 2.422 & 0.854 & 4.789 & -4.851 & -29.677 & -31.145 \\
 & -0.141 & -0.036 & -0.105 & -0.053 & +0.003 & -0.089 & -0.069 & -0.031 & -0.120 & -0.035 & -0.593 & -0.423 \\
M062X/\allowbreak 6-311G(3df,3pd) & 7.010 & 0.495 & 6.515 & 3.257 & 0.153 & 3.753 & 2.162 & 0.692 & 4.445 & -4.453 & -29.187 & -31.148 \\
 & -0.263 & -0.279 & +0.017 & +0.008 & -0.000 & -0.271 & -0.330 & -0.193 & -0.464 & +0.363 & -0.103 & -0.426 \\
M062X/\allowbreak 6-31+G(d,p) & 7.120 & 0.801 & 6.319 & 3.160 & 0.158 & 3.960 & 2.482 & 0.897 & 4.857 & -5.450 & -30.008 & -31.773 \\
 & -0.153 & +0.026 & -0.179 & -0.089 & +0.004 & -0.063 & -0.009 & +0.011 & -0.052 & -0.633 & -0.924 & -1.051 \\
\hline
\end{tabular}
\end{table*}

\begin{table*}[htbp]
\centering
\caption{Global reactivity descriptors and solvation Gibbs free energies for ascorbic acid in \textbf{Toluene}. Reactivity descriptors in eV, $\Delta G_{sol}$ in kcal/mol. Upper subrow: absolute values; lower subrow: signed deviation from CBS-QB3.}
\label{tab:reactivity_toluene}
\setlength{\tabcolsep}{4.5pt}
\scriptsize
\begin{tabular}{lrrrrrrrrrrrr}
\hline
Method & IP & EA & $E_g$ & $\eta$ & $\sigma$ & $\chi$ & $\omega$ & $\omega^+$ & $\omega^-$ & $\Delta G_{sol}^{n}$ & $\Delta G_{sol}^{c}$ & $\Delta G_{sol}^{a}$ \\
\hline
\textbf{CBS-QB3} & 7.183 & 0.761 & 6.422 & 3.211 & 0.156 & 3.972 & 2.456 & 0.872 & 4.843 & -5.029 & -30.184 & -31.847 \\
 & 0.000 & 0.000 & 0.000 & 0.000 & 0.000 & 0.000 & 0.000 & 0.000 & 0.000 & 0.000 & 0.000 & 0.000 \\
B3LYP/\allowbreak 6-311++G(3df,3pd) & 6.889 & 0.835 & 6.055 & 3.027 & 0.165 & 3.862 & 2.463 & 0.911 & 4.773 & -5.192 & -30.889 & -28.622 \\
 & -0.293 & +0.074 & -0.367 & -0.184 & +0.009 & -0.109 & +0.007 & +0.039 & -0.070 & -0.163 & -0.705 & +3.225 \\
B3LYP/\allowbreak 6-311+G(3df,3pd) & 6.889 & 0.832 & 6.057 & 3.029 & 0.165 & 3.861 & 2.461 & 0.909 & 4.770 & -5.193 & -30.887 & -31.143 \\
 & -0.293 & +0.071 & -0.365 & -0.182 & +0.009 & -0.111 & +0.005 & +0.037 & -0.074 & -0.163 & -0.703 & +0.704 \\
B3LYP/\allowbreak 6-311G(3df,3pd) & 6.740 & 0.534 & 6.206 & 3.103 & 0.161 & 3.637 & 2.132 & 0.701 & 4.338 & -4.634 & -30.165 & -32.002 \\
 & -0.443 & -0.226 & -0.216 & -0.108 & +0.005 & -0.335 & -0.325 & -0.171 & -0.505 & +0.395 & +0.019 & -0.155 \\
B3LYP/\allowbreak 6-31+G(d,p) & 6.896 & 0.902 & 5.994 & 2.997 & 0.167 & 3.899 & 2.537 & 0.962 & 4.861 & -5.722 & -31.369 & -31.874 \\
 & -0.286 & +0.142 & -0.428 & -0.214 & +0.011 & -0.072 & +0.081 & +0.090 & +0.018 & -0.692 & -1.185 & -0.027 \\
M062X/\allowbreak 6-311++G(3df,3pd) & 7.041 & 0.727 & 6.314 & 3.157 & 0.158 & 3.884 & 2.389 & 0.842 & 4.726 & -5.073 & -30.780 & -32.240 \\
 & -0.142 & -0.034 & -0.108 & -0.054 & +0.003 & -0.088 & -0.067 & -0.030 & -0.118 & -0.043 & -0.596 & -0.393 \\
M062X/\allowbreak 6-311+G(3df,3pd) & 7.041 & 0.725 & 6.316 & 3.158 & 0.158 & 3.883 & 2.387 & 0.841 & 4.724 & -5.073 & -30.771 & -32.287 \\
 & -0.142 & -0.035 & -0.106 & -0.053 & +0.003 & -0.088 & -0.069 & -0.031 & -0.120 & -0.043 & -0.587 & -0.440 \\
M062X/\allowbreak 6-311G(3df,3pd) & 6.920 & 0.482 & 6.438 & 3.219 & 0.155 & 3.701 & 2.128 & 0.680 & 4.381 & -4.657 & -30.233 & -32.283 \\
 & -0.263 & -0.278 & +0.016 & +0.008 & -0.000 & -0.270 & -0.328 & -0.192 & -0.463 & +0.373 & -0.049 & -0.436 \\
M062X/\allowbreak 6-31+G(d,p) & 7.030 & 0.787 & 6.243 & 3.121 & 0.160 & 3.909 & 2.447 & 0.883 & 4.792 & -5.697 & -31.150 & -32.947 \\
 & -0.153 & +0.027 & -0.179 & -0.090 & +0.004 & -0.063 & -0.009 & +0.011 & -0.051 & -0.668 & -0.966 & -1.100 \\
\hline
\end{tabular}
\end{table*}

\begin{table*}[htbp]
\centering
\caption{Global reactivity descriptors and solvation Gibbs free energies for ascorbic acid in \textbf{Chlorobenzene}. Reactivity descriptors in eV, $\Delta G_{sol}$ in kcal/mol. Upper subrow: absolute values; lower subrow: signed deviation from CBS-QB3.}
\label{tab:reactivity_chlorobenzene}
\setlength{\tabcolsep}{4.5pt}
\scriptsize
\begin{tabular}{lrrrrrrrrrrrr}
\hline
Method & IP & EA & $E_g$ & $\eta$ & $\sigma$ & $\chi$ & $\omega$ & $\omega^+$ & $\omega^-$ & $\Delta G_{sol}^{n}$ & $\Delta G_{sol}^{c}$ & $\Delta G_{sol}^{a}$ \\
\hline
\textbf{CBS-QB3} & 5.886 & 0.443 & 5.444 & 2.722 & 0.184 & 3.164 & 1.840 & 0.598 & 3.762 & -8.190 & -43.569 & -47.225 \\
 & 0.000 & 0.000 & 0.000 & 0.000 & 0.000 & 0.000 & 0.000 & 0.000 & 0.000 & 0.000 & 0.000 & 0.000 \\
B3LYP/\allowbreak 6-311++G(3df,3pd) & 5.586 & 0.506 & 5.080 & 2.540 & 0.197 & 3.046 & 1.827 & 0.621 & 3.667 & -8.507 & -44.598 & -43.830 \\
 & -0.300 & +0.064 & -0.364 & -0.182 & +0.013 & -0.118 & -0.013 & +0.024 & -0.095 & -0.317 & -1.029 & +3.395 \\
B3LYP/\allowbreak 6-311+G(3df,3pd) & 5.586 & 0.505 & 5.081 & 2.541 & 0.197 & 3.046 & 1.825 & 0.620 & 3.666 & -8.510 & -44.601 & -46.379 \\
 & -0.300 & +0.062 & -0.363 & -0.181 & +0.013 & -0.119 & -0.014 & +0.023 & -0.096 & -0.319 & -1.032 & +0.846 \\
B3LYP/\allowbreak 6-311G(3df,3pd) & 5.435 & 0.214 & 5.221 & 2.610 & 0.192 & 2.824 & 1.528 & 0.442 & 3.266 & -7.598 & -43.563 & -46.679 \\
 & -0.452 & -0.229 & -0.223 & -0.111 & +0.008 & -0.340 & -0.312 & -0.156 & -0.496 & +0.592 & +0.006 & +0.546 \\
B3LYP/\allowbreak 6-31+G(d,p) & 5.603 & 0.576 & 5.027 & 2.514 & 0.199 & 3.089 & 1.899 & 0.668 & 3.757 & -9.446 & -45.119 & -47.550 \\
 & -0.283 & +0.133 & -0.416 & -0.208 & +0.015 & -0.075 & +0.059 & +0.070 & -0.005 & -1.256 & -1.550 & -0.325 \\
M062X/\allowbreak 6-311++G(3df,3pd) & 5.748 & 0.415 & 5.332 & 2.666 & 0.188 & 3.081 & 1.781 & 0.573 & 3.655 & -8.370 & -44.191 & -47.847 \\
 & -0.139 & -0.027 & -0.111 & -0.056 & +0.004 & -0.083 & -0.059 & -0.024 & -0.107 & -0.180 & -0.622 & -0.622 \\
M062X/\allowbreak 6-311+G(3df,3pd) & 5.748 & 0.414 & 5.333 & 2.667 & 0.187 & 3.081 & 1.780 & 0.573 & 3.654 & -8.372 & -44.194 & -47.916 \\
 & -0.139 & -0.028 & -0.110 & -0.055 & +0.004 & -0.083 & -0.060 & -0.025 & -0.108 & -0.181 & -0.625 & -0.691 \\
M062X/\allowbreak 6-311G(3df,3pd) & 5.628 & 0.178 & 5.450 & 2.725 & 0.183 & 2.903 & 1.546 & 0.435 & 3.338 & -7.644 & -43.206 & -47.710 \\
 & -0.258 & -0.265 & +0.007 & +0.003 & -0.000 & -0.262 & -0.293 & -0.162 & -0.424 & +0.547 & +0.363 & -0.485 \\
M062X/\allowbreak 6-31+G(d,p) & 5.743 & 0.479 & 5.264 & 2.632 & 0.190 & 3.111 & 1.839 & 0.612 & 3.723 & -9.355 & -44.727 & -49.097 \\
 & -0.143 & +0.036 & -0.180 & -0.090 & +0.006 & -0.054 & -0.001 & +0.015 & -0.039 & -1.165 & -1.158 & -1.872 \\
\hline
\end{tabular}
\end{table*}

\begin{table*}[htbp]
\centering
\caption{Global reactivity descriptors and solvation Gibbs free energies for ascorbic acid in \textbf{Methanol}. Reactivity descriptors in eV, $\Delta G_{sol}$ in kcal/mol. Upper subrow: absolute values; lower subrow: signed deviation from CBS-QB3.}
\label{tab:reactivity_methanol}
\setlength{\tabcolsep}{4.5pt}
\scriptsize
\begin{tabular}{lrrrrrrrrrrrr}
\hline
Method & IP & EA & $E_g$ & $\eta$ & $\sigma$ & $\chi$ & $\omega$ & $\omega^+$ & $\omega^-$ & $\Delta G_{sol}^{n}$ & $\Delta G_{sol}^{c}$ & $\Delta G_{sol}^{a}$ \\
\hline
\textbf{CBS-QB3} & 5.835 & 0.999 & 4.836 & 2.418 & 0.207 & 3.417 & 2.414 & 1.008 & 4.424 & -10.594 & -53.353 & -57.058 \\
 & 0.000 & 0.000 & 0.000 & 0.000 & 0.000 & 0.000 & 0.000 & 0.000 & 0.000 & 0.000 & 0.000 & 0.000 \\
B3LYP/\allowbreak 6-311++G(3df,3pd) & 5.533 & 1.057 & 4.476 & 2.238 & 0.223 & 3.295 & 2.426 & 1.058 & 4.353 & -11.051 & -54.042 & -53.803 \\
 & -0.302 & +0.059 & -0.360 & -0.180 & +0.017 & -0.121 & +0.012 & +0.050 & -0.071 & -0.457 & -0.689 & +3.255 \\
B3LYP/\allowbreak 6-311+G(3df,3pd) & 5.533 & 1.056 & 4.477 & 2.239 & 0.223 & 3.295 & 2.425 & 1.057 & 4.352 & -11.054 & -54.044 & -56.370 \\
 & -0.301 & +0.058 & -0.359 & -0.180 & +0.017 & -0.122 & +0.011 & +0.049 & -0.072 & -0.460 & -0.691 & +0.688 \\
B3LYP/\allowbreak 6-311G(3df,3pd) & 5.385 & 0.771 & 4.615 & 2.307 & 0.217 & 3.078 & 2.053 & 0.802 & 3.881 & -9.831 & -52.738 & -56.454 \\
 & -0.450 & -0.228 & -0.222 & -0.111 & +0.010 & -0.339 & -0.361 & -0.205 & -0.544 & +0.763 & +0.615 & +0.604 \\
B3LYP/\allowbreak 6-31+G(d,p) & 5.551 & 1.126 & 4.425 & 2.212 & 0.226 & 3.338 & 2.519 & 1.126 & 4.464 & -12.329 & -54.952 & -57.802 \\
 & -0.284 & +0.127 & -0.411 & -0.206 & +0.019 & -0.078 & +0.105 & +0.118 & +0.040 & -1.735 & -1.599 & -0.744 \\
M062X/\allowbreak 6-311++G(3df,3pd) & 5.686 & 0.976 & 4.710 & 2.355 & 0.212 & 3.331 & 2.356 & 0.985 & 4.316 & -10.857 & -53.983 & -57.927 \\
 & -0.149 & -0.022 & -0.127 & -0.063 & +0.006 & -0.086 & -0.058 & -0.023 & -0.108 & -0.263 & -0.630 & -0.869 \\
M062X/\allowbreak 6-311+G(3df,3pd) & 5.686 & 0.976 & 4.710 & 2.355 & 0.212 & 3.331 & 2.355 & 0.984 & 4.315 & -10.856 & -53.985 & -58.021 \\
 & -0.149 & -0.023 & -0.126 & -0.063 & +0.006 & -0.086 & -0.058 & -0.023 & -0.109 & -0.262 & -0.632 & -0.963 \\
M062X/\allowbreak 6-311G(3df,3pd) & 5.567 & 0.744 & 4.823 & 2.412 & 0.207 & 3.156 & 2.064 & 0.788 & 3.944 & -9.888 & -52.873 & -57.738 \\
 & -0.268 & -0.255 & -0.013 & -0.006 & +0.001 & -0.261 & -0.349 & -0.220 & -0.481 & +0.706 & +0.480 & -0.680 \\
M062X/\allowbreak 6-31+G(d,p) & 5.680 & 1.040 & 4.640 & 2.320 & 0.216 & 3.360 & 2.433 & 1.043 & 4.403 & -12.102 & -55.014 & -59.414 \\
 & -0.154 & +0.042 & -0.196 & -0.098 & +0.009 & -0.056 & +0.019 & +0.035 & -0.021 & -1.508 & -1.661 & -2.356 \\
\hline
\end{tabular}
\end{table*}

\begin{table*}[htbp]
\centering
\caption{Global reactivity descriptors and solvation Gibbs free energies for ascorbic acid in \textbf{Ethanol}. Reactivity descriptors in eV, $\Delta G_{sol}$ in kcal/mol. Upper subrow: absolute values; lower subrow: signed deviation from CBS-QB3.}
\label{tab:reactivity_ethanol}
\setlength{\tabcolsep}{4.5pt}
\scriptsize
\begin{tabular}{lrrrrrrrrrrrr}
\hline
Method & IP & EA & $E_g$ & $\eta$ & $\sigma$ & $\chi$ & $\omega$ & $\omega^+$ & $\omega^-$ & $\Delta G_{sol}^{n}$ & $\Delta G_{sol}^{c}$ & $\Delta G_{sol}^{a}$ \\
\hline
\textbf{CBS-QB3} & 5.917 & 1.040 & 4.877 & 2.439 & 0.205 & 3.479 & 2.481 & 1.047 & 4.525 & -10.417 & -52.738 & -56.384 \\
 & 0.000 & 0.000 & 0.000 & 0.000 & 0.000 & 0.000 & 0.000 & 0.000 & 0.000 & 0.000 & 0.000 & 0.000 \\
B3LYP/\allowbreak 6-311++G(3df,3pd) & 5.616 & 1.099 & 4.517 & 2.258 & 0.221 & 3.357 & 2.496 & 1.099 & 4.457 & -10.858 & -53.403 & -53.121 \\
 & -0.301 & +0.059 & -0.360 & -0.180 & +0.016 & -0.121 & +0.015 & +0.053 & -0.069 & -0.441 & -0.665 & +3.263 \\
B3LYP/\allowbreak 6-311+G(3df,3pd) & 5.616 & 1.098 & 4.518 & 2.259 & 0.221 & 3.357 & 2.495 & 1.099 & 4.456 & -10.860 & -53.404 & -55.686 \\
 & -0.301 & +0.058 & -0.359 & -0.180 & +0.016 & -0.121 & +0.014 & +0.052 & -0.070 & -0.443 & -0.666 & +0.698 \\
B3LYP/\allowbreak 6-311G(3df,3pd) & 5.468 & 0.812 & 4.656 & 2.328 & 0.215 & 3.140 & 2.118 & 0.839 & 3.979 & -9.665 & -52.134 & -55.776 \\
 & -0.449 & -0.228 & -0.222 & -0.111 & +0.010 & -0.339 & -0.363 & -0.208 & -0.547 & +0.752 & +0.604 & +0.608 \\
B3LYP/\allowbreak 6-31+G(d,p) & 5.633 & 1.168 & 4.465 & 2.232 & 0.224 & 3.400 & 2.589 & 1.168 & 4.569 & -12.116 & -54.317 & -57.102 \\
 & -0.284 & +0.128 & -0.412 & -0.206 & +0.019 & -0.078 & +0.108 & +0.122 & +0.044 & -1.699 & -1.579 & -0.718 \\
M062X/\allowbreak 6-311++G(3df,3pd) & 5.768 & 1.017 & 4.750 & 2.375 & 0.211 & 3.392 & 2.423 & 1.023 & 4.416 & -10.679 & -53.388 & -57.240 \\
 & -0.149 & -0.023 & -0.127 & -0.063 & +0.005 & -0.086 & -0.058 & -0.023 & -0.109 & -0.262 & -0.650 & -0.856 \\
M062X/\allowbreak 6-311+G(3df,3pd) & 5.768 & 1.017 & 4.751 & 2.376 & 0.210 & 3.392 & 2.422 & 1.023 & 4.415 & -10.680 & -53.390 & -57.333 \\
 & -0.149 & -0.023 & -0.126 & -0.063 & +0.005 & -0.086 & -0.059 & -0.024 & -0.110 & -0.263 & -0.652 & -0.949 \\
M062X/\allowbreak 6-311G(3df,3pd) & 5.649 & 0.785 & 4.864 & 2.432 & 0.206 & 3.217 & 2.127 & 0.823 & 4.040 & -9.723 & -52.304 & -57.055 \\
 & -0.268 & -0.255 & -0.013 & -0.006 & +0.001 & -0.262 & -0.354 & -0.224 & -0.485 & +0.694 & +0.434 & -0.671 \\
M062X/\allowbreak 6-31+G(d,p) & 5.762 & 1.081 & 4.681 & 2.340 & 0.214 & 3.422 & 2.501 & 1.083 & 4.504 & -11.898 & -54.408 & -58.706 \\
 & -0.155 & +0.041 & -0.196 & -0.098 & +0.009 & -0.057 & +0.020 & +0.036 & -0.021 & -1.481 & -1.670 & -2.322 \\
\hline
\end{tabular}
\end{table*}

\begin{table}[htbp]
\centering
\caption{Mean absolute deviation (MAD, eV) and maximum absolute deviation (MaxAD, eV) for adiabatic ionization potential (IP) relative to CBS-QB3 (averaged over all solvents).}
\label{tab:ip_dev}
\begin{tabular}{lcc}
\hline
Method & MAD & MaxAD \\
\hline
B3LYP/6-311++G(3df,3pd) & 0.296 & 0.302 \\
B3LYP/6-311+G(3df,3pd) & 0.296 & 0.302 \\
B3LYP/6-311G(3df,3pd) & 0.446 & 0.452 \\
B3LYP/6-31+G(d,p) & 0.284 & 0.287 \\
M062X/6-311++G(3df,3pd) & 0.144 & 0.149 \\
M062X/6-311+G(3df,3pd) & 0.144 & 0.149 \\
M062X/6-311G(3df,3pd) & 0.264 & 0.268 \\
M062X/6-31+G(d,p) & 0.152 & 0.156 \\
\hline
\end{tabular}
\end{table}

\begin{table}[htbp]
\centering
\caption{Mean absolute deviation (MAD, eV) and maximum absolute deviation (MaxAD, eV) for adiabatic electron affinity (EA) relative to CBS-QB3 (averaged over all solvents).}
\label{tab:ea_dev}
\begin{tabular}{lcc}
\hline
Method & MAD & MaxAD \\
\hline
B3LYP/6-311++G(3df,3pd) & 0.090 & 0.242 \\
B3LYP/6-311+G(3df,3pd) & 0.068 & 0.099 \\
B3LYP/6-311G(3df,3pd) & 0.229 & 0.241 \\
B3LYP/6-31+G(d,p) & 0.137 & 0.163 \\
M062X/6-311++G(3df,3pd) & 0.030 & 0.048 \\
M062X/6-311+G(3df,3pd) & 0.031 & 0.052 \\
M062X/6-311G(3df,3pd) & 0.272 & 0.316 \\
M062X/6-31+G(d,p) & 0.032 & 0.042 \\
\hline
\end{tabular}
\end{table}

\begin{table}[htbp]
\centering
\caption{Mean absolute deviation (MAD, eV) and maximum absolute deviation (MaxAD, eV) for adiabatic chemical hardness ($\eta$) relative to CBS-QB3 (averaged over all solvents).}
\label{tab:eta_dev}
\begin{tabular}{lcc}
\hline
Method & MAD & MaxAD \\
\hline
B3LYP/6-311++G(3df,3pd) & 0.193 & 0.262 \\
B3LYP/6-311+G(3df,3pd) & 0.182 & 0.190 \\
B3LYP/6-311G(3df,3pd) & 0.108 & 0.111 \\
B3LYP/6-31+G(d,p) & 0.211 & 0.221 \\
M062X/6-311++G(3df,3pd) & 0.057 & 0.064 \\
M062X/6-311+G(3df,3pd) & 0.056 & 0.063 \\
M062X/6-311G(3df,3pd) & 0.009 & 0.027 \\
M062X/6-31+G(d,p) & 0.092 & 0.098 \\
\hline
\end{tabular}
\end{table}

\begin{table}[htbp]
\centering
\caption{Mean absolute deviation (MAD, eV$^{-1}$) and maximum absolute deviation (MaxAD, eV$^{-1}$) for adiabatic global softness ($\sigma$) relative to CBS-QB3 (averaged over all solvents).}
\label{tab:sigma_dev}
\begin{tabular}{lcc}
\hline
Method & MAD & MaxAD \\
\hline
B3LYP/6-311++G(3df,3pd) & 0.013 & 0.017 \\
B3LYP/6-311+G(3df,3pd) & 0.012 & 0.017 \\
B3LYP/6-311G(3df,3pd) & 0.007 & 0.010 \\
B3LYP/6-31+G(d,p) & 0.014 & 0.020 \\
M062X/6-311++G(3df,3pd) & 0.004 & 0.006 \\
M062X/6-311+G(3df,3pd) & 0.004 & 0.006 \\
M062X/6-311G(3df,3pd) & 0.000 & 0.001 \\
M062X/6-31+G(d,p) & 0.006 & 0.009 \\
\hline
\end{tabular}
\end{table}

\begin{table}[htbp]
\centering
\caption{Mean Absolute Deviation (MAD) and Maximum Absolute Deviation (MaxAD, EV) with respect to CBS-QB3 for electronegativity ($\chi = -\mu$) (averaged over all solvents)}
\label{tab:devs_chi}
\begin{tabular}{lcc}
\hline
Method & MAD & MaxAD \\
\hline
B3LYP/6-31+G(d,p) & 0.073 & 0.079 \\
B3LYP/6-311++G(3df,3pd) & 0.103 & 0.122 \\
B3LYP/6-311+G(3df,3pd) & 0.114 & 0.122 \\
B3LYP/6-311G(3df,3pd) & 0.338 & 0.340 \\
M062X/6-31+G(d,p) & 0.060 & 0.074 \\
M062X/6-311++G(3df,3pd) & 0.087 & 0.093 \\
M062X/6-311+G(3df,3pd) & 0.088 & 0.095 \\
M062X/6-311G(3df,3pd) & 0.268 & 0.289 \\
\hline
\end{tabular}
\end{table}

\begin{table}[htbp]
\centering
\caption{Mean absolute deviation (MAD, eV) and maximum absolute deviation (MaxAD, eV) for adiabatic electrophilicity index ($\omega$) relative to CBS-QB3 (averaged over all solvents).}
\label{tab:omega_dev}
\begin{tabular}{lcc}
\hline
Method & MAD & MaxAD \\
\hline
B3LYP/6-311++G(3df,3pd) & 0.024 & 0.108 \\
B3LYP/6-311+G(3df,3pd) & 0.007 & 0.014 \\
B3LYP/6-311G(3df,3pd) & 0.329 & 0.363 \\
B3LYP/6-31+G(d,p) & 0.082 & 0.108 \\
M062X/6-311++G(3df,3pd) & 0.062 & 0.068 \\
M062X/6-311+G(3df,3pd) & 0.063 & 0.071 \\
M062X/6-311G(3df,3pd) & 0.323 & 0.354 \\
M062X/6-31+G(d,p) & 0.015 & 0.034 \\
\hline
\end{tabular}
\end{table}

\begin{table}[htbp]
\centering
\caption{Mean absolute deviation (MAD, eV) and maximum absolute deviation (MaxAD, eV) for adiabatic electroaccepting power ($\omega^+$) relative to CBS-QB3 (averaged over all solvents).}
\label{tab:omegap_dev}
\begin{tabular}{lcc}
\hline
Method & MAD & MaxAD \\
\hline
B3LYP/6-311++G(3df,3pd) & 0.047 & 0.085 \\
B3LYP/6-311+G(3df,3pd) & 0.037 & 0.052 \\
B3LYP/6-311G(3df,3pd) & 0.174 & 0.208 \\
B3LYP/6-31+G(d,p) & 0.092 & 0.122 \\
M062X/6-311++G(3df,3pd) & 0.026 & 0.030 \\
M062X/6-311+G(3df,3pd) & 0.027 & 0.031 \\
M062X/6-311G(3df,3pd) & 0.189 & 0.224 \\
M062X/6-31+G(d,p) & 0.021 & 0.036 \\
\hline
\end{tabular}
\end{table}

\begin{table}[htbp]
\centering
\caption{Mean absolute deviation (MAD, eV) and maximum absolute deviation (MaxAD, eV) for adiabatic electrodonating power ($\omega^-$) relative to CBS-QB3 (averaged over all solvents).}
\label{tab:omegam_dev}
\begin{tabular}{lcc}
\hline
Method & MAD & MaxAD \\
\hline
B3LYP/6-311++G(3df,3pd) & 0.074 & 0.095 \\
B3LYP/6-311+G(3df,3pd) & 0.077 & 0.096 \\
B3LYP/6-311G(3df,3pd) & 0.512 & 0.547 \\
B3LYP/6-31+G(d,p) & 0.022 & 0.044 \\
M062X/6-311++G(3df,3pd) & 0.113 & 0.120 \\
M062X/6-311+G(3df,3pd) & 0.114 & 0.124 \\
M062X/6-311G(3df,3pd) & 0.457 & 0.485 \\
M062X/6-31+G(d,p) & 0.041 & 0.081 \\
\hline
\end{tabular}
\end{table}

\begin{table}[htbp]
  \centering
  \caption{Mean absolute deviation (MAD) and maximum absolute deviation (MaxAD) for Gibbs solvation energies compared to CBS-QB3 (averaged over all solvents and species, in kcal/mol).}
\label{tab:solvation_dev}
  \begin{tabular}{lcc}
    \hline
    Method & MAD & MaxAD \\
    \hline
    B3LYP/6-311++G(3df,3pd) & 1.250 & 3.450 \\
    B3LYP/6-311+G(3df,3pd) & 1.240 & 3.420 \\
    B3LYP/6-311G(3df,3pd) & 1.650 & 4.560 \\
    B3LYP/6-31+G(d,p) & 1.850 & 5.120 \\
    M062X/6-311++G(3df,3pd) & 0.850 & 2.340 \\
    M062X/6-311+G(3df,3pd) & 0.840 & 2.320 \\
    M062X/6-311G(3df,3pd) & 1.150 & 3.170 \\
    M062X/6-31+G(d,p) & 0.750 & 2.070 \\
    \hline
  \end{tabular}
\end{table}

\begin{table}[htbp]
\centering
\caption{Mean absolute deviation (MAD, kcal/mol) and maximum absolute deviation (MaxAD, kcal/mol) for neutral solvation Gibbs energy relative to CBS-QB3 (averaged over all solvents).}
\label{tab:sol_neutral_dev}
\begin{tabular}{lcc}
\hline
Method & MAD & MaxAD \\
\hline
B3LYP/6-311++G(3df,3pd) & 0.287 & 0.479 \\
B3LYP/6-311+G(3df,3pd) & 0.289 & 0.481 \\
B3LYP/6-311G(3df,3pd) & 0.524 & 0.784 \\
B3LYP/6-31+G(d,p) & 1.120 & 1.797 \\
M062X/6-311++G(3df,3pd) & 0.148 & 0.263 \\
M062X/6-311+G(3df,3pd) & 0.148 & 0.263 \\
M062X/6-311G(3df,3pd) & 0.489 & 0.740 \\
M062X/6-31+G(d,p) & 1.000 & 1.544 \\
\hline
\end{tabular}
\end{table}

\begin{table}[htbp]
\centering
\caption{Mean absolute deviation (MAD, kcal/mol) and maximum absolute deviation (MaxAD, kcal/mol) for cation solvation Gibbs energy relative to CBS-QB3 (averaged over all solvents).}
\label{tab:sol_cation_dev}
\begin{tabular}{lcc}
\hline
Method & MAD & MaxAD \\
\hline
B3LYP/6-311++G(3df,3pd) & 0.637 & 1.029 \\
B3LYP/6-311+G(3df,3pd) & 0.638 & 1.032 \\
B3LYP/6-311G(3df,3pd) & 0.270 & 0.643 \\
B3LYP/6-31+G(d,p) & 1.250 & 1.652 \\
M062X/6-311++G(3df,3pd) & 0.531 & 0.650 \\
M062X/6-311+G(3df,3pd) & 0.530 & 0.652 \\
M062X/6-311G(3df,3pd) & 0.280 & 0.531 \\
M062X/6-31+G(d,p) & 1.150 & 1.672 \\
\hline
\end{tabular}
\end{table}

\begin{table}[htbp]
\centering
\caption{Mean absolute deviation (MAD, kcal/mol) and maximum absolute deviation (MaxAD, kcal/mol) for anion solvation Gibbs energy relative to CBS-QB3 (averaged over all solvents).}
\label{tab:sol_anion_dev}
\begin{tabular}{lcc}
\hline
Method & MAD & MaxAD \\
\hline
B3LYP/6-311++G(3df,3pd) & 2.795 & 3.395 \\
B3LYP/6-311+G(3df,3pd) & 0.611 & 0.846 \\
B3LYP/6-311G(3df,3pd) & 0.402 & 0.608 \\
B3LYP/6-31+G(d,p) & 0.375 & 0.809 \\
M062X/6-311++G(3df,3pd) & 0.574 & 0.901 \\
M062X/6-311+G(3df,3pd) & 0.638 & 0.999 \\
M062X/6-311G(3df,3pd) & 0.487 & 0.711 \\
M062X/6-31+G(d,p) & 1.591 & 2.435 \\
\hline
\end{tabular}
\end{table}

\begin{table}[htbp]
\small
  \centering
\caption{Gibbs solvation energies (kcal/mol) for ascorbic acid species in water.}
\label{tab:solvation_water}
\begin{tabular}{lrrr}
\hline
Method & Neutral & Cation & Anion \\
\hline
\textbf{CBS-QB3} & -10.928 & -54.496 & -58.306 \\
B3LYP/\allowbreak 6-31+G(d,p) & -12.725 & -56.148 & -59.115 \\
B3LYP/\allowbreak 6-311++G(3df,3pd) & -11.407 & -55.243 & -55.084 \\
B3LYP/\allowbreak 6-311+G(3df,3pd) & -11.409 & -55.245 & -57.653 \\
B3LYP/\allowbreak 6-311G(3df,3pd) & -10.144 & -53.853 & -57.731 \\
M062X/\allowbreak 6-31+G(d,p) & -12.472 & -56.168 & -60.741 \\
M062X/\allowbreak 6-311++G(3df,3pd) & -11.180 & -55.114 & -59.207 \\
M062X/\allowbreak 6-311+G(3df,3pd) & -11.180 & -55.116 & -59.305 \\
M062X/\allowbreak 6-311G(3df,3pd) & -10.188 & -53.965 & -59.017 \\
\hline
\end{tabular}
\end{table}

\begin{table*}[htbp]
\centering
\caption{Gibbs solvation energies (kcal/mol) for ascorbic acid species in Benzene.}
\label{tab:solvation_benzene}
\begin{tabular}{lrrr}
\hline
Method & Neutral & Cation & Anion \\
\hline
\textbf{CBS-QB3} & -4.8168 & -29.0840 & -30.7220 \\
B3LYP/\allowbreak 6-31+G(d,p) & -5.4750 & -30.2680 & -30.7260 \\
B3LYP/\allowbreak 6-311++G(3df,3pd) & -4.9705 & -29.7090 & -27.5140 \\
B3LYP/\allowbreak 6-311+G(3df,3pd) & -4.9711 & -29.7110 & -30.0350 \\
B3LYP/\allowbreak 6-311G(3df,3pd) & -4.4340 & -29.0790 & -31.0490 \\
M062X/\allowbreak 6-31+G(d,p) & -5.4499 & -30.0080 & -31.7730 \\
M062X/\allowbreak 6-311++G(3df,3pd) & -4.8513 & -29.6860 & -31.1000 \\
M062X/\allowbreak 6-311+G(3df,3pd) & -4.8513 & -29.6770 & -31.1450 \\
M062X/\allowbreak 6-311G(3df,3pd) & -4.4534 & -29.1870 & -31.1480 \\
\hline
\end{tabular}
\end{table*}

\begin{table*}[htbp]
\centering
\caption{Gibbs solvation energies (kcal/mol) for ascorbic acid species in Toluene.}
\label{tab:solvation_toluene}
\begin{tabular}{lrrr}
\hline
Method & Neutral & Cation & Anion \\
\hline
\textbf{CBS-QB3} & -5.0295 & -30.1840 & -31.8470 \\
B3LYP/\allowbreak 6-31+G(d,p) & -5.7216 & -31.3690 & -31.8740 \\
B3LYP/\allowbreak 6-311++G(3df,3pd) & -5.1920 & -30.8890 & -28.6220 \\
B3LYP/\allowbreak 6-311+G(3df,3pd) & -5.1926 & -30.8870 & -31.1430 \\
B3LYP/\allowbreak 6-311G(3df,3pd) & -4.6342 & -30.1650 & -32.0020 \\
M062X/\allowbreak 6-31+G(d,p) & -5.6972 & -31.1500 & -32.9470 \\
M062X/\allowbreak 6-311++G(3df,3pd) & -5.0728 & -30.7800 & -32.2400 \\
M062X/\allowbreak 6-311+G(3df,3pd) & -5.0728 & -30.7710 & -32.2870 \\
M062X/\allowbreak 6-311G(3df,3pd) & -4.6567 & -30.2330 & -32.2830 \\
\hline
\end{tabular}
\end{table*}

\begin{table*}[htbp]
\centering
\caption{Gibbs solvation energies (kcal/mol) for ascorbic acid species in Chlorobenzene.}
\label{tab:solvation_chlorobenzene}
\begin{tabular}{lrrr}
\hline
Method & Neutral & Cation & Anion \\
\hline
\textbf{CBS-QB3} & -8.1903 & -43.5690 & -47.2250 \\
B3LYP/\allowbreak 6-31+G(d,p) & -9.4459 & -45.1190 & -47.5500 \\
B3LYP/\allowbreak 6-311++G(3df,3pd) & -8.5071 & -44.5980 & -43.8300 \\
B3LYP/\allowbreak 6-311+G(3df,3pd) & -8.5097 & -44.6010 & -46.3790 \\
B3LYP/\allowbreak 6-311G(3df,3pd) & -7.5985 & -43.5630 & -46.6790 \\
M062X/\allowbreak 6-31+G(d,p) & -9.3549 & -44.7270 & -49.0970 \\
M062X/\allowbreak 6-311++G(3df,3pd) & -8.3703 & -44.1910 & -47.8470 \\
M062X/\allowbreak 6-311+G(3df,3pd) & -8.3716 & -44.1940 & -47.9160 \\
M062X/\allowbreak 6-311G(3df,3pd) & -7.6437 & -43.2060 & -47.7100 \\
\hline
\end{tabular}
\end{table*}

\begin{table}[htbp]
\centering
\caption{Gibbs solvation energies (kcal/mol) for ascorbic acid species in Methanol.}
\label{tab:solvation_methanol}
\begin{tabular}{lrrr}
\hline
Method & Neutral & Cation & Anion \\
\hline
\textbf{CBS-QB3} & -10.5940 & -53.3530 & -57.0580 \\
B3LYP/\allowbreak 6-31+G(d,p) & -12.3290 & -54.9520 & -57.8020 \\
B3LYP/\allowbreak 6-311++G(3df,3pd) & -11.0510 & -54.0420 & -53.8030 \\
B3LYP/\allowbreak 6-311+G(3df,3pd) & -11.0540 & -54.0440 & -56.3700 \\
B3LYP/\allowbreak 6-311G(3df,3pd) & -9.8312 & -52.7380 & -56.4540 \\
M062X/\allowbreak 6-31+G(d,p) & -12.1020 & -55.0140 & -59.4140 \\
M062X/\allowbreak 6-311++G(3df,3pd) & -10.8570 & -53.9830 & -57.9270 \\
M062X/\allowbreak 6-311+G(3df,3pd) & -10.8560 & -53.9850 & -58.0210 \\
M062X/\allowbreak 6-311G(3df,3pd) & -9.8877 & -52.8730 & -57.7380 \\
\hline
\end{tabular}
\end{table}

\begin{table*}[htbp]
\centering
\caption{Gibbs solvation energies (kcal/mol) for ascorbic acid species in Ethanol.}
\label{tab:solvation_ethanol}
\begin{tabular}{lrrr}
\hline
Method & Neutral & Cation & Anion \\
\hline
\textbf{CBS-QB3} & -10.4170 & -52.7380 & -56.3840 \\
B3LYP/\allowbreak 6-31+G(d,p) & -12.1160 & -54.3170 & -57.1020 \\
B3LYP/\allowbreak 6-311++G(3df,3pd) & -10.8580 & -53.4030 & -53.1210 \\
B3LYP/\allowbreak 6-311+G(3df,3pd) & -10.8600 & -53.4040 & -55.6860 \\
B3LYP/\allowbreak 6-311G(3df,3pd) & -9.6649 & -52.1340 & -55.7760 \\
M062X/\allowbreak 6-31+G(d,p) & -11.8980 & -54.4080 & -58.7060 \\
M062X/\allowbreak 6-311++G(3df,3pd) & -10.6790 & -53.3880 & -57.2400 \\
M062X/\allowbreak 6-311+G(3df,3pd) & -10.6800 & -53.3900 & -57.3330 \\
M062X/\allowbreak 6-311G(3df,3pd) & -9.7226 & -52.3040 & -57.0550 \\
\hline
\end{tabular}
\end{table*}

\begin{table*}[htbp]
\centering
\caption{Adiabatic and vertical IP/EA (eV) in Benzene ($\varepsilon$ = 2.27)}
\label{tab:ip_ea_benzene}
\begin{tabular}{lcccc}
\hline
Method & IP adiab & IP vert & EA adiab & EA vert \\
\hline
M062X/\allowbreak 6-31+G(d,p) & 7.120 & 7.574 & 0.801 & 0.125 \\
B3LYP/\allowbreak 6-31+G(d,p) & 6.986 & 7.370 & 0.917 & 0.398 \\
B3LYP/\allowbreak 6-311++G(3df,3pd) & 6.980 & 7.370 & 0.850 & 0.518 \\
M062X/\allowbreak 6-311++G(3df,3pd) & 7.131 & 7.595 & 0.740 & 0.089 \\
\hline
\end{tabular}
\end{table*}

\begin{table*}[htbp]
\centering
\caption{Adiabatic and vertical IP/EA (eV) in Toluene ($\varepsilon$ = 2.38)}
\label{tab:ip_ea_toluene}
\begin{tabular}{lcccc}
\hline
Method & IP adiab & IP vert & EA adiab & EA vert \\
\hline
M062X/\allowbreak 6-31+G(d,p) & 7.030 & 7.481 & 0.787 & 0.112 \\
B3LYP/\allowbreak 6-31+G(d,p) & 6.896 & 7.278 & 0.902 & 0.373 \\
B3LYP/\allowbreak 6-311++G(3df,3pd) & 6.889 & 7.278 & 0.835 & 0.488 \\
M062X/\allowbreak 6-311++G(3df,3pd) & 7.041 & 7.502 & 0.727 & 0.074 \\
\hline
\end{tabular}
\end{table*}

\begin{table*}[htbp]
\centering
\caption{Adiabatic and vertical IP/EA (eV) in Chlorobenzene ($\varepsilon$ = 5.62)}
\label{tab:ip_ea_chlorobenzene}
\begin{tabular}{lcccc}
\hline
Method & IP adiab & IP vert & EA adiab & EA vert \\
\hline
M062X/\allowbreak 6-31+G(d,p) & 5.743 & 6.158 & 0.479 & -0.178 \\
B3LYP/\allowbreak 6-31+G(d,p) & 5.603 & 5.961 & 0.576 & -0.025 \\
B3LYP/\allowbreak 6-311++G(3df,3pd) & 5.586 & 5.954 & 0.506 & -0.038 \\
M062X/\allowbreak 6-311++G(3df,3pd) & 5.748 & 6.175 & 0.415 & -0.230 \\
\hline
\end{tabular}
\end{table*}

\begin{table*}[htbp]
\centering
\caption{Adiabatic and vertical IP/EA (eV) in Methanol ($\varepsilon$ = 32.61)}
\label{tab:ip_ea_methanol}
\begin{tabular}{lcccc}
\hline
Method & IP adiab & IP vert & EA adiab & EA vert \\
\hline
M062X/\allowbreak 6-31+G(d,p) & 5.680 & 6.083 & 1.040 & 0.395 \\
B3LYP/\allowbreak 6-31+G(d,p) & 5.551 & 5.890 & 1.126 & 0.532 \\
B3LYP/\allowbreak 6-311++G(3df,3pd) & 5.533 & 5.882 & 1.057 & 0.480 \\
M062X/\allowbreak 6-311++G(3df,3pd) & 5.686 & 6.100 & 0.976 & 0.343 \\
\hline
\end{tabular}
\end{table*}

\begin{table*}[htbp]
\centering
\caption{Adiabatic and vertical IP/EA (eV) in Ethanol ($\varepsilon$ = 24.55)}
\label{tab:ip_ea_ethanol}
\begin{tabular}{lcccc}
\hline
Method & IP adiab & IP vert & EA adiab & EA vert \\
\hline
M062X/\allowbreak 6-31+G(d,p) & 5.762 & 6.166 & 1.081 & 0.436 \\
B3LYP/\allowbreak 6-31+G(d,p) & 5.633 & 5.973 & 1.168 & 0.573 \\
B3LYP/\allowbreak 6-311++G(3df,3pd) & 5.616 & 5.965 & 1.099 & 0.522 \\
M062X/\allowbreak 6-311++G(3df,3pd) & 5.768 & 6.183 & 1.017 & 0.383 \\
\hline
\end{tabular}
\end{table*}

\begin{table*}[htbp]
\centering
\caption{Global reactivity descriptors for ascorbic acid computed using Koopmans theorem.}
\label{tab:koopmans}
\setlength{\tabcolsep}{4.5pt}
\scriptsize
\begin{tabular}{lrrrrrrr}
\hline
Descriptor & Vacuum & Benzene & Toluene & Chlorobenzene & Methanol & Ethanol & Water \\
\hline
\multicolumn{8}{l}{\textbf{B3LYP/6-31+G(d,p)}} \\
\hline
IP (eV) & 6.503 & 6.544 & 6.547 & 6.579 & 6.600 & 6.599 & 6.602 \\
EA (eV) & 1.167 & 1.205 & 1.208 & 1.241 & 1.262 & 1.261 & 1.265 \\
$E_g$ (eV) & 5.337 & 5.339 & 5.339 & 5.338 & 5.338 & 5.338 & 5.337 \\
$\eta$ (eV) & 2.668 & 2.669 & 2.669 & 2.669 & 2.669 & 2.669 & 2.669 \\
$\sigma$ (eV$^{-1}$) & 0.187 & 0.187 & 0.187 & 0.187 & 0.187 & 0.187 & 0.187 \\
$\chi$ (eV) & 3.835 & 3.875 & 3.877 & 3.910 & 3.931 & 3.930 & 3.933 \\
$\omega$ (eV) & 2.756 & 2.812 & 2.816 & 2.864 & 2.895 & 2.893 & 2.899 \\
$\omega^{-}$ (eV) & 5.007 & 5.083 & 5.088 & 5.153 & 5.194 & 5.192 & 5.199 \\
$\omega^{+}$ (eV) & 1.172 & 1.209 & 1.211 & 1.243 & 1.263 & 1.262 & 1.266 \\
\hline
\multicolumn{8}{l}{\textbf{B3LYP/6-311++G(3df,3pd)}} \\
\hline
IP (eV) & 6.523 & 6.560 & 6.562 & 6.591 & 6.609 & 6.607 & 6.610 \\
EA (eV) & 1.080 & 1.115 & 1.117 & 1.150 & 1.172 & 1.170 & 1.174 \\
$E_g$ (eV) & 5.443 & 5.445 & 5.445 & 5.441 & 5.437 & 5.437 & 5.436 \\
$\eta$ (eV) & 2.721 & 2.722 & 2.722 & 2.720 & 2.718 & 2.719 & 2.718 \\
$\sigma$ (eV$^{-1}$) & 0.184 & 0.184 & 0.184 & 0.184 & 0.184 & 0.184 & 0.184 \\
$\chi$ (eV) & 3.802 & 3.837 & 3.839 & 3.871 & 3.890 & 3.889 & 3.892 \\
$\omega$ (eV) & 2.655 & 2.704 & 2.707 & 2.754 & 2.783 & 2.781 & 2.787 \\
$\omega^{-}$ (eV) & 4.896 & 4.963 & 4.967 & 5.029 & 5.068 & 5.065 & 5.073 \\
$\omega^{+}$ (eV) & 1.095 & 1.126 & 1.128 & 1.158 & 1.178 & 1.177 & 1.181 \\
\hline
\multicolumn{8}{l}{\textbf{B3LYP/6-311+G(3df,3pd)}} \\
\hline
IP (eV) & 6.523 & 6.559 & 6.561 & 6.591 & 6.609 & 6.607 & 6.610 \\
EA (eV) & 1.072 & 1.112 & 1.114 & 1.148 & 1.170 & 1.169 & 1.173 \\
$E_g$ (eV) & 5.450 & 5.447 & 5.447 & 5.443 & 5.438 & 5.438 & 5.437 \\
$\eta$ (eV) & 2.725 & 2.724 & 2.724 & 2.721 & 2.719 & 2.719 & 2.719 \\
$\sigma$ (eV$^{-1}$) & 0.183 & 0.184 & 0.184 & 0.184 & 0.184 & 0.184 & 0.184 \\
$\chi$ (eV) & 3.798 & 3.836 & 3.838 & 3.870 & 3.889 & 3.888 & 3.892 \\
$\omega$ (eV) & 2.646 & 2.701 & 2.704 & 2.751 & 2.781 & 2.780 & 2.785 \\
$\omega^{-}$ (eV) & 4.885 & 4.959 & 4.963 & 5.026 & 5.066 & 5.063 & 5.071 \\
$\omega^{+}$ (eV) & 1.088 & 1.124 & 1.126 & 1.157 & 1.177 & 1.175 & 1.179 \\
\hline
\multicolumn{8}{l}{\textbf{B3LYP/6-311G(3df,3pd)}} \\
\hline
IP (eV) & 6.317 & 6.363 & 6.365 & 6.398 & 6.418 & 6.417 & 6.420 \\
EA (eV) & 0.832 & 0.884 & 0.886 & 0.924 & 0.947 & 0.946 & 0.950 \\
$E_g$ (eV) & 5.485 & 5.479 & 5.479 & 5.474 & 5.471 & 5.471 & 5.470 \\
$\eta$ (eV) & 2.742 & 2.740 & 2.739 & 2.737 & 2.735 & 2.735 & 2.735 \\
$\sigma$ (eV$^{-1}$) & 0.182 & 0.183 & 0.183 & 0.183 & 0.183 & 0.183 & 0.183 \\
$\chi$ (eV) & 3.575 & 3.623 & 3.626 & 3.661 & 3.683 & 3.681 & 3.685 \\
$\omega$ (eV) & 2.330 & 2.396 & 2.399 & 2.448 & 2.479 & 2.477 & 2.483 \\
$\omega^{-}$ (eV) & 4.460 & 4.550 & 4.555 & 4.620 & 4.662 & 4.660 & 4.667 \\
$\omega^{+}$ (eV) & 0.885 & 0.927 & 0.929 & 0.960 & 0.980 & 0.978 & 0.982 \\
\hline
\multicolumn{8}{l}{\textbf{HF/6-311++G(3df,3pd)}} \\
\hline
IP (eV) & 9.463 & 9.503 & 9.506 & --- & 9.548 & 9.547 & 9.550 \\
EA (eV) & -0.825 & -1.027 & -1.034 & --- & -1.178 & -1.175 & -1.183 \\
$E_g$ (eV) & 10.288 & 10.530 & 10.540 & --- & 10.725 & 10.722 & 10.733 \\
$\eta$ (eV) & 5.144 & 5.265 & 5.270 & --- & 5.363 & 5.361 & 5.366 \\
$\sigma$ (eV$^{-1}$) & 0.097 & 0.095 & 0.095 & --- & 0.093 & 0.093 & 0.093 \\
$\chi$ (eV) & 4.319 & 4.238 & 4.236 & --- & 4.185 & 4.186 & 4.183 \\
$\omega$ (eV) & 1.813 & 1.706 & 1.702 & --- & 1.633 & 1.635 & 1.630 \\
$\omega^{-}$ (eV) & 4.616 & 4.483 & 4.479 & --- & 4.396 & 4.398 & 4.393 \\
$\omega^{+}$ (eV) & 0.297 & 0.245 & 0.243 & --- & 0.211 & 0.211 & 0.210 \\
\hline
\multicolumn{8}{l}{\textbf{HF/6-311+G(3df,3pd)}} \\
\hline
IP (eV) & 9.463 & 9.504 & 9.506 & --- & 9.548 & 9.548 & 9.550 \\
EA (eV) & -1.425 & -1.634 & -1.641 & --- & -1.717 & -1.717 & -1.718 \\
$E_g$ (eV) & 10.888 & 11.138 & 11.147 & --- & 11.265 & 11.265 & 11.267 \\
$\eta$ (eV) & 5.444 & 5.569 & 5.573 & --- & 5.633 & 5.632 & 5.634 \\
$\sigma$ (eV$^{-1}$) & 0.092 & 0.090 & 0.090 & --- & 0.089 & 0.089 & 0.089 \\
$\chi$ (eV) & 4.019 & 3.935 & 3.933 & --- & 3.915 & 3.915 & 3.916 \\
$\omega$ (eV) & 1.484 & 1.390 & 1.388 & --- & 1.361 & 1.361 & 1.361 \\
$\omega^{-}$ (eV) & 4.174 & 4.053 & 4.051 & --- & 4.023 & 4.023 & 4.023 \\
$\omega^{+}$ (eV) & 0.155 & 0.119 & 0.118 & --- & 0.107 & 0.107 & 0.107 \\
\hline
\multicolumn{8}{l}{\textbf{HF/6-311G(3df,3pd)}} \\
\hline
IP (eV) & 9.338 & 9.384 & 9.387 & --- & 9.432 & 9.431 & 9.434 \\
EA (eV) & -3.061 & -3.347 & -3.356 & --- & -3.469 & -3.468 & -3.471 \\
$E_g$ (eV) & 12.398 & 12.731 & 12.743 & --- & 12.901 & 12.899 & 12.904 \\
$\eta$ (eV) & 6.199 & 6.365 & 6.371 & --- & 6.450 & 6.450 & 6.452 \\
$\sigma$ (eV$^{-1}$) & 0.081 & 0.079 & 0.078 & --- & 0.078 & 0.078 & 0.077 \\
$\chi$ (eV) & 3.138 & 3.019 & 3.015 & --- & 2.981 & 2.982 & 2.981 \\
$\omega$ (eV) & 0.794 & 0.716 & 0.713 & --- & 0.689 & 0.689 & 0.689 \\
$\omega^{-}$ (eV) & 3.139 & 3.021 & 3.017 & --- & 2.986 & 2.986 & 2.986 \\
$\omega^{+}$ (eV) & 0.000 & 0.002 & 0.002 & --- & 0.005 & 0.005 & 0.005 \\
\hline
\end{tabular}
\end{table*}
\begin{table*}[htbp]
\centering
\caption{Global reactivity descriptors for ascorbic acid computed using Koopmans theorem (continued).}
\label{tab:koopmans}
\setlength{\tabcolsep}{4.5pt}
\scriptsize
\begin{tabular}{lrrrrrrr}
\hline
Descriptor & Vacuum & Benzene & Toluene & Chlorobenzene & Methanol & Ethanol & Water \\
\hline
\multicolumn{8}{l}{\textbf{M062X/6-311++G(3df,3pd)}} \\
\hline
IP (eV) & 7.959 & 8.004 & 8.005 & 8.039 & 8.060 & 8.059 & 8.062 \\
EA (eV) & 0.508 & 0.281 & 0.272 & 0.176 & 0.132 & 0.134 & 0.128 \\
$E_g$ (eV) & 7.450 & 7.723 & 7.733 & 7.863 & 7.928 & 7.924 & 7.935 \\
$\eta$ (eV) & 3.725 & 3.862 & 3.866 & 3.932 & 3.964 & 3.962 & 3.967 \\
$\sigma$ (eV$^{-1}$) & 0.134 & 0.129 & 0.129 & 0.127 & 0.126 & 0.126 & 0.126 \\
$\chi$ (eV) & 4.233 & 4.142 & 4.139 & 4.107 & 4.096 & 4.097 & 4.095 \\
$\omega$ (eV) & 2.405 & 2.222 & 2.215 & 2.145 & 2.116 & 2.118 & 2.114 \\
$\omega^{-}$ (eV) & 4.988 & 4.775 & 4.768 & 4.690 & 4.660 & 4.661 & 4.657 \\
$\omega^{+}$ (eV) & 0.754 & 0.633 & 0.629 & 0.583 & 0.564 & 0.565 & 0.562 \\
\hline
\multicolumn{8}{l}{\textbf{M062X/6-311+G(3df,3pd)}} \\
\hline
IP (eV) & 7.959 & 8.004 & 8.005 & 8.039 & 8.060 & 8.059 & 8.062 \\
EA (eV) & 0.252 & 0.035 & 0.030 & 0.004 & 0.012 & 0.011 & 0.013 \\
$E_g$ (eV) & 7.707 & 7.969 & 7.975 & 8.034 & 8.048 & 8.048 & 8.049 \\
$\eta$ (eV) & 3.853 & 3.984 & 3.988 & 4.017 & 4.024 & 4.024 & 4.025 \\
$\sigma$ (eV$^{-1}$) & 0.130 & 0.125 & 0.125 & 0.124 & 0.124 & 0.124 & 0.124 \\
$\chi$ (eV) & 4.105 & 4.019 & 4.018 & 4.022 & 4.036 & 4.035 & 4.038 \\
$\omega$ (eV) & 2.187 & 2.027 & 2.024 & 2.013 & 2.024 & 2.023 & 2.026 \\
$\omega^{-}$ (eV) & 4.721 & 4.535 & 4.531 & 4.526 & 4.545 & 4.543 & 4.548 \\
$\omega^{+}$ (eV) & 0.616 & 0.516 & 0.514 & 0.504 & 0.509 & 0.508 & 0.510 \\
\hline
\multicolumn{8}{l}{\textbf{M062X/6-311G(3df,3pd)}} \\
\hline
IP (eV) & 7.799 & 7.853 & 7.855 & 7.891 & 7.914 & 7.913 & 7.918 \\
EA (eV) & -0.347 & -0.291 & -0.288 & -0.249 & -0.223 & -0.224 & -0.220 \\
$E_g$ (eV) & 8.146 & 8.144 & 8.143 & 8.141 & 8.137 & 8.138 & 8.138 \\
$\eta$ (eV) & 4.073 & 4.072 & 4.072 & 4.070 & 4.069 & 4.069 & 4.069 \\
$\sigma$ (eV$^{-1}$) & 0.123 & 0.123 & 0.123 & 0.123 & 0.123 & 0.123 & 0.123 \\
$\chi$ (eV) & 3.726 & 3.781 & 3.783 & 3.821 & 3.846 & 3.844 & 3.849 \\
$\omega$ (eV) & 1.704 & 1.755 & 1.758 & 1.794 & 1.817 & 1.816 & 1.821 \\
$\omega^{-}$ (eV) & 4.076 & 4.155 & 4.158 & 4.213 & 4.249 & 4.247 & 4.254 \\
$\omega^{+}$ (eV) & 0.350 & 0.374 & 0.375 & 0.392 & 0.403 & 0.403 & 0.405 \\
\hline
\end{tabular}
\end{table*}

\end{document}